%% file: mag-stream.tex
\documentclass{aastex63}
\usepackage{gensymb}
\usepackage{graphicx}
\usepackage{longtable}
\usepackage{amssymb}
\usepackage[caption=false]{subfig}

\newcommand{\hi}{\ion{H}{1}}     \newcommand{\cf}{\ion{C}{4}} 
\newcommand{\sif}{\ion{Si}{4}}   \newcommand{\sit}{\ion{Si}{3}}
\newcommand{\cw}{\ion{C}{2}}     \newcommand{\siw}{\ion{Si}{2}} 
\newcommand{\hst}{\emph{HST}}    \newcommand{\kms}{\,km\,s$^{-1}$}
\newcommand{\tm}{\tablenotemark} \newcommand{\tn}{\tablenotetext}
\newcommand{\ha}{H$\alpha$}
\newcommand{\nsamp}{31} 
\newcommand{\nms}{21} \newcommand{\nla}{10} 
\newcommand{\nlow}{7} 
\defcitealias{Fo14}{F14}

\graphicspath{{./}{figures/}}

\received{February 4, 2020}
\revised{May 11, 2020}
\accepted{May 11, 2020}
\submitjournal{ApJ}

\shorttitle{Magellanic Stream Kinematics}
\shortauthors{Fox, Frazer, Bland-Hawthorn et al.}

\begin{document}

\title{Kinematics of the Magellanic Stream and Implications for its
Ionization\footnote{Based on observations made with the NASA/ESA Hubble Space Telescope, obtained from the Data Archive at the Space Telescope Science Institute, which is operated by the Association of Universities for Research in Astronomy, Inc., under NASA contract NAS5-26555. These observations are associated with program 11541, 11632, 11524, 11585, 11598, 11686, 12025, 12029, 12038, 12212, 12248, 12275, 12569, 12604, and 14687.}}

\correspondingauthor{Andrew Fox}
\email{afox@stsci.edu}

\author[0000-0003-0724-4115]{Andrew J. Fox}
\affil{AURA for ESA, Space Telescope Science Institute, 3700 San Martin Drive, Baltimore, MD 21218}

\author[0000-0001-9352-9023]{Elaine M. Frazer}
\affil{Space Telescope Science Institute, 3700 San Martin Drive, Baltimore, MD 21218}

\author[0000-0001-7516-4016]{Joss Bland-Hawthorn}
\affil{Sydney Institute for Astronomy, School of Physics A28, University of Sydney, NSW 2006, Australia}
\affil{ARC Centre of Excellence for All Sky Astrophysics in Three Dimensions (ASTRO-3D), Sydney, Australia}

\author[0000-0002-0507-7096]{Bart P. Wakker}
\affil{Department of Astronomy, University of Wisconsin-Madison, 475 N. Charter St., Madison, WI 53706}

\author[0000-0001-5817-0932]{Kathleen A. Barger}
\affil{Department of Physics and Astronomy, Texas Christian University, 
TCU Box 298840, Fort Worth, TX 76129}

\author[0000-0002-1188-1435]{Philipp Richter}
\affil{Institut f\"ur Physik und Astronomie, Universit\"at Potsdam, Haus 28, Karl-Liebknecht-Str. 24/25, D-14476, Potsdam, Germany}

\begin{abstract}
The Magellanic Stream and the Leading Arm form a massive, 
filamentary system of gas clouds surrounding the Large and 
Small Magellanic Clouds. Here we present a new component-level
analysis of their ultraviolet (UV) kinematic properties 
using a sample of \nsamp\ sightlines through the Magellanic 
System observed with the {\it Hubble Space Telescope}/Cosmic 
Origins Spectrograph. Using Voigt profile fits to UV 
metal-line absorption, we quantify the kinematic differences 
between the low-ion (\siw\ and \cw), intermediate-ion (\sit), 
and high-ion (\sif\ and \cf) absorption lines and 
compare the kinematics between the Stream and Leading Arm.
We find that the Stream shows generally simple, single-phase kinematics,
with statistically indistinguishable $b$-value distributions
for the low-, intermediate-, and high-ion components,
all dominated by narrow ($b\la25$\kms) components that are well aligned in velocity. 
In contrast, we find tentative evidence that the Leading Arm shows complex, 
multi-phase kinematics, with broader high ions than low ions. 
These results suggest that the Stream is photoionized up to \cf\ by a 
hard ionizing radiation field. 
This can be naturally explained by the Seyfert-flare model 
of Bland-Hawthorn et al. (2013, 2019), in which a burst of 
ionizing radiation from the Galactic Center photoionized the Stream 
as it passed below the south Galactic pole.
The Seyfert flare is the only known source of radiation that is both
powerful enough to explain the H$\alpha$ intensity of the Stream and hard enough
to photoionize \sif\ and \cf\ to the observed levels.
The flare's timescale of a few Myr suggests it is the same event that created
the giant X-ray/$\gamma$-ray Fermi Bubbles at the Galactic Center.
\end{abstract}


\section{Introduction}\label{sec:intro}

The Milky Way provides an unmatched opportunity to dissect the gas flows around a 
star-forming spiral galaxy. By combining radio 21\,cm emission measurements of 
neutral gas with ultraviolet (UV) absorption measurements of ionized gas and other 
tracers, we can build an all-sky picture of the multi-phase halo gas and conduct a
spatially-resolved analysis of the baryon cycle. Furthermore, we can compare the 
location of halo clouds with the positions of dwarf satellites and known structures 
in the Galactic disk, such as spiral arms and the giant Fermi Bubbles at the 
Galactic Center (GC). This additional knowledge makes the Galactic halo an ideal 
location for studying gas flows and their role in galaxy evolution.

By a considerable margin, the largest and most massive gaseous structure in the 
Galactic halo is the Magellanic Stream (hereafter the Stream), which together with 
the Leading Arm (LA) extends over 200 degrees across the sky \citep{Ni10}. The 
Stream is an interwoven tail of filaments stripped out of the Magellanic Clouds 
and trailing them in their orbit around the Milky Way 
\citep[see][]{Ma74, Pu03a, Br05, Ni08, DF16}. 
Thought to be created by a combination of tidal forces, ram pressure, and halo 
interactions, the Stream and Leading Arm form a benchmark for dynamical 
models of the Magellanic System 
\citep[e.g.][]{MD94, Ma05, Be10, Gu14, Ha15, Pa18, Wa19, Lu20}
and a probe of many astrophysical processes. 

UV absorption-line studies with the spectrographs on board the 
{\it Hubble Space Telescope} ({\it HST}) have led to considerable progress 
in our knowledge of the Stream's physical and chemical properties 
\citep{Lu94, Gi00, Fo10, Fo13, Fo14, Ri13, Ku15, Ho17}. 
These studies indicate that the Stream has a dual origin 
\citep[as also indicated by the \hi\ kinematics;][]{Ni08}, 
with one filament showing LMC-like kinematics and chemical abundances and the 
other showing SMC-like kinematics and abundances \citep{Fo13, Ri13}. 
In the LA, only gas with SMC-like abundance patterns has been observed 
\citep{Fo18, Ri18}, though with considerable variation in metallicity 
between different regions, implying a complex creation history.

Despite this progress, the \emph{kinematics} of the UV metal-line absorption 
from the Magellanic Stream and LA have never been addressed in detail. These 
kinematics contain important information on the phase structure, temperature, 
and non-thermal motions of the gas in the Stream, and therefore provide clues 
to its origin and history. In this paper we present the first detailed UV 
kinematic analysis of the Magellanic System. Our study is partly motivated 
by the results of \citet{BH13, BH19}, who discuss evidence from H$\alpha$ and 
UV studies for a GC flare several Myr ago. Such a ``Seyfert flare'' would have 
flash-ionized the Stream in the Galactic polar regions directly below the GC, 
where the flux of escaping ionizing radiation is highest, but not the Leading Arm, 
which is located in a lower-latitude region shielded from the ionization cone. 
In the Seyfert-flare scenario, the high ions in the polar regions of the Stream 
are photoionized by the escaping ionizing radiation, and so we expect them to 
show similar velocity centroids and line widths as the low ions (a single-phase model). 
On the other hand, if the high ions in the Stream are produced by collisional 
ionization via processes such as shocks, conductive interfaces, or turbulent 
mixing layers, then they should show different kinematics than the low ions 
(a multi-phase model). Indeed, these collisional processes are often invoked for other 
high-velocity clouds (HVCs) in the Galactic halo \citep{Se03, Fo04, Fo05, Ga05, Co05}.

A comparative kinematic study of the high and low ions and their variation across 
the Stream therefore has the potential to directly address whether an ionizing 
flare from the GC occurred. 
This question is given additional relevance by the compelling evidence that now exists 
for recent (Myr-timescale) activity at the GC, including the giant $\gamma$-ray 
emitting Fermi Bubbles \citep{Su10, Do10, Ac14}, their counterparts in X-rays 
\citep{BH03, MB16}, microwaves \citep{Fi04}, and polarized radio emission 
\citep{Ca13}, the presence of smaller-scale ($\approx$400\,pc) radio lobes 
within the Bubbles \citep{Hy19}, and the tentative detection of radio jets
\citep{SF12}, though see \citet{Ac14}. Furthermore, AGN activity on 
$\sim$Myr timescales has recently been inferred from X-ray studies of M31 
\citep{Zh19}, suggesting that such processes are common in MW-mass spiral galaxies. 
Our UV kinematic analysis of the Magellanic Stream brings a new method for 
gauging the GC activity: assessing its impact on the extended gaseous environment.

This paper is structured as follows. In \autoref{sec:data} we describe the 
creation of our sample and the data-modeling procedures 
and present our new Voigt-profile fits to the \hst/Cosmic Origins Spectrograph (COS) 
spectra. We analyze the UV 
kinematics of the Stream and Leading Arm in \autoref{sec:kinematics}. We then 
present a discussion in \autoref{sec:discussion}, where we interpret the 
kinematics in light of origin models for the Stream and the Leading Arm. 
We summarize our conclusions in \autoref{sec:conclusions}.

\section{Observations and Data Handling}\label{sec:data}

\subsection{The Sample}\label{subsec:sample}
To form our sample, we began with the 69 \hst/COS Magellanic sightlines 
compiled in \citet[][hereafter F14]{Fo14}, and added one 
recently-observed LA sightline from \citet{Fo18}. The \citetalias{Fo14} 
sample was defined to include AGN that: 
(1) lie within 30\degree\ from the 21\,cm emission of the Magellanic System, 
as defined using the \citet{Mo00} \hi\ contours; 
(2) lie in regions where the Magellanic absorption is at 
$|v_{\rm LSR}|>100$ km s$^{-1}$,
to avoid blending with interstellar absorption from the Milky Way, and 
(3) have COS/FUV data with the G130M grating with a signal-to-noise (S/N) 
ratio of $\gtrsim5$ per resolution element at 1250\AA. The G130M grating spans
the wavelength range $\approx$1150--1450\AA, covering \cw\ $\lambda$ 1334, \siw\
$\lambda$1260,1190,1193, \sit\ $\lambda$1206, and \sif\ $\lambda$1393,1402. 
We also include G160M observations if they exist, covering the wavelength range
$\approx$1405--1775\AA\ and so including the \cf\ doublet $\lambda\lambda$1548,1550. 
The design and performance of the COS spectrograph is described in \citet{Gr12}.

\begin{figure*}[!ht]
\centering 
\includegraphics[width=0.8\textwidth]{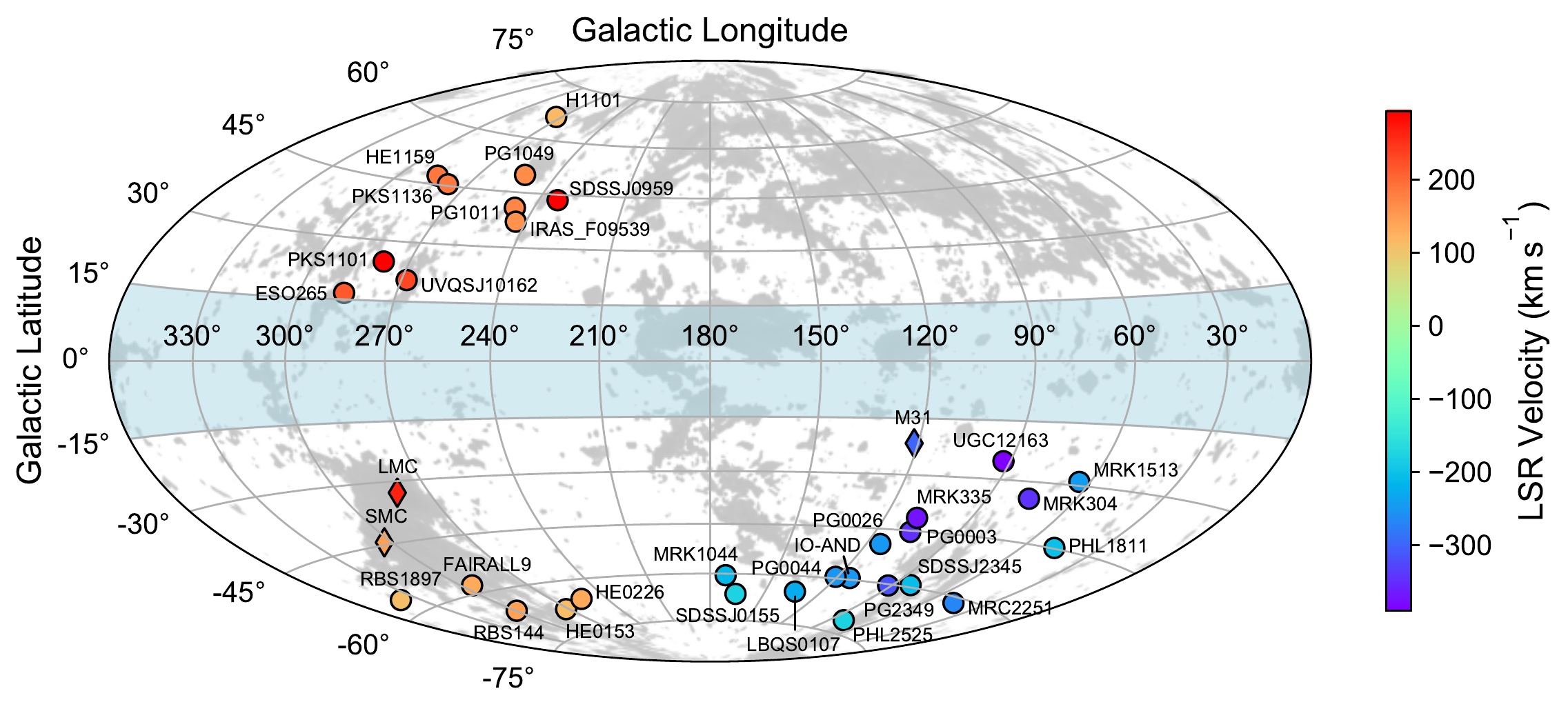} 
\caption{All-sky map in Galactic coordinates showing the location of our COS AGN sightlines with respect to the Stream and Leading Arm. The symbols are color-coded by the LSR velocity of absorption. The grayscale shows the all-sky 21\,cm HVC map from \citet{We18} based on HI4PI data \citep{Hi16}. The locations of the LMC, SMC, and M31 are marked and the Galactic disk is shaded.}
\label{fig:map1}
\end{figure*}

We then down-selected the sample to only include sightlines behind the Stream and LA, 
and not the additional sightlines in \citetalias{Fo14} passing through the 
Magellanic Bridge, the LMC halo, and Compact HVCs. 
This choice was made to keep the analysis focused on two specific spatial regions (Stream and LA). 
We also removed \nlow\ low-S/N sightlines (S/N$<$10 per resolution element), 
since any kinematic information (particularly $b$-values)
extracted from these data is unreliable.
These are the sightlines toward \object{LBQS0107--0233}, \object{RX\_J0209.5--0438}, 
\object{SDSS\,J001224.01--102226.5}, \object{SDSS\,J225738.20+134045.0}, 
\object{SDSS\,J094331.60+053131.0}, \object{ESO\,267--G13}, and \object{NGC\,3125}.
Together, these steps led to a final sample of \nsamp\ Magellanic directions 
(\nms\ MS and \nla\ LA). The data can be accessed at MAST via the following link:  \url{https://doi.org/10.17909/t9-94ka-p284.}

In our earlier work \citepalias{Fo14} we drew a distinction between ``On-Stream" 
and ``Off-Stream" sightlines, 
and between ``On-Leading Arm" and ``Off-Leading Arm" sightlines, based on whether \hi\ 
21\,cm emission is detected from the Magellanic component in each direction. 
The reason for the On-Off distinction is that a considerable fraction of the total cross-section of 
the Magellanic System 
has an \hi\ column density too low to be detected in 21 cm, 
and can only be detected in UV absorption \citep[][F14]{Se03} or
H$\alpha$ emission \citep{WW96, Pu03b, Ba17}.
However, in order to preserve a sample size large enough to draw statistically 
significant conclusions, we do not make the On-Off distinction in this paper, instead 
leaving our sample as \nms\ MS directions and \nla\ LA directions. 
This means we are covering both On-Stream and Off-Stream directions.
The sky distribution of our sample is shown in \autoref{fig:map1}.

Finally, because our sample is defined by an extended spatial region on the sky,
some components might have alternative, non-Magellanic origins.
A few specific cases are worthy of mention:\\
(i) Several of our LA components 
were previously cataloged as tracing other HVC Complexes. These are the components 
at 80 and 130\kms\ toward \object{PG1011-040}, which lies behind Complex 
WA/WB \citep{WW91}, and the components at 140 and 190\kms\ toward 
\object{ESO265-G25}, which lies behind Complex WD \citep{WW91}. 
The association of \object{H1101-232} and \object{PG1049-055} with the LA is also 
unconfirmed, because they lie off the side of the main \hi\ regions.
Despite these complications, 
we retain these absorbers in the LA sample for two reasons. 
First, they have high positive LSR velocities broadly consistent with the LA, 
and so it is possible that they represent detached fragments of the LA 
regardless of their historical classifications.
Second, they are located in 
the LA region of the halo and are thus exposed to a similar gaseous environment and a 
similar ionizing radiation field. \\
(ii) Two absorbers detected in outer-Stream directions (the high-negative-velocity components
toward \object{IO\,And} and \object{Mrk 335}) 
might be associated with the halo of M31, since the velocity fields of the Stream 
and M31 overlap \citep{Le15, Le20}.\\
Despite the presence of these few ambiguous cases, the 
good general agreement between the kinematics of our 
UV sample with the 21\,cm kinematics of the Magellanic System \citepalias{Fo14} supports 
our treatment of the sample as Magellanic,
and suggests that the number of non-Magellanic components is small. 

\subsection{Voigt-Component Fitting} \label{subsec:fitting}

\begin{figure*}[!ht]
    \includegraphics[width=0.30\textwidth]{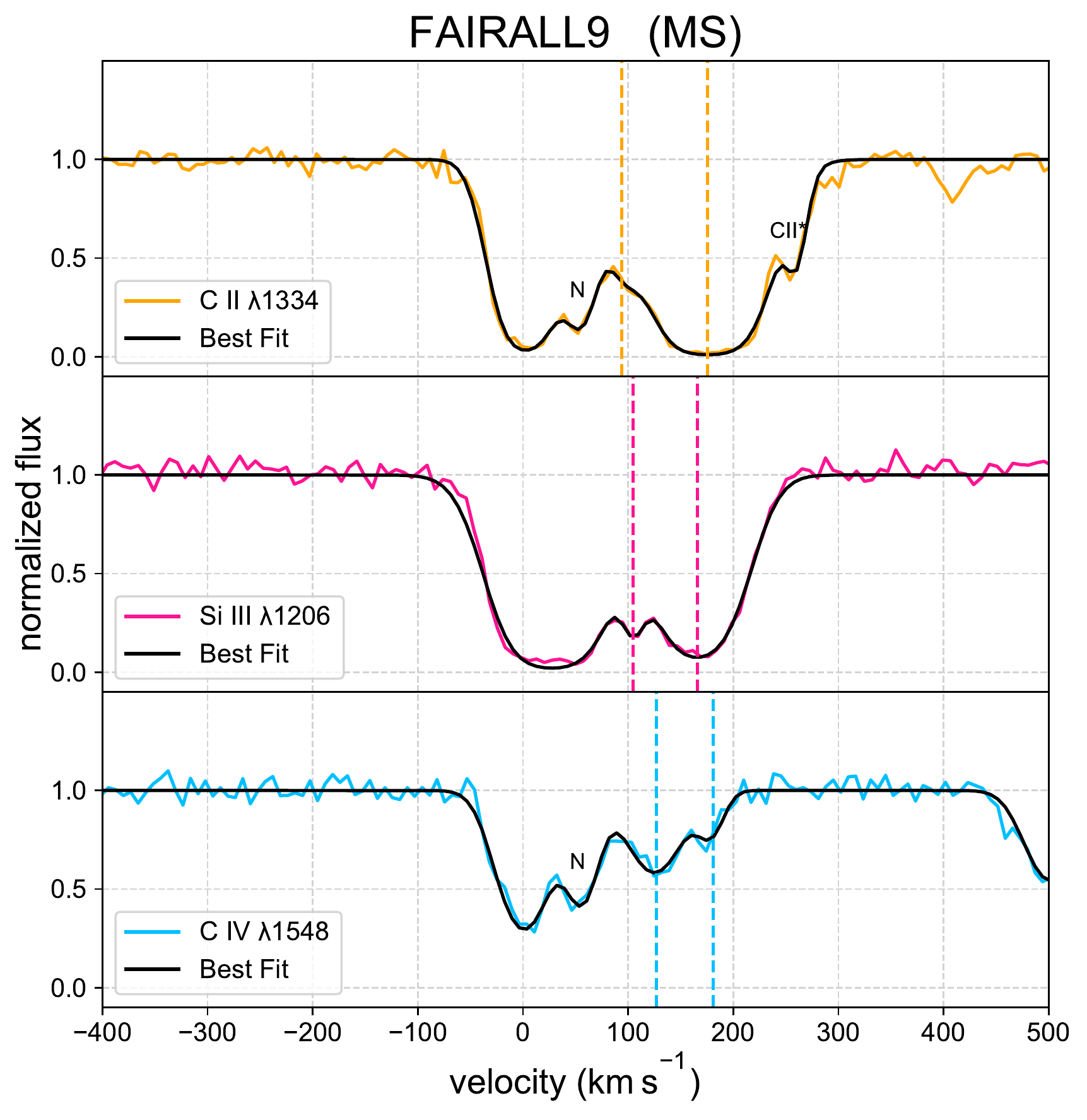}
    \includegraphics[width=0.30\textwidth]{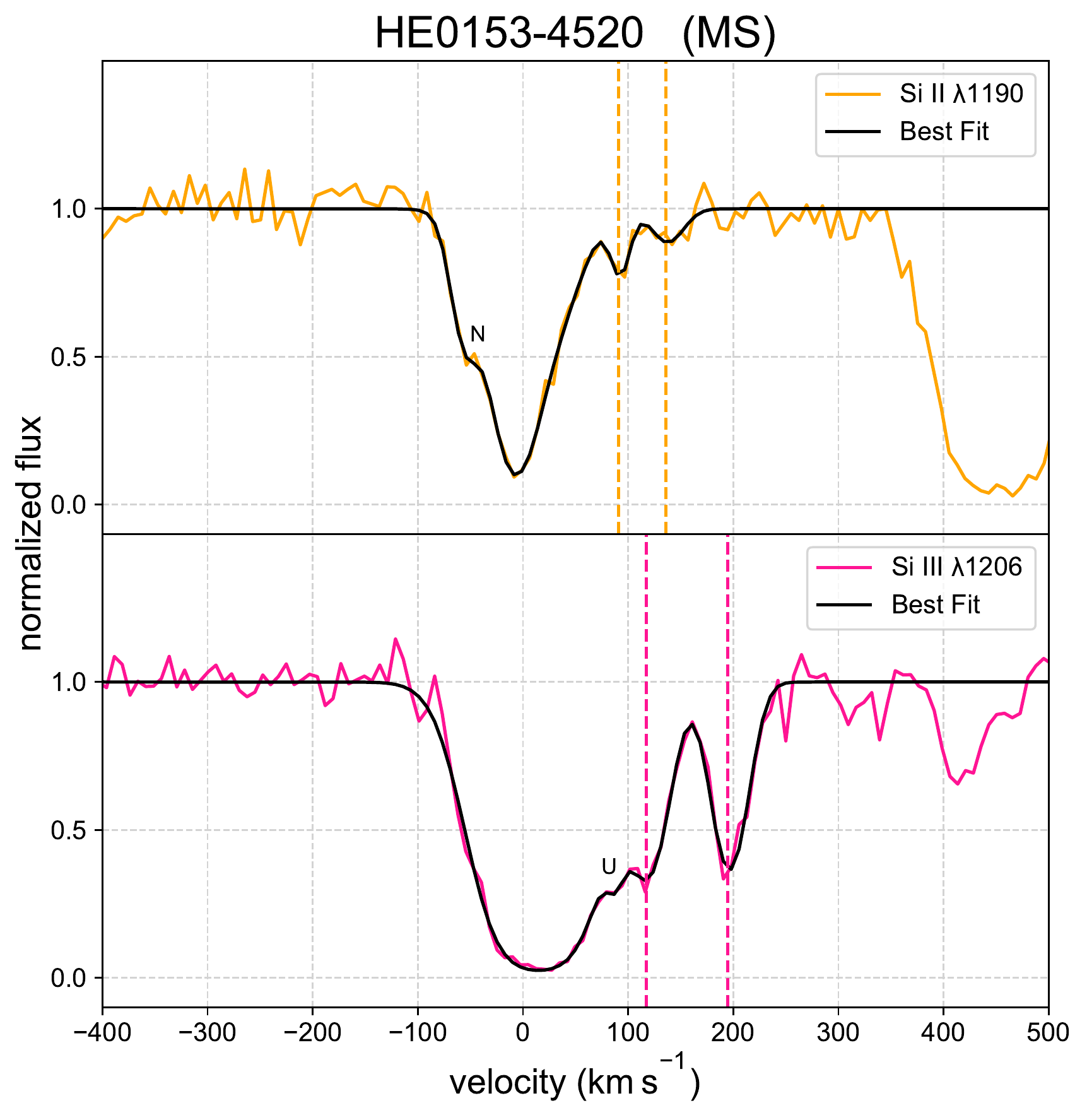}
    \includegraphics[width=0.30\textwidth]{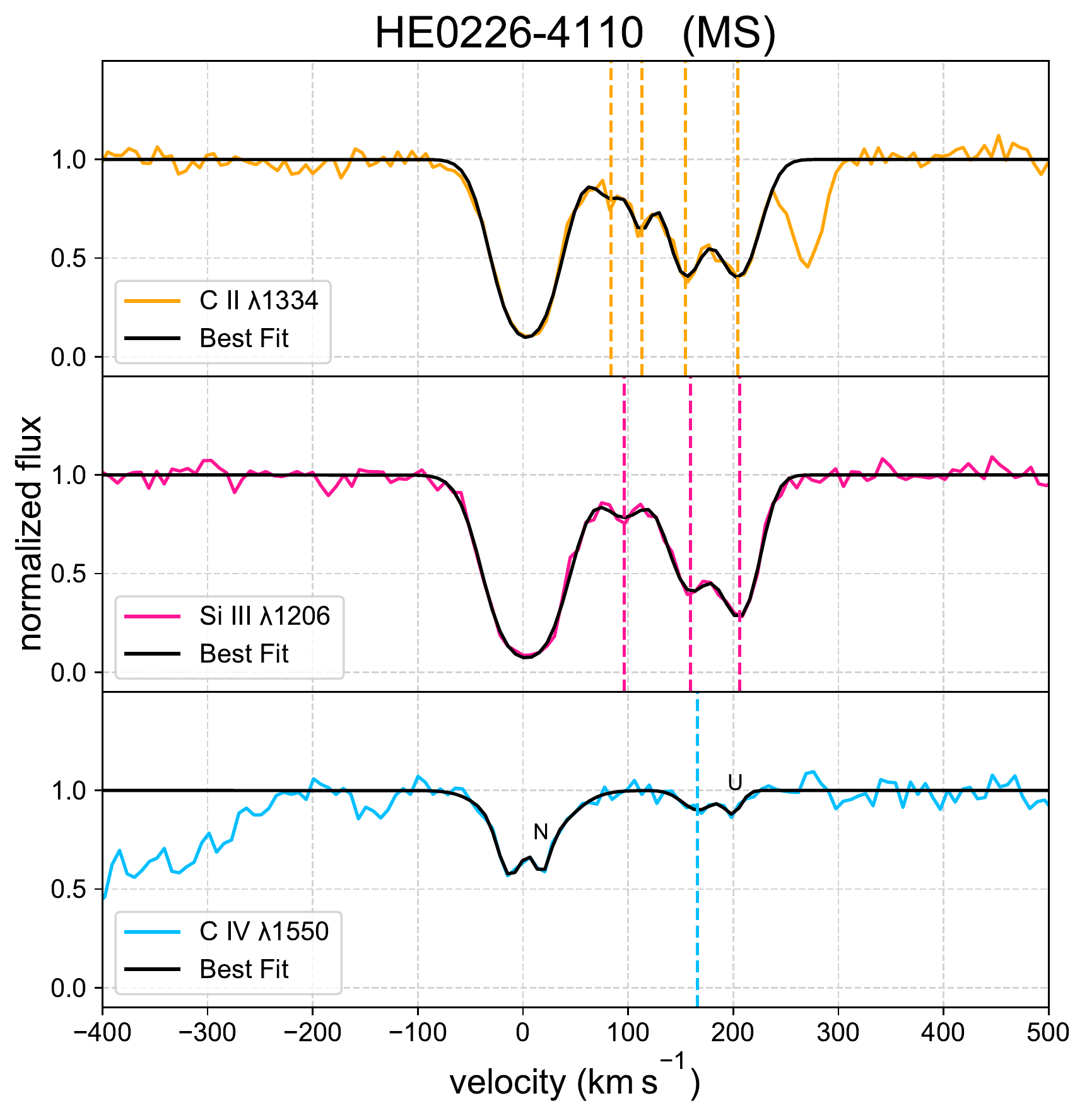}
    
    \includegraphics[width=0.30\textwidth]{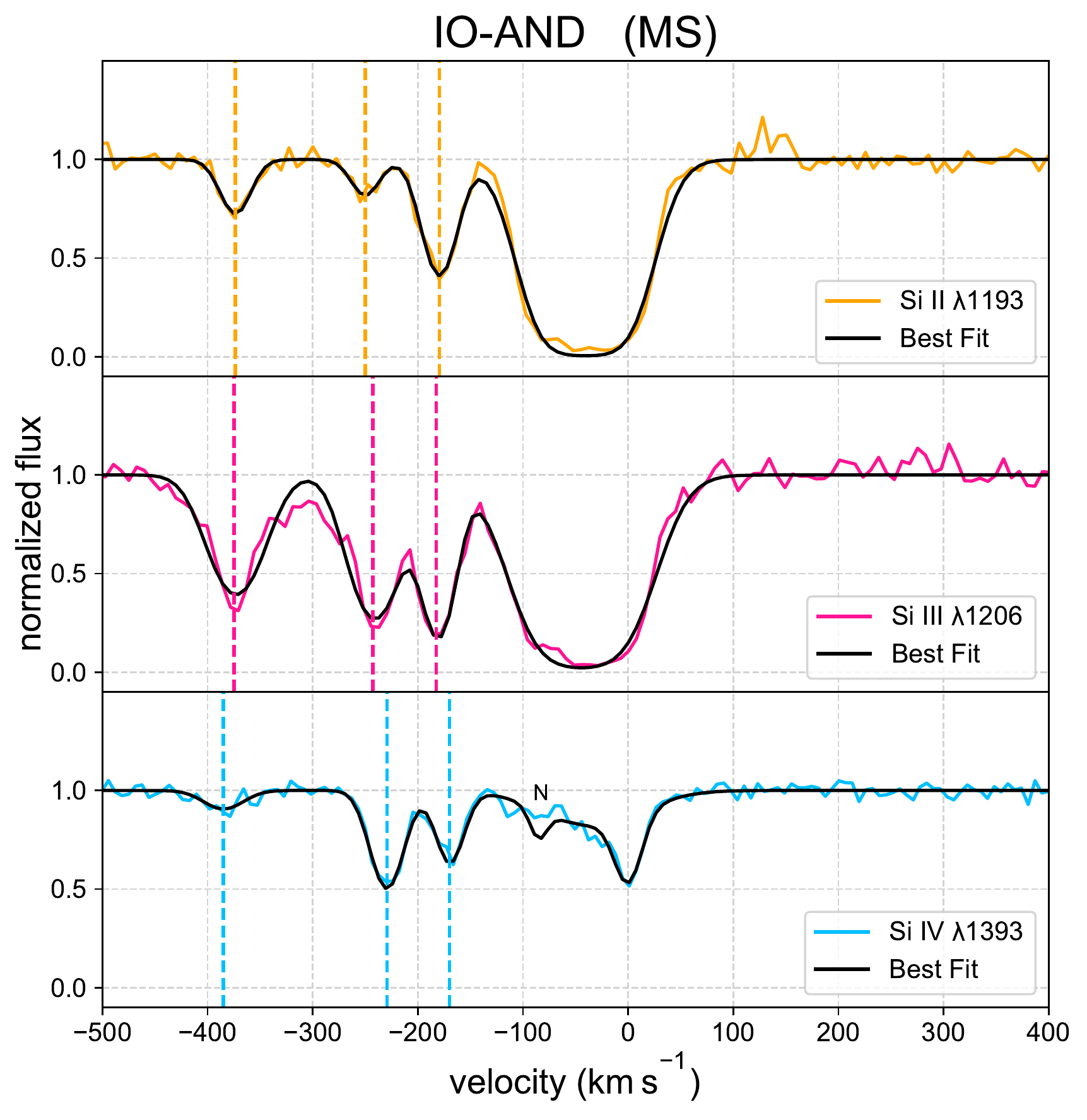}
    \includegraphics[width=0.30\textwidth]{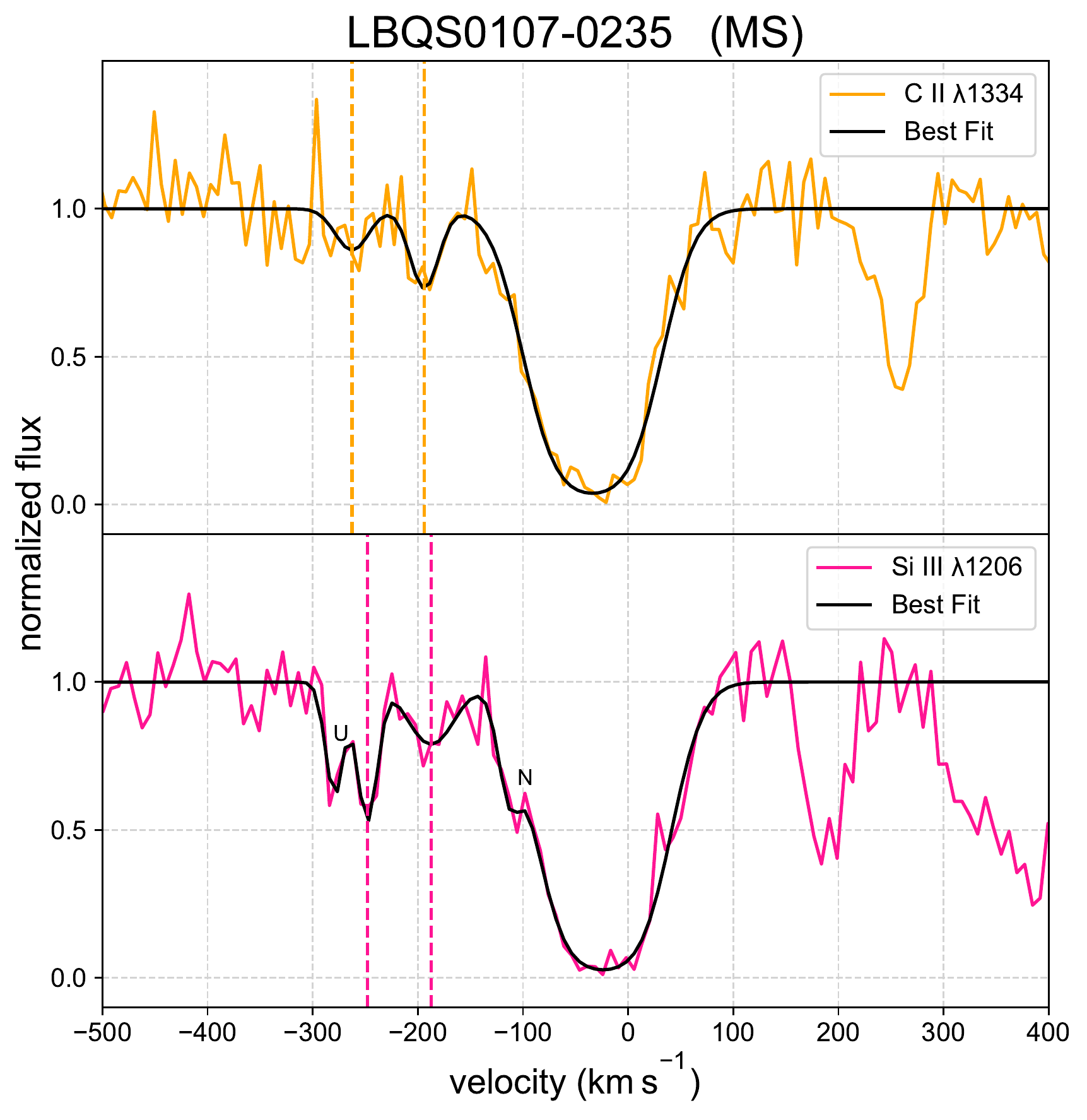}
    \includegraphics[width=0.30\textwidth]{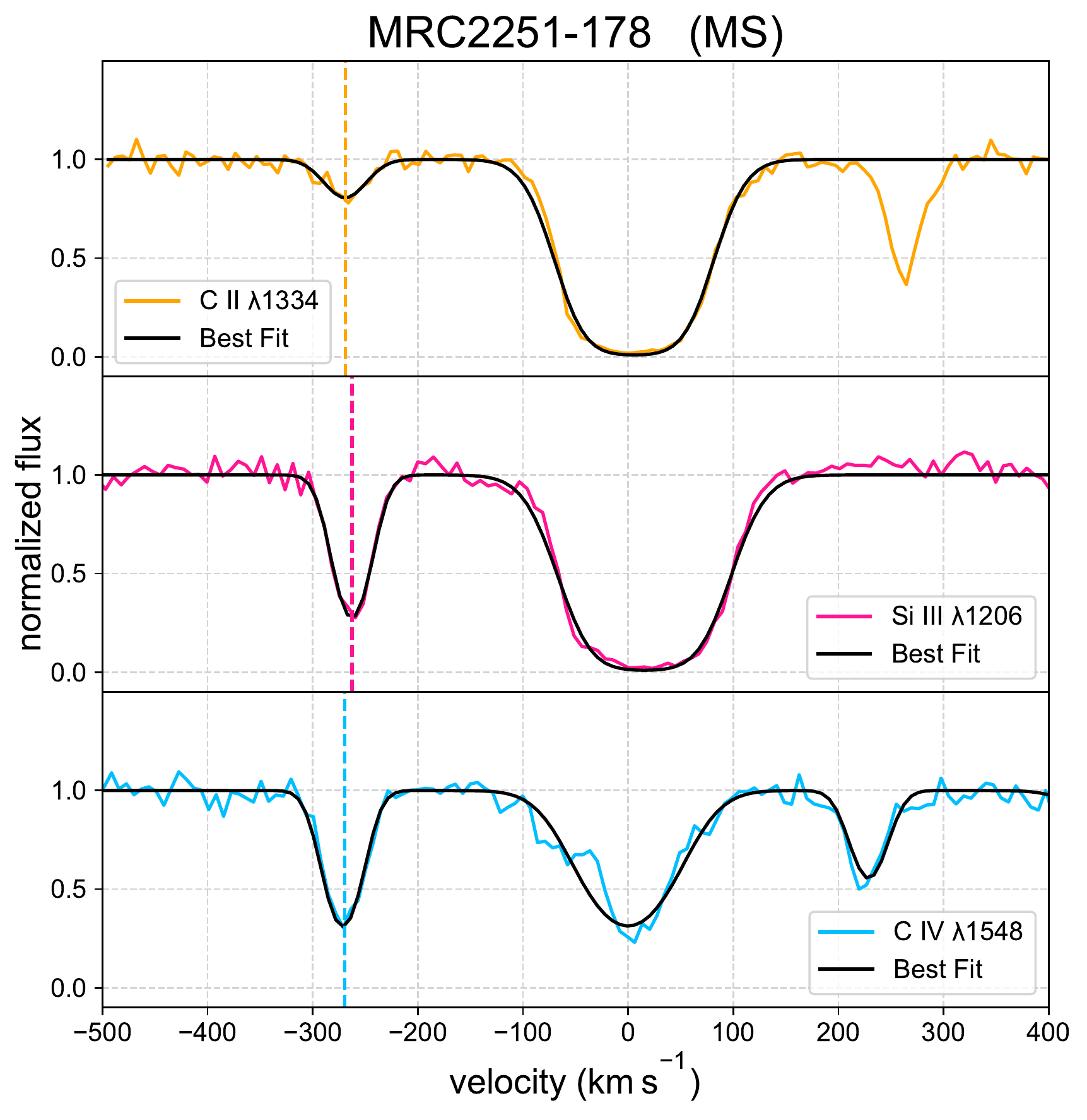}
    
    \includegraphics[width=0.30\textwidth]{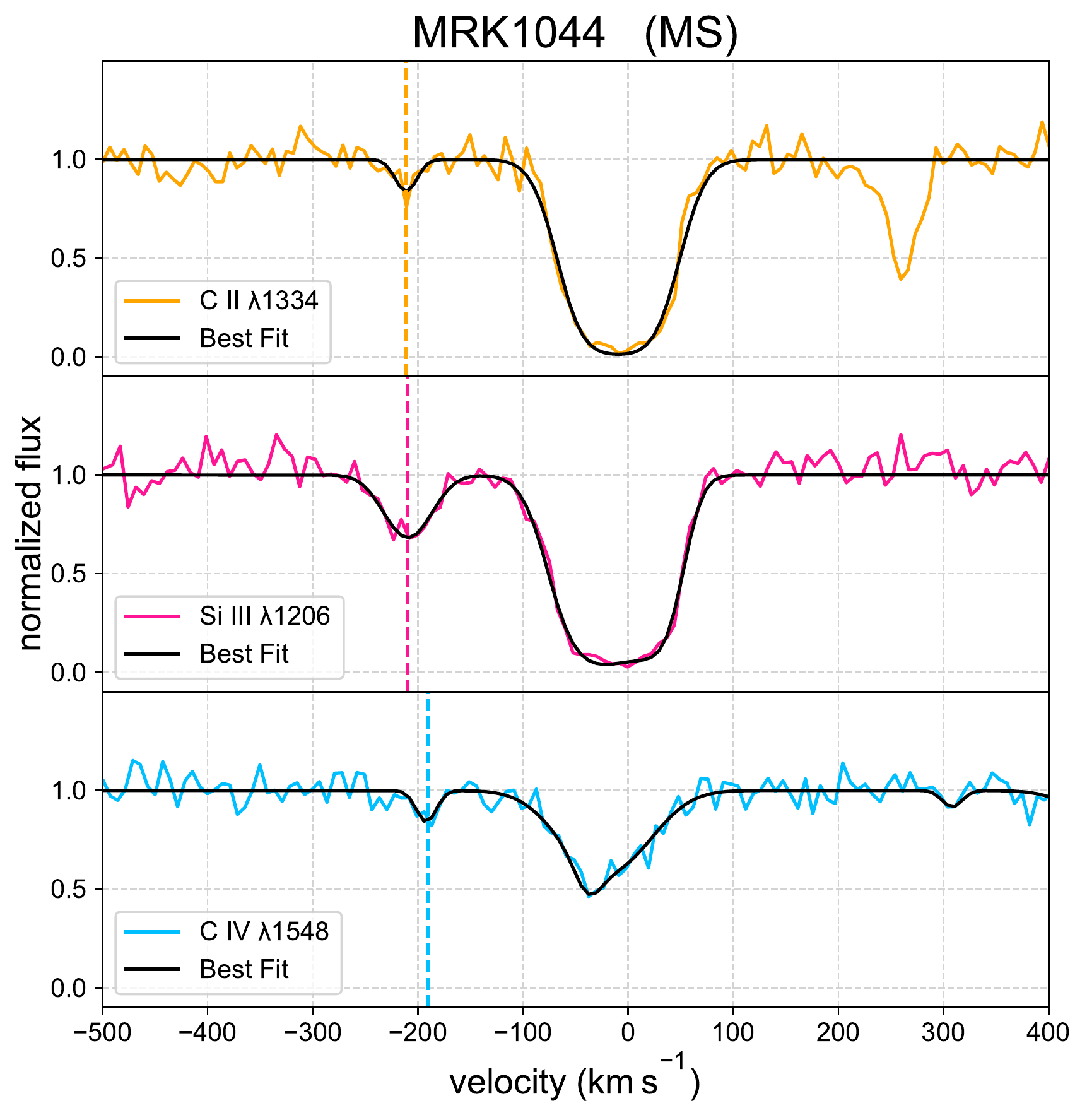}
    \includegraphics[width=0.30\textwidth]{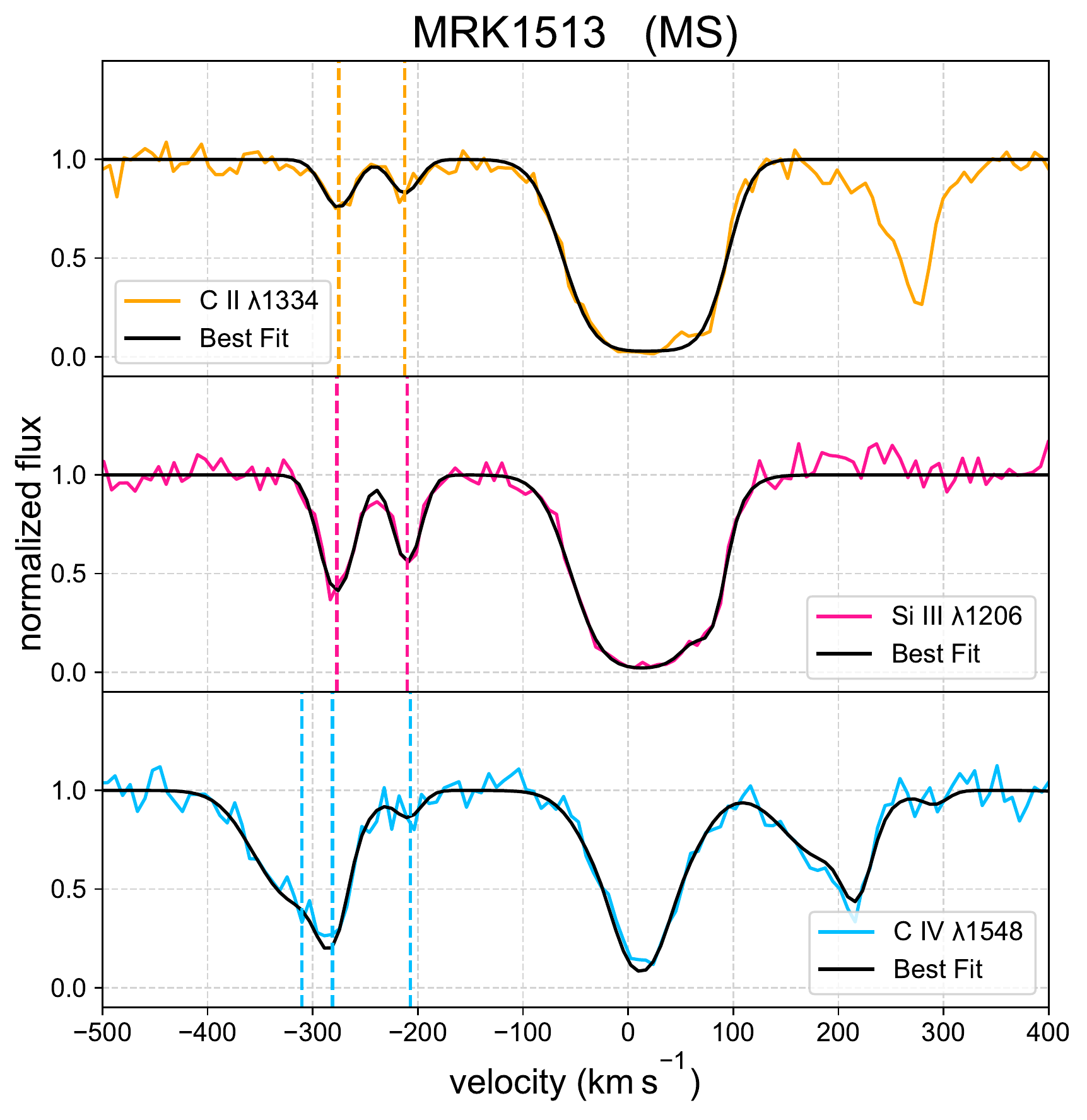}
    \includegraphics[width=0.30\textwidth]{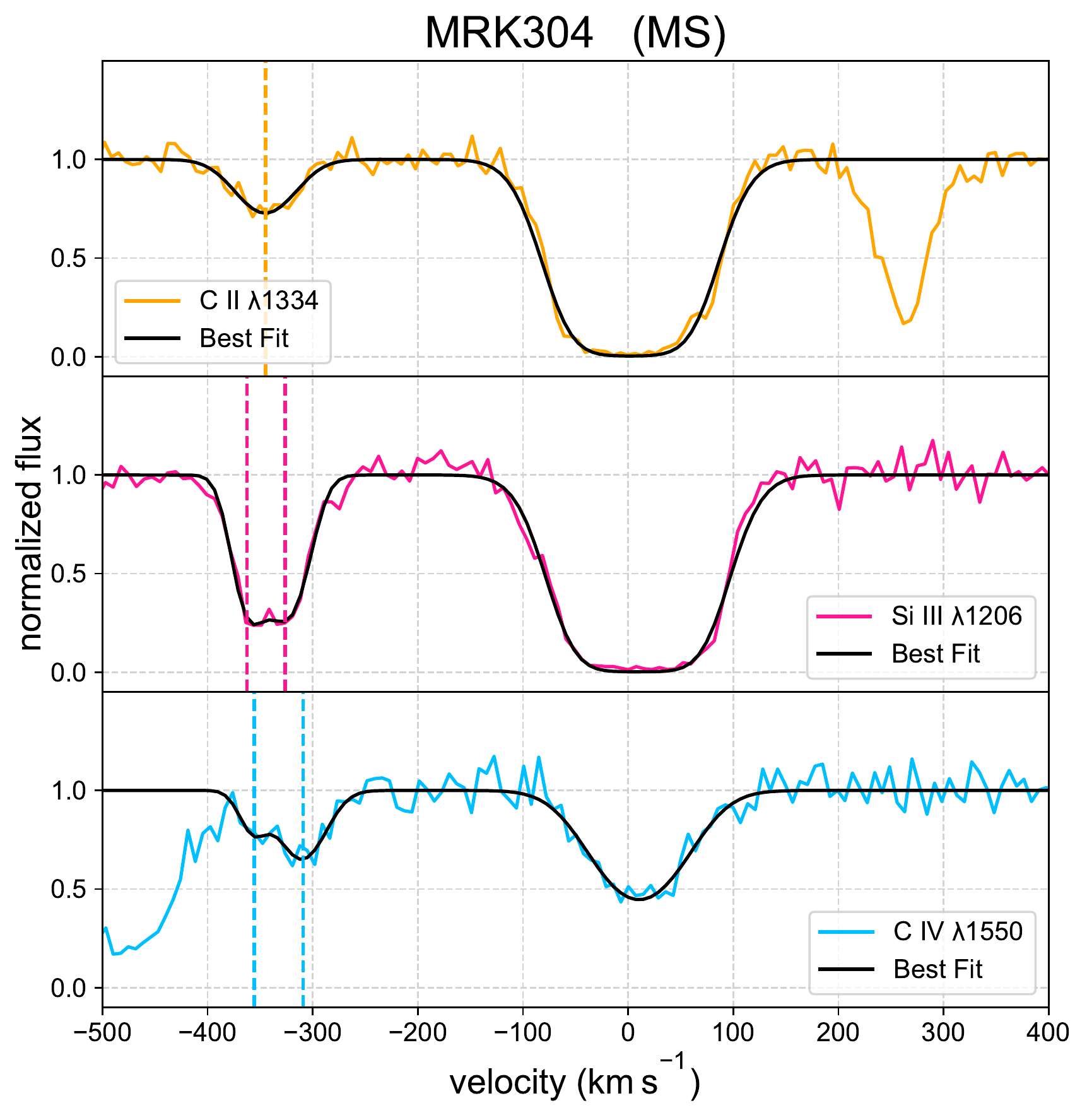}
    
    \includegraphics[width=0.30\textwidth]{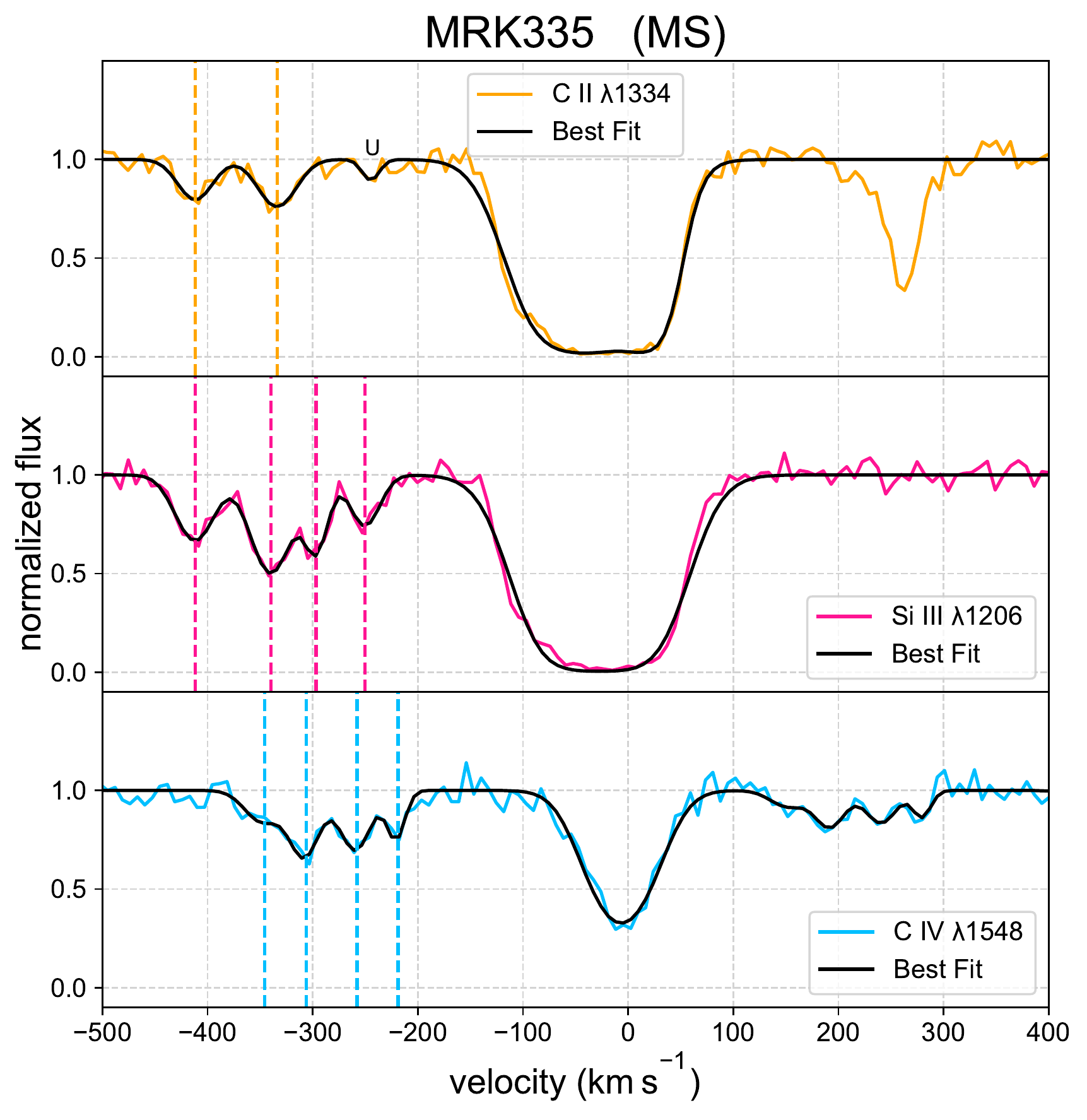}
    \includegraphics[width=0.30\textwidth]{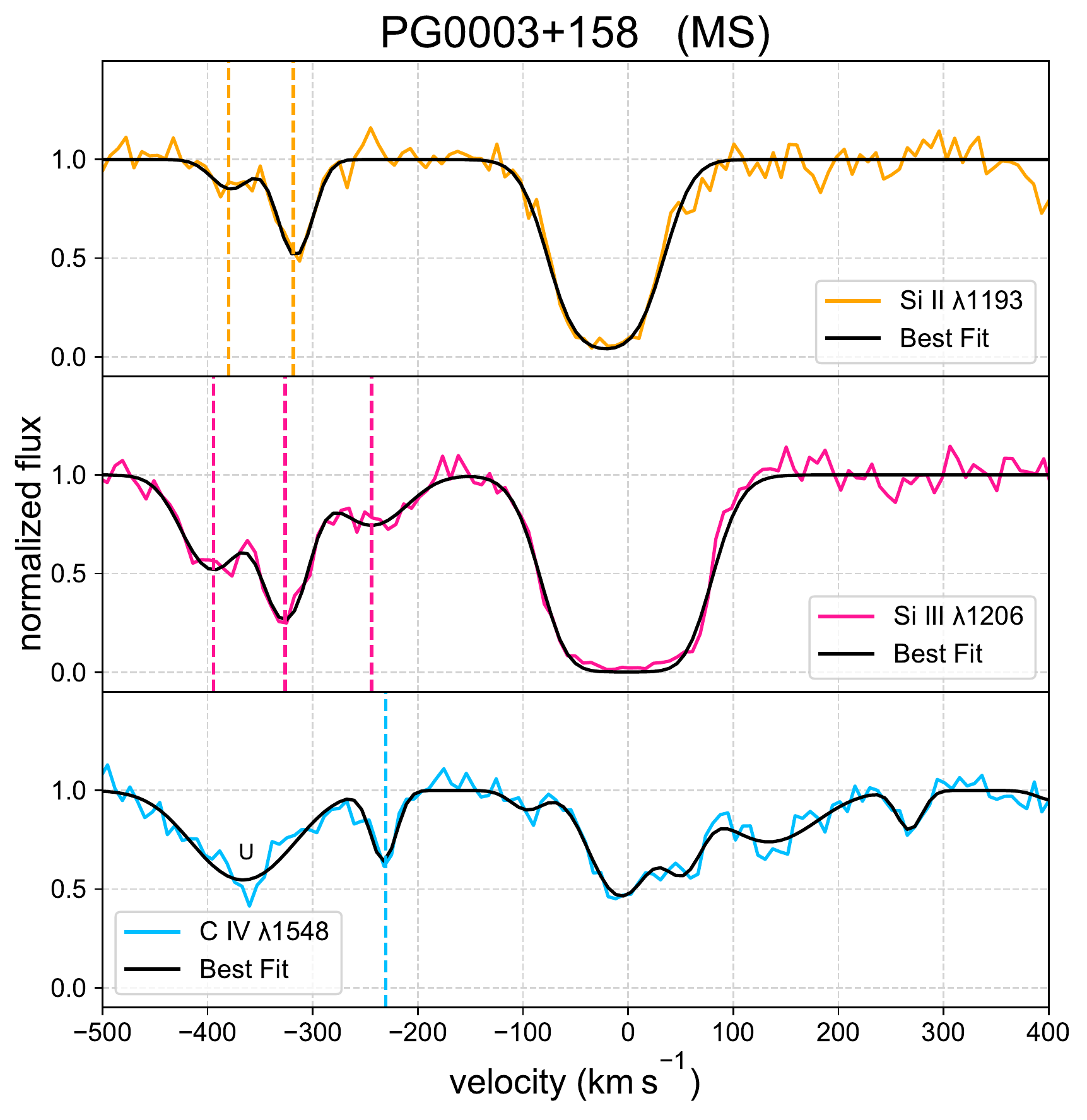}
    \includegraphics[width=0.30\textwidth]{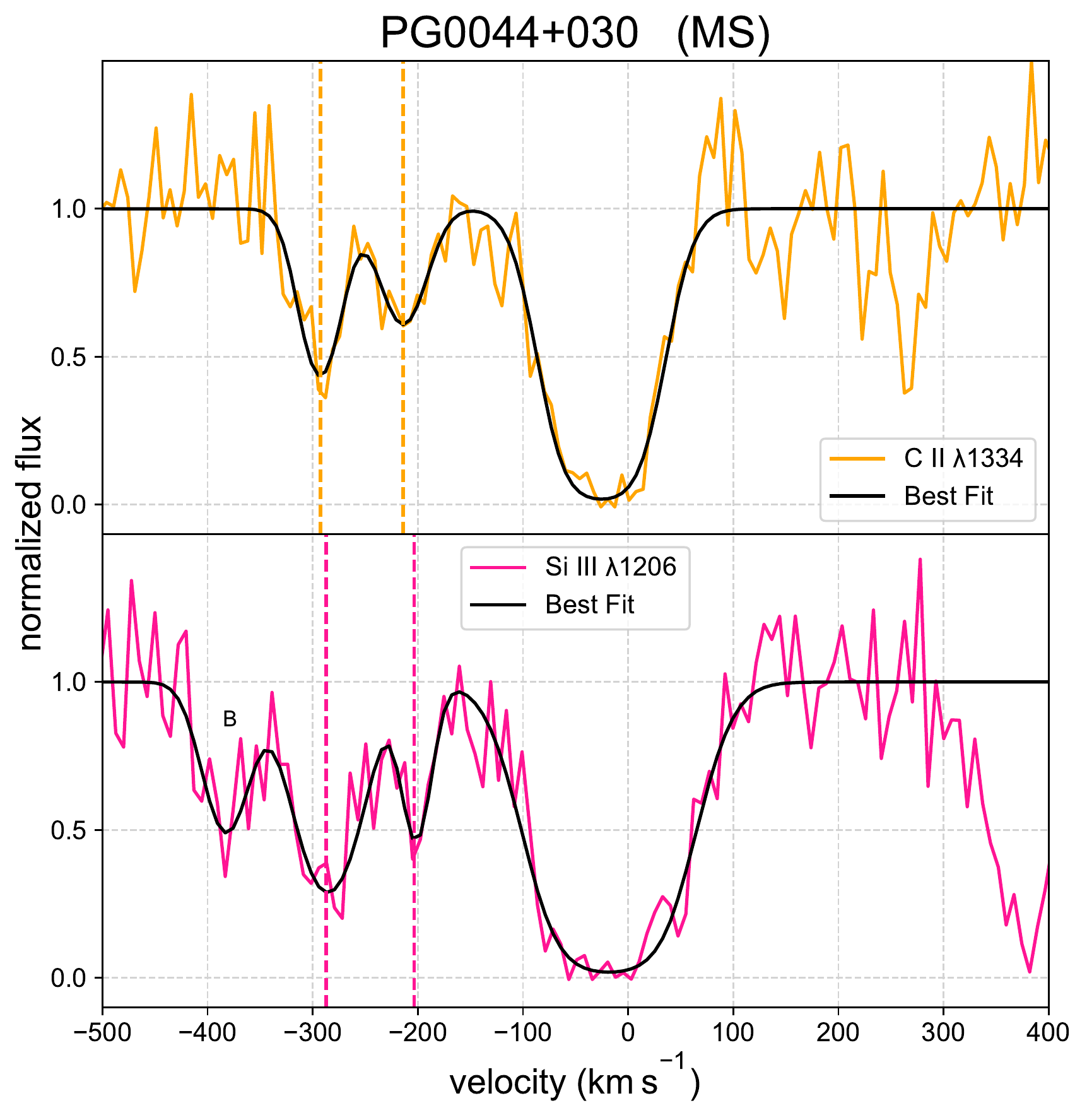}
    
    \caption{Comparison of low-ion (\siw\ or \cw), intermediate-ion (\sit), 
    and high-ion (\sif\ or \cf) absorption profiles for each sightline in the sample. 
    In each panel, normalized flux is plotted against LSR velocity, 
    with the data shown as solid colored lines and the best-fit 
    {\it Voigtfit} model shown in black. The dashed vertical lines show 
    the velocity centroids of each Magellanic component detected 
    (non-Magellanic and low-significance HVCs are included in the fits but 
    do not have vertical tick marks).
    Blends are marked with the letter ``B", non-Magellanic HVCs are marked with ``N", and uncertain (low-significance) HVCs are marked with the letter ``U". These lettered components are not included in the analysis.}
    \label{fig:montage_ms}
\end{figure*}

\setcounter{figure}{1}
\begin{figure*}[!ht]
    \includegraphics[width=0.32\textwidth]{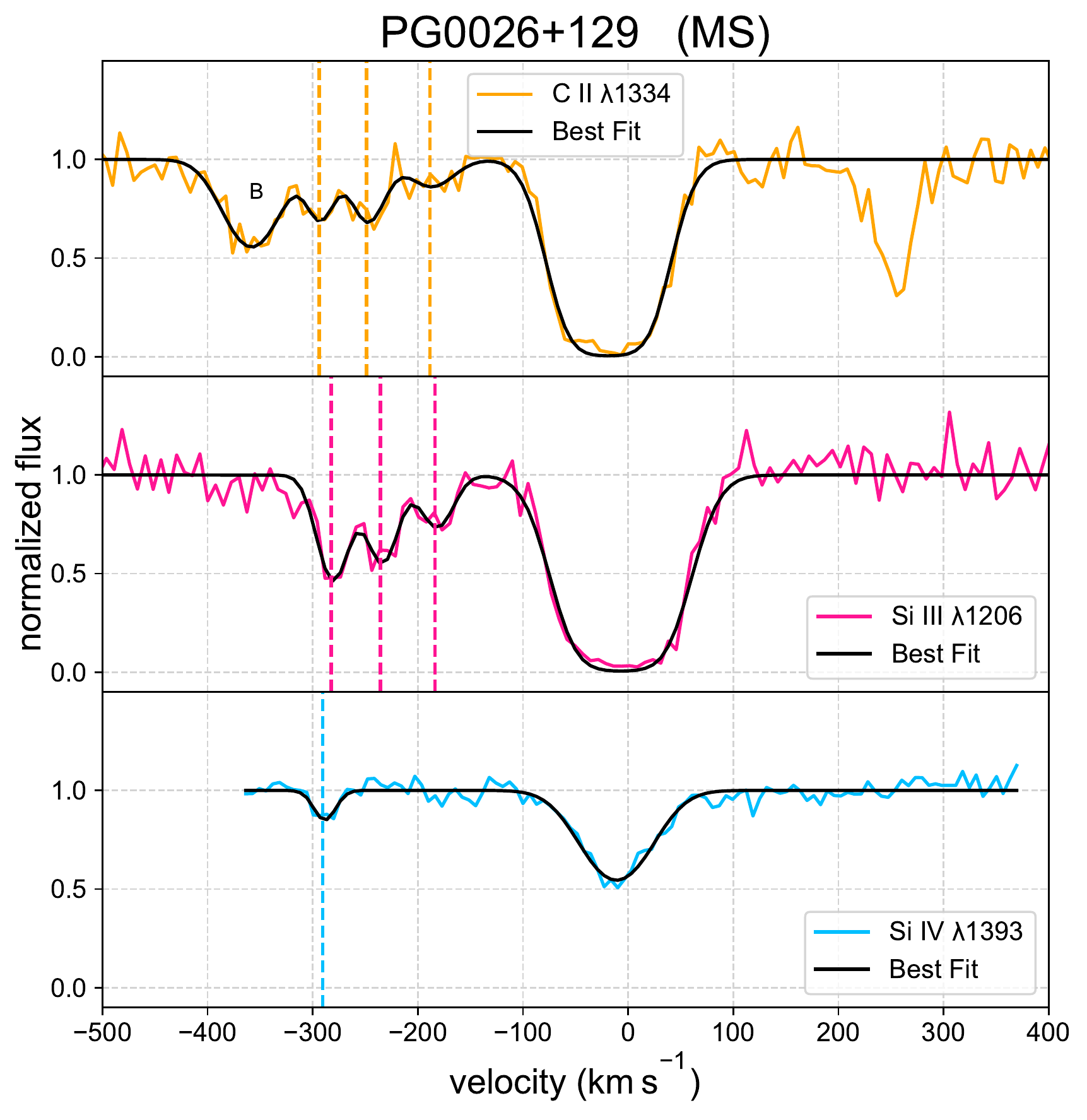}
    \includegraphics[width=0.32\textwidth]{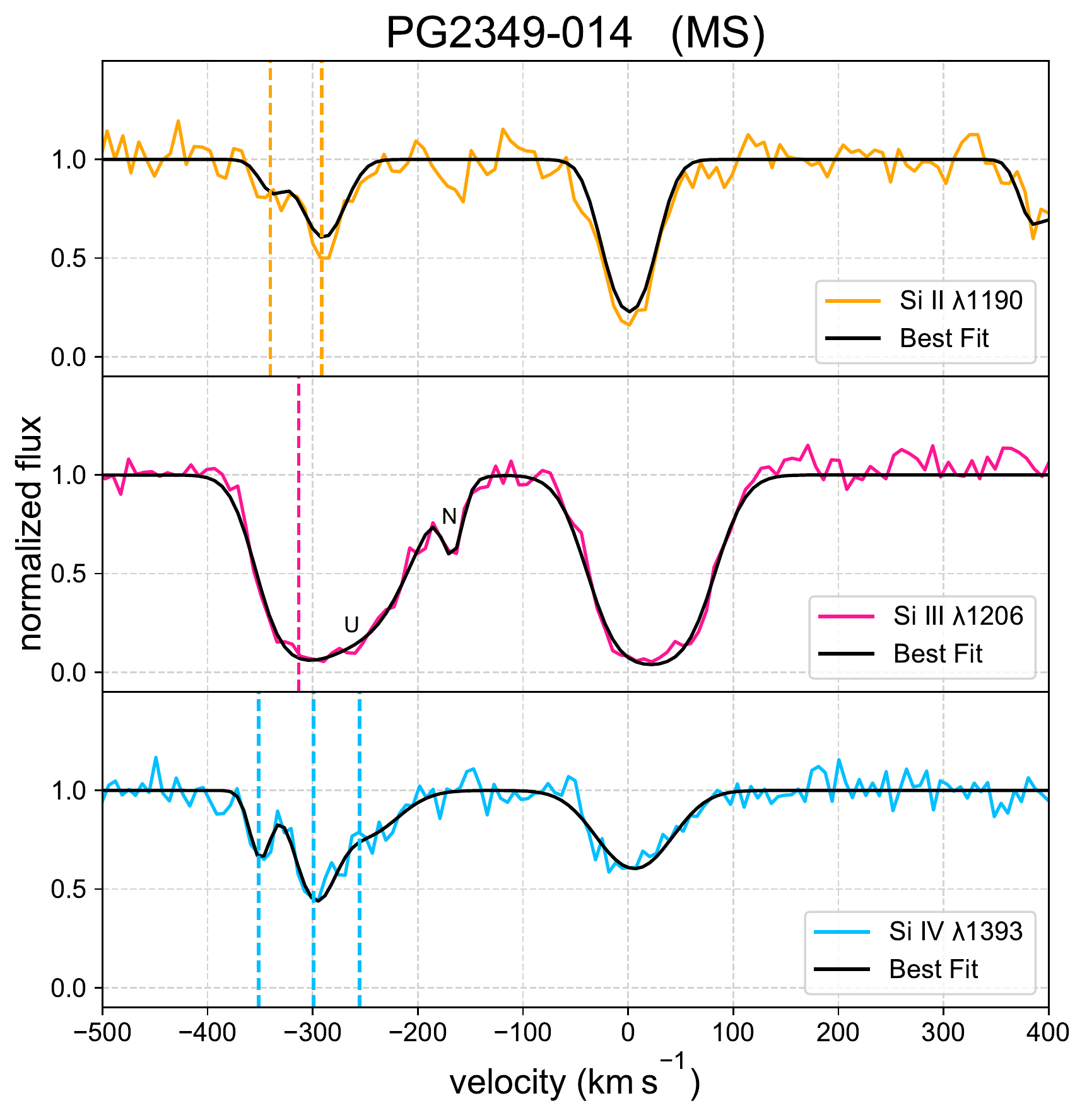}
    \includegraphics[width=0.32\textwidth]{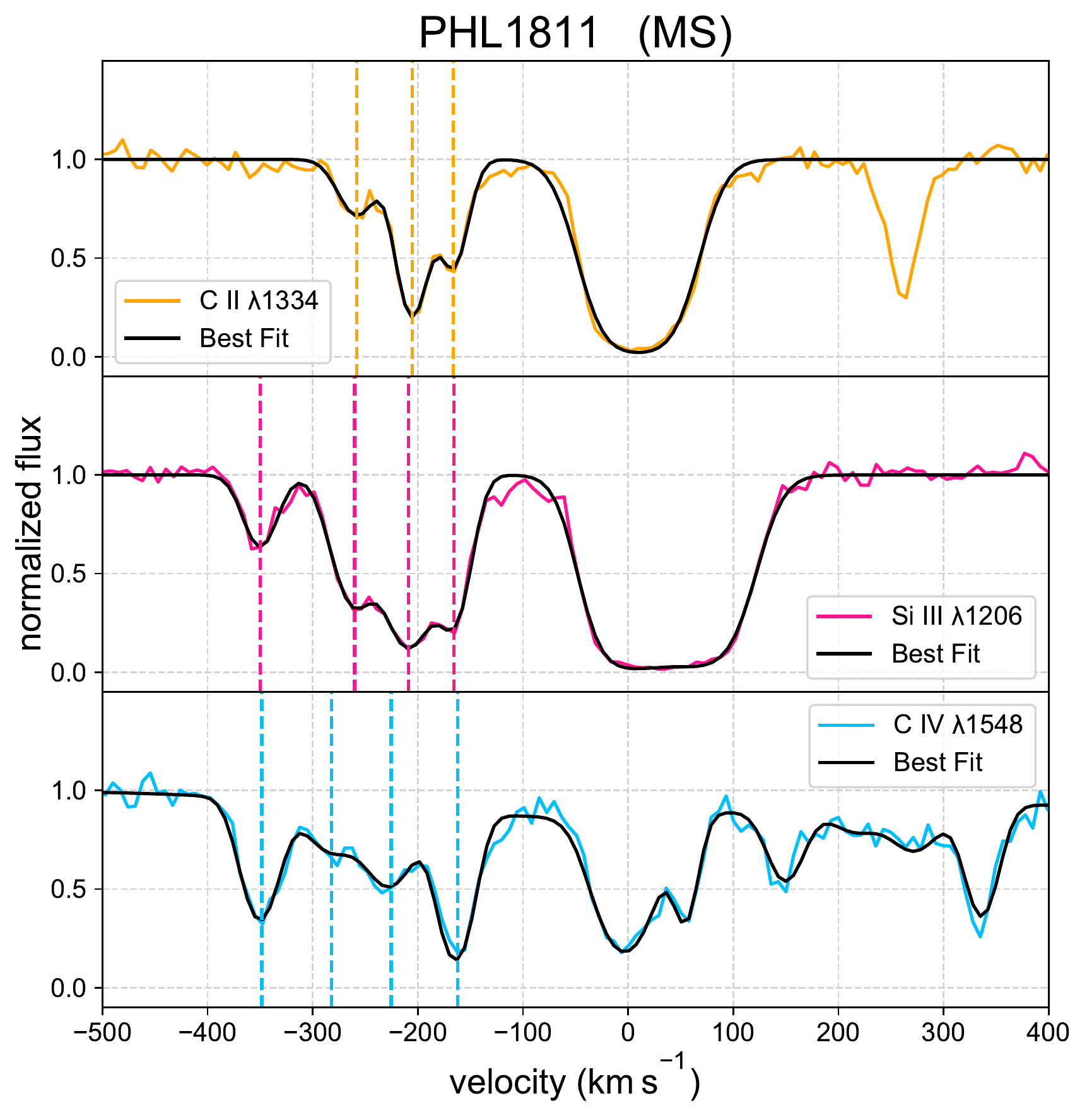}
    
    \includegraphics[width=0.32\textwidth]{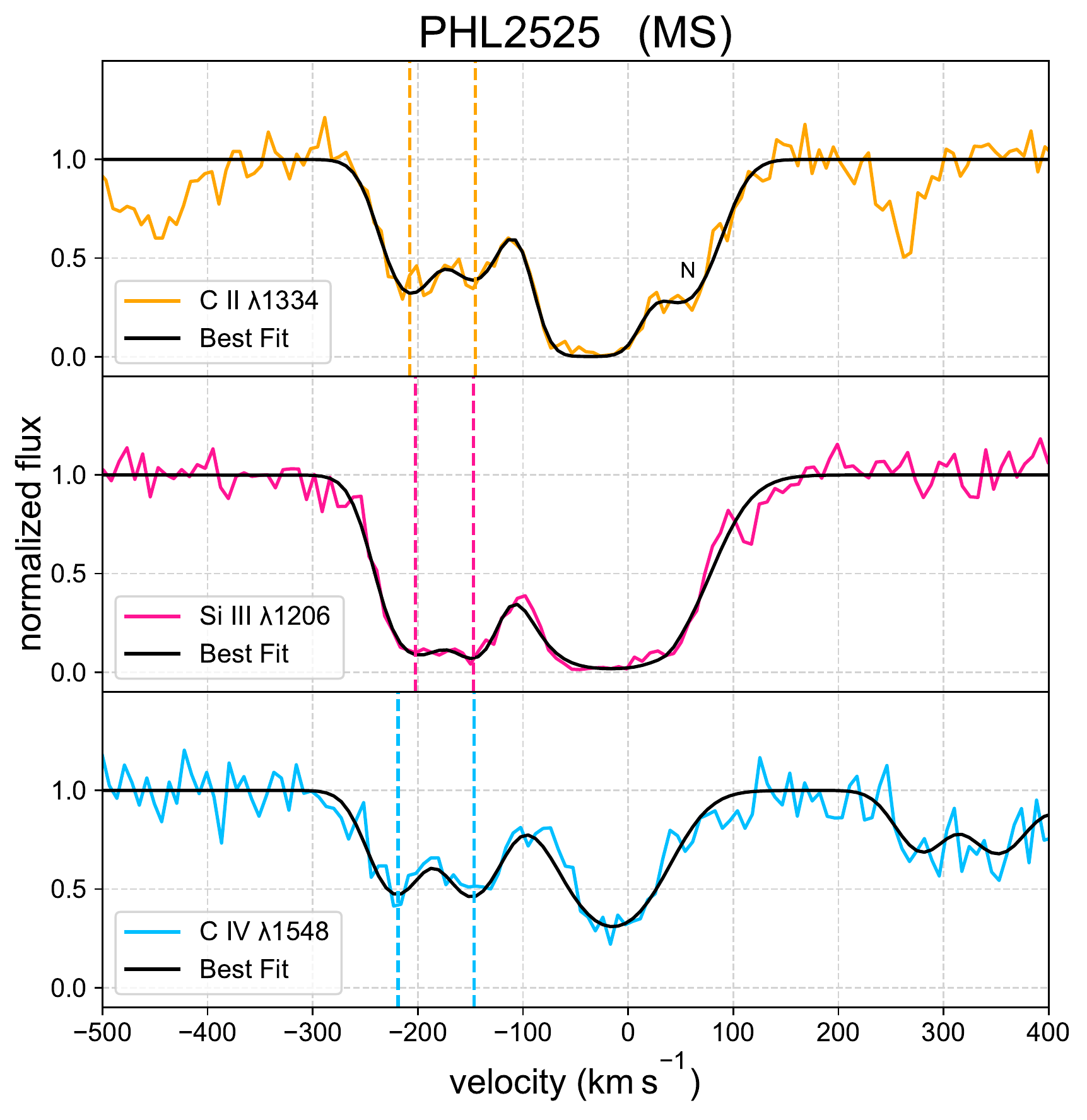}
    \includegraphics[width=0.32\textwidth]{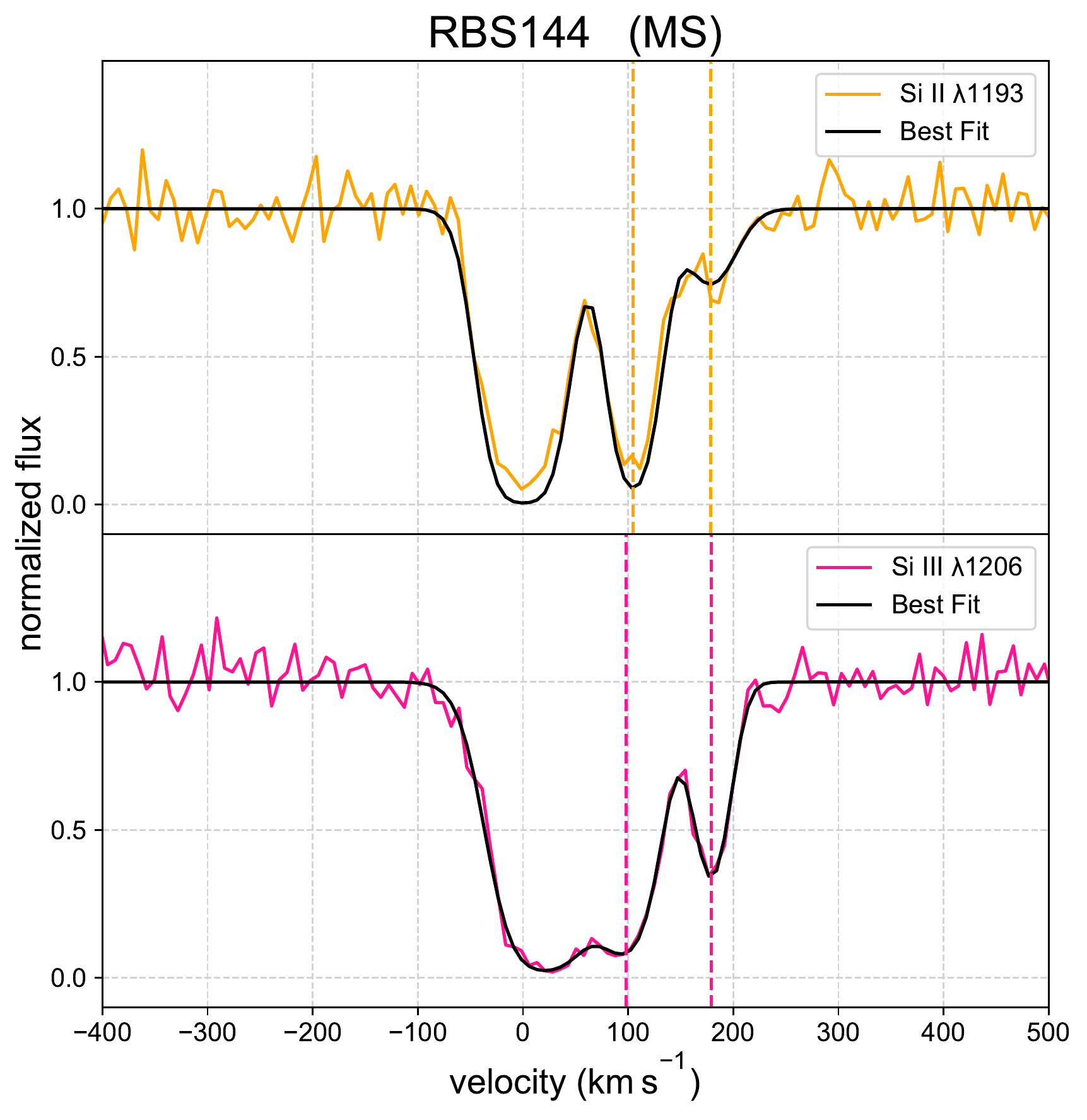}
    \includegraphics[width=0.32\textwidth]{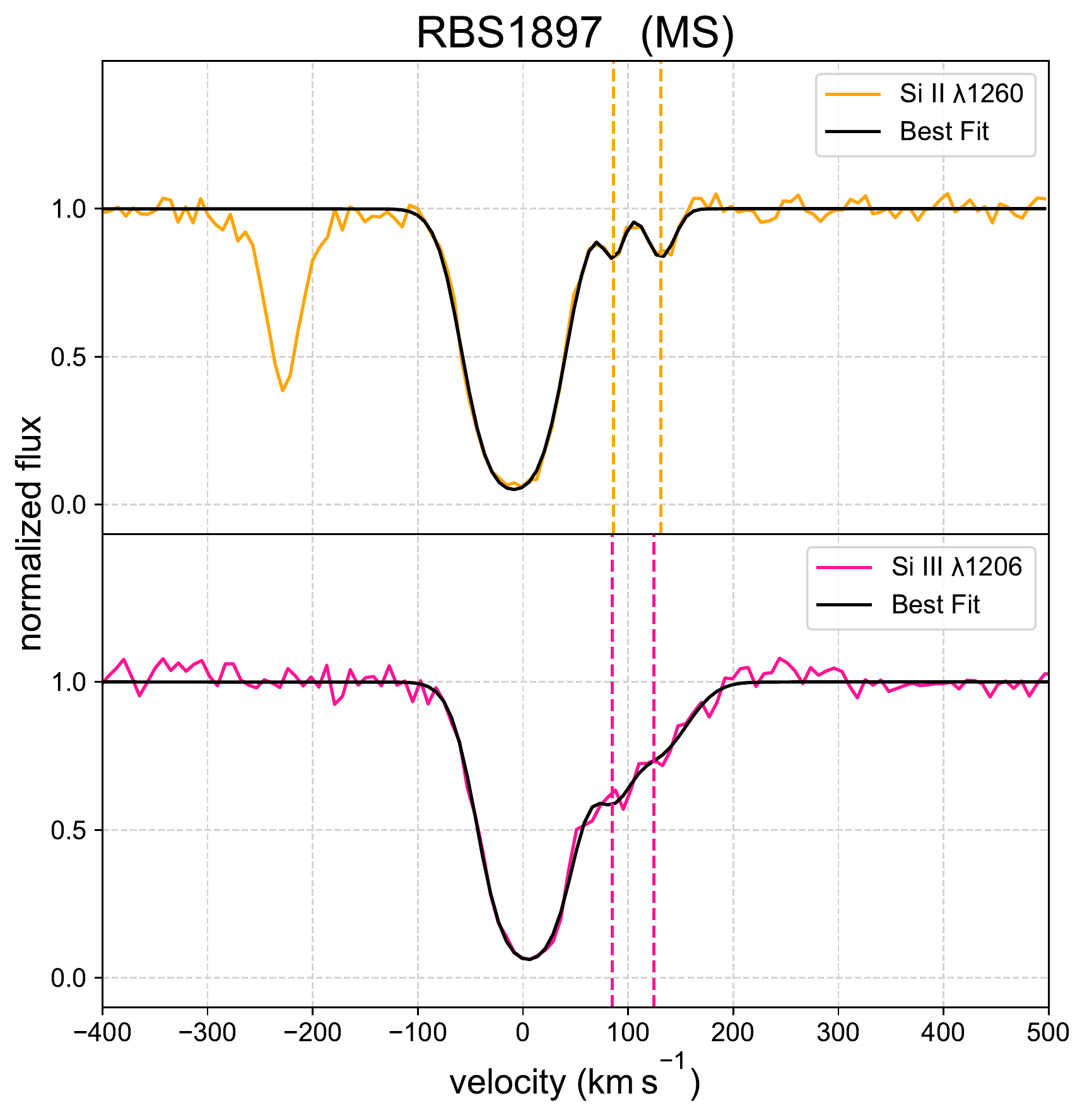}

    \includegraphics[width=0.32\textwidth]{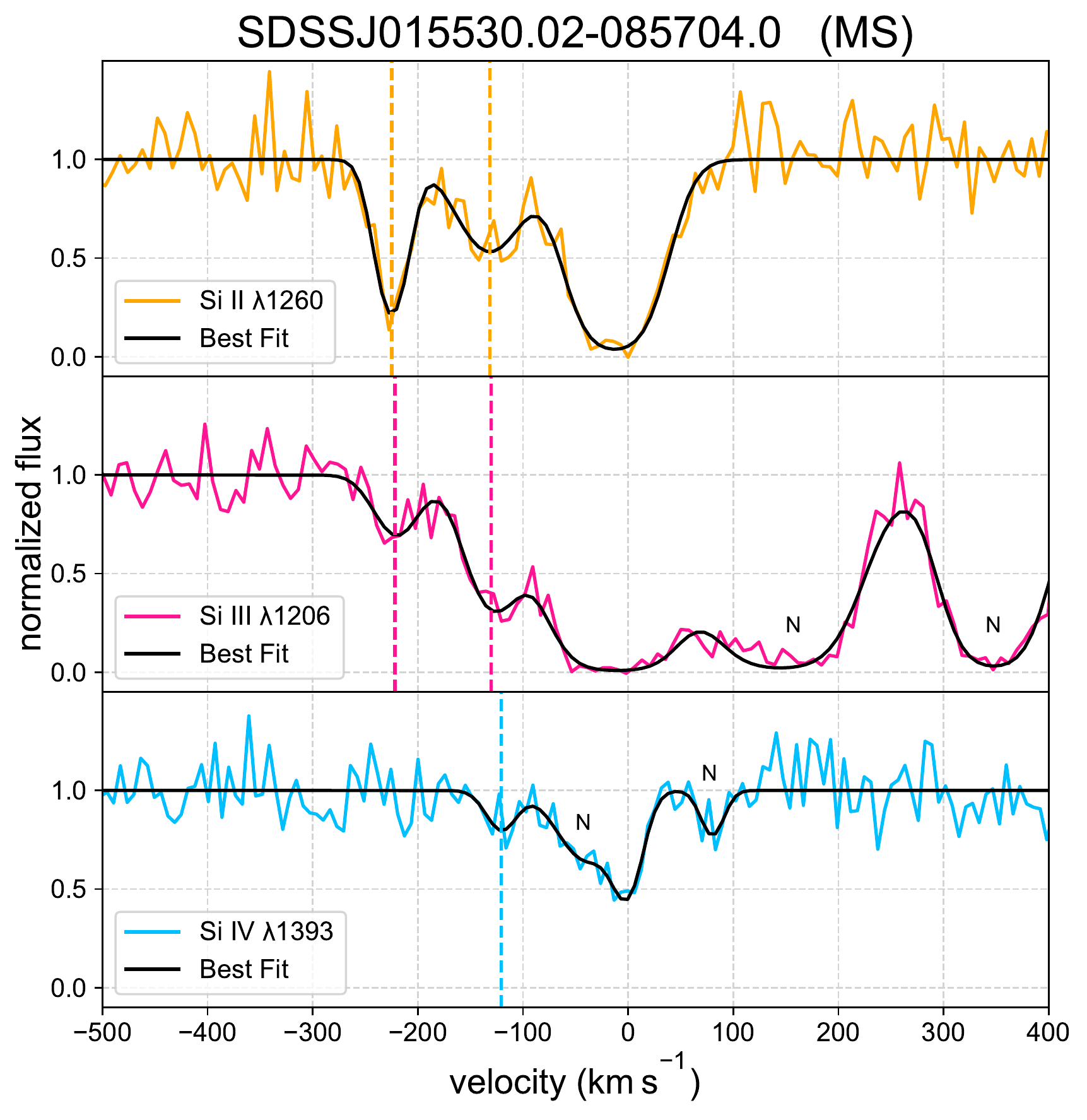}
    \includegraphics[width=0.32\textwidth]{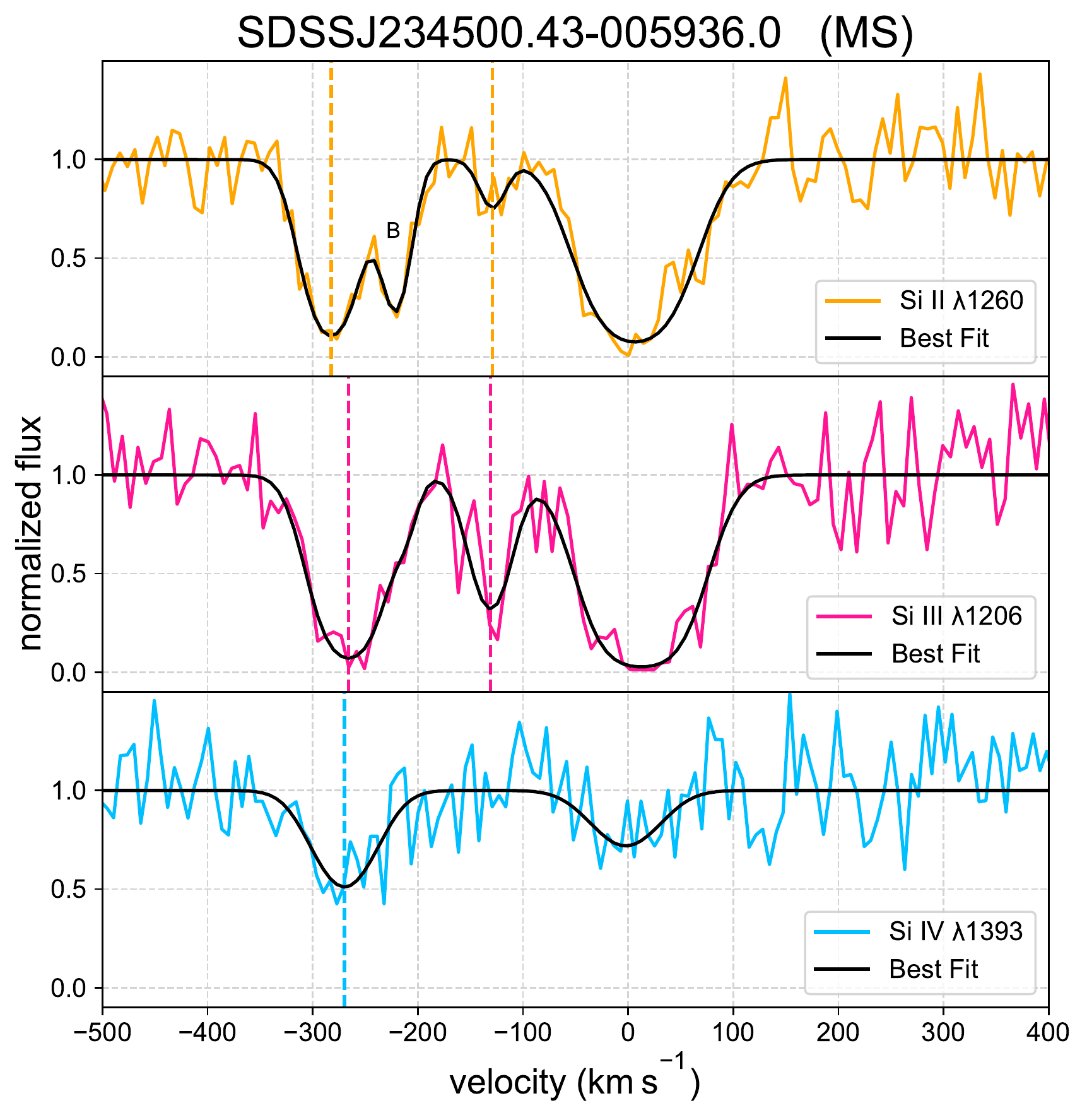}
    \includegraphics[width=0.32\textwidth]{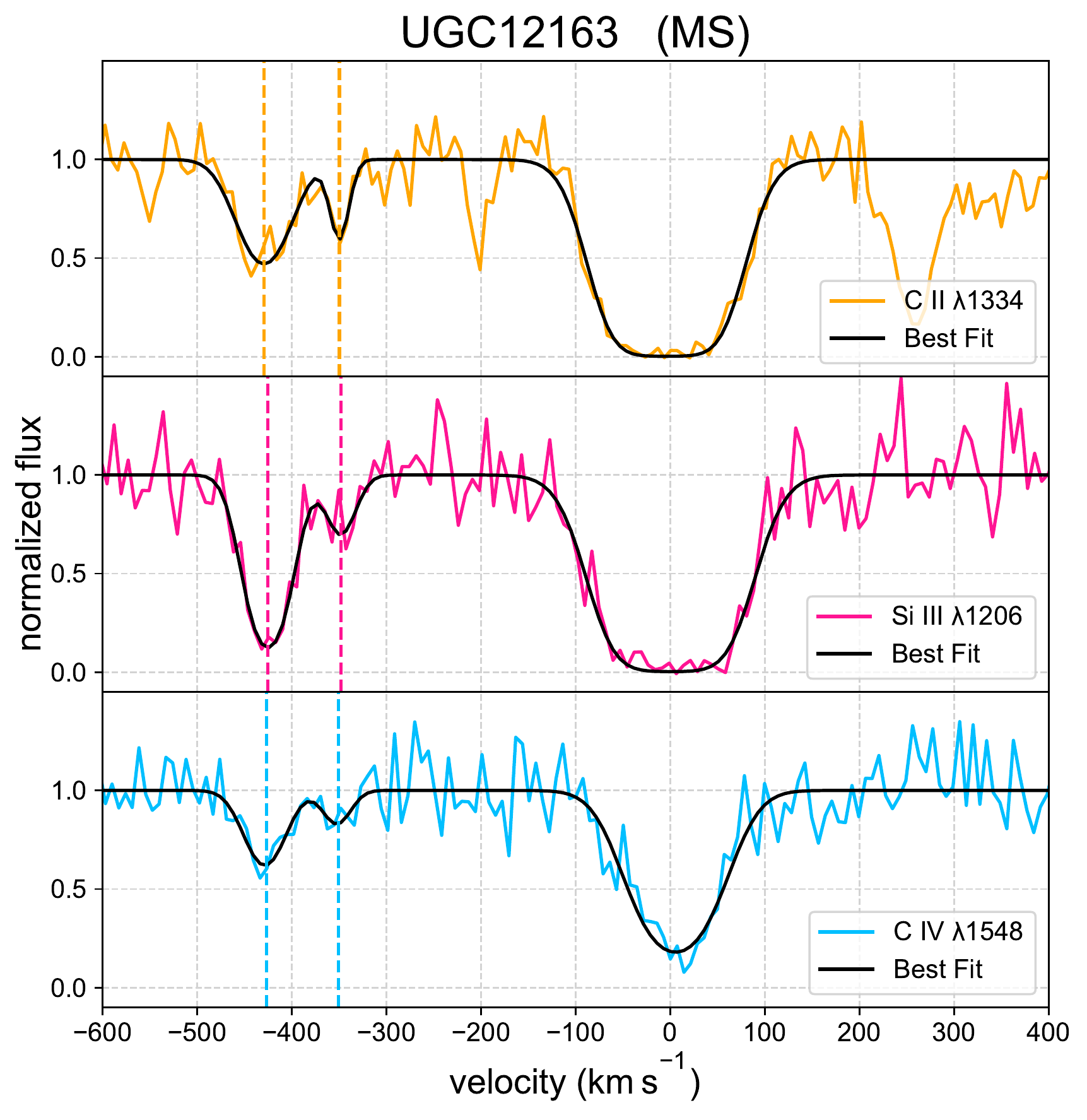}
    
    \includegraphics[width=0.32\textwidth]{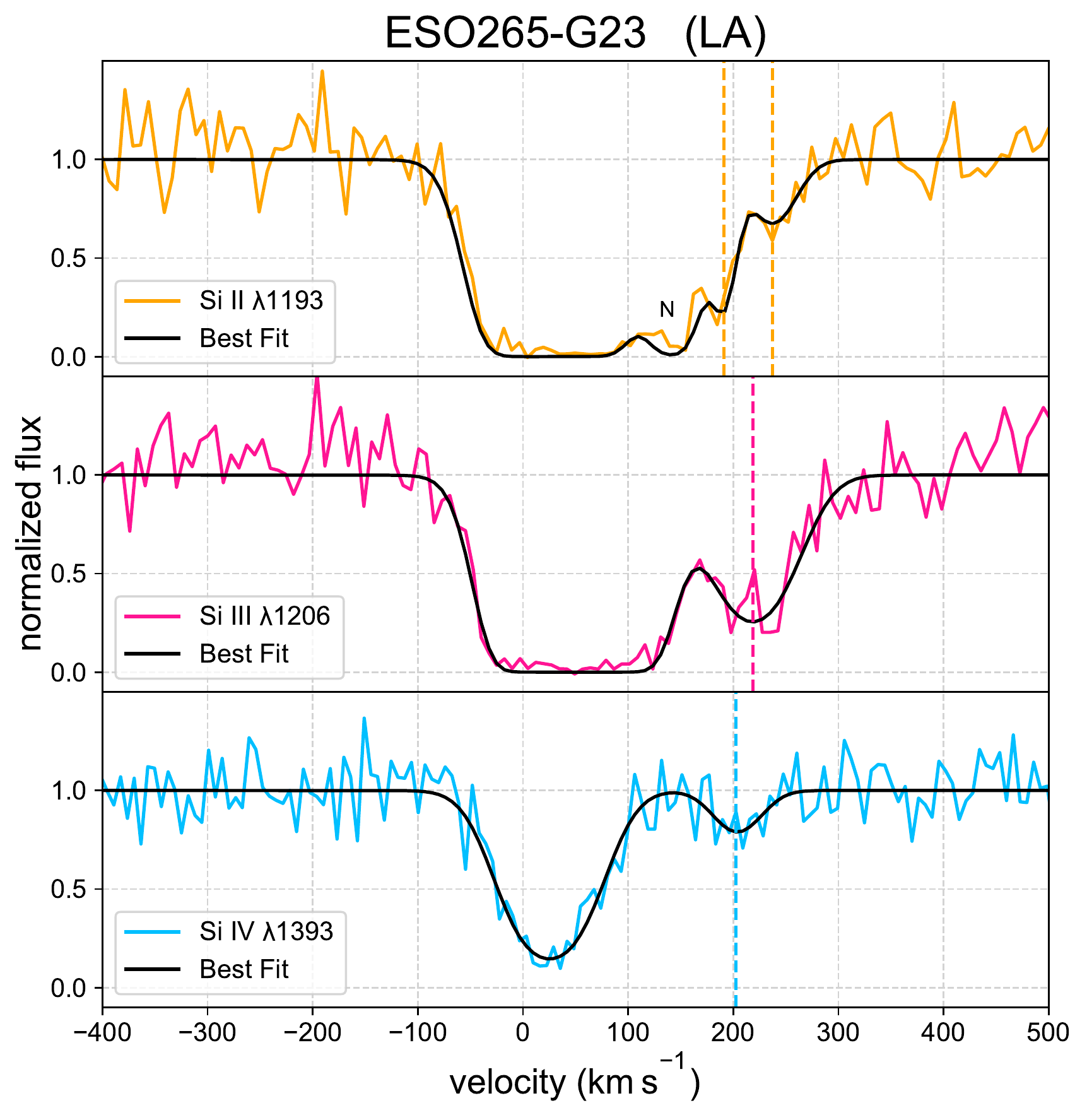}
    \includegraphics[width=0.32\textwidth]{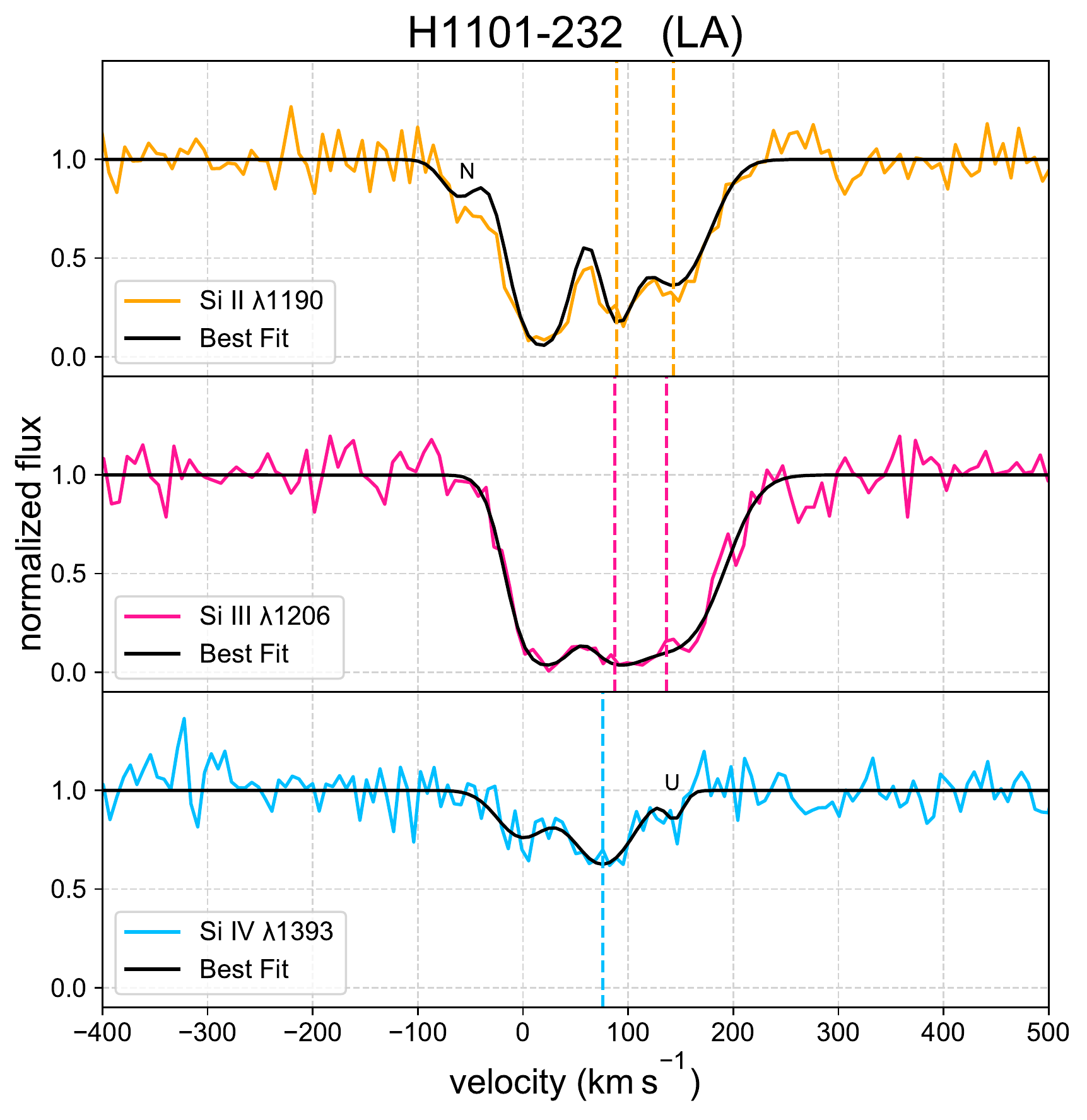}
    \includegraphics[width=0.32\textwidth]{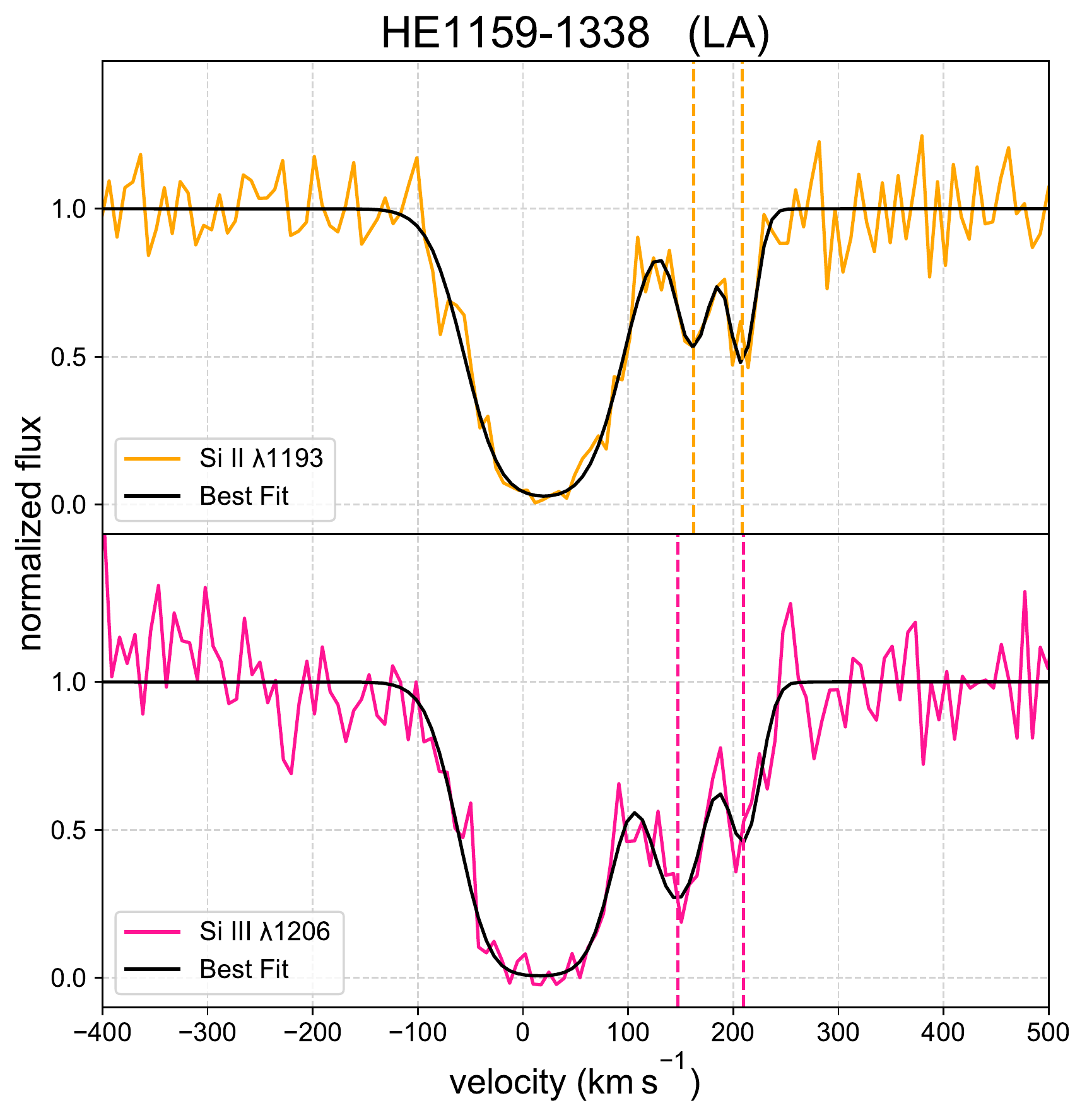}
    
    \caption{(continued). In some sightlines, only two metal lines are 
    shown, depending on which lines are covered and detected.}
    \label{fig:montage_ms2}
\end{figure*}

\setcounter{figure}{1}
\begin{figure*}[!ht]
    
    \includegraphics[width=0.32\textwidth]{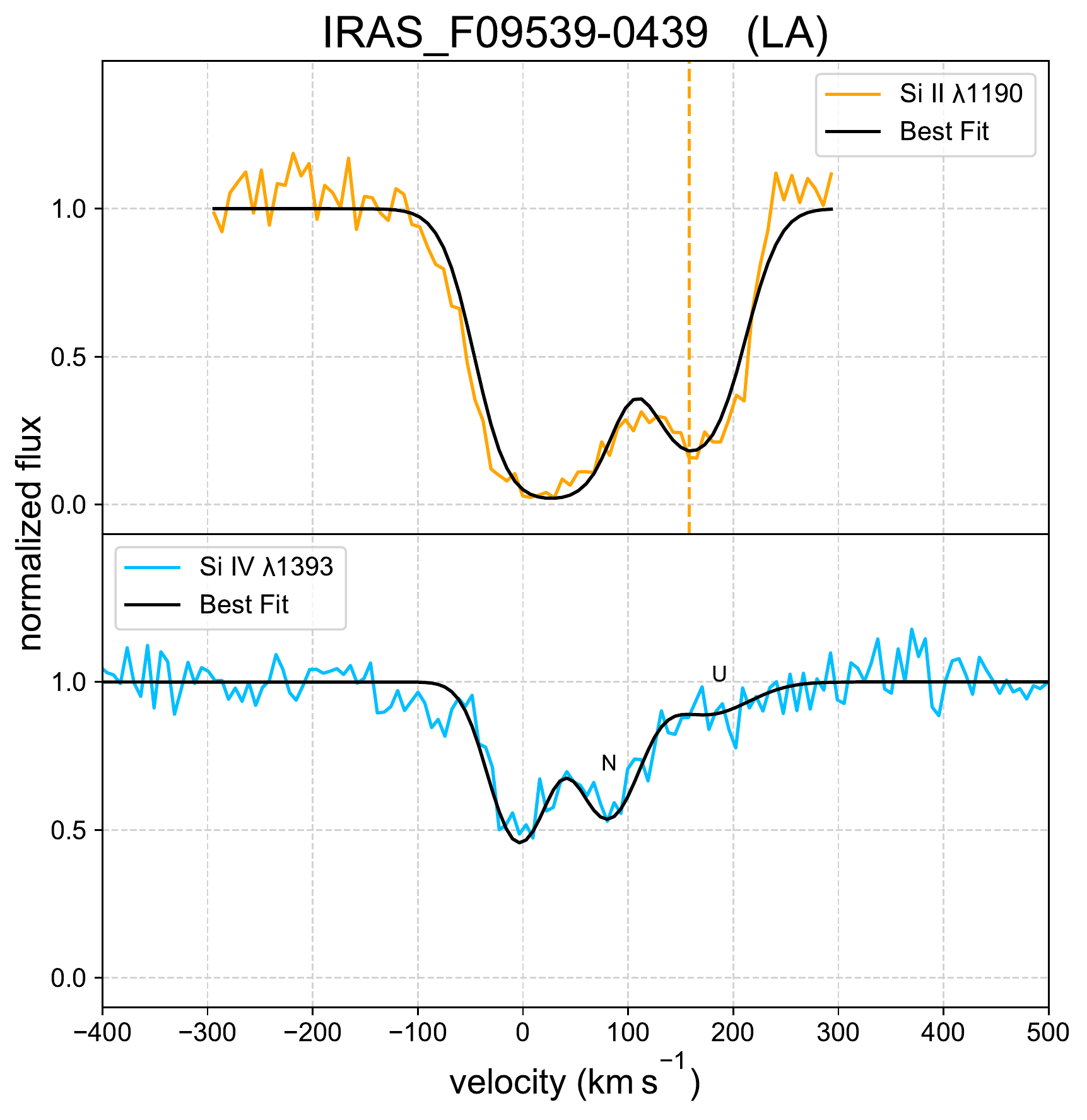}
    \includegraphics[width=0.32\textwidth]{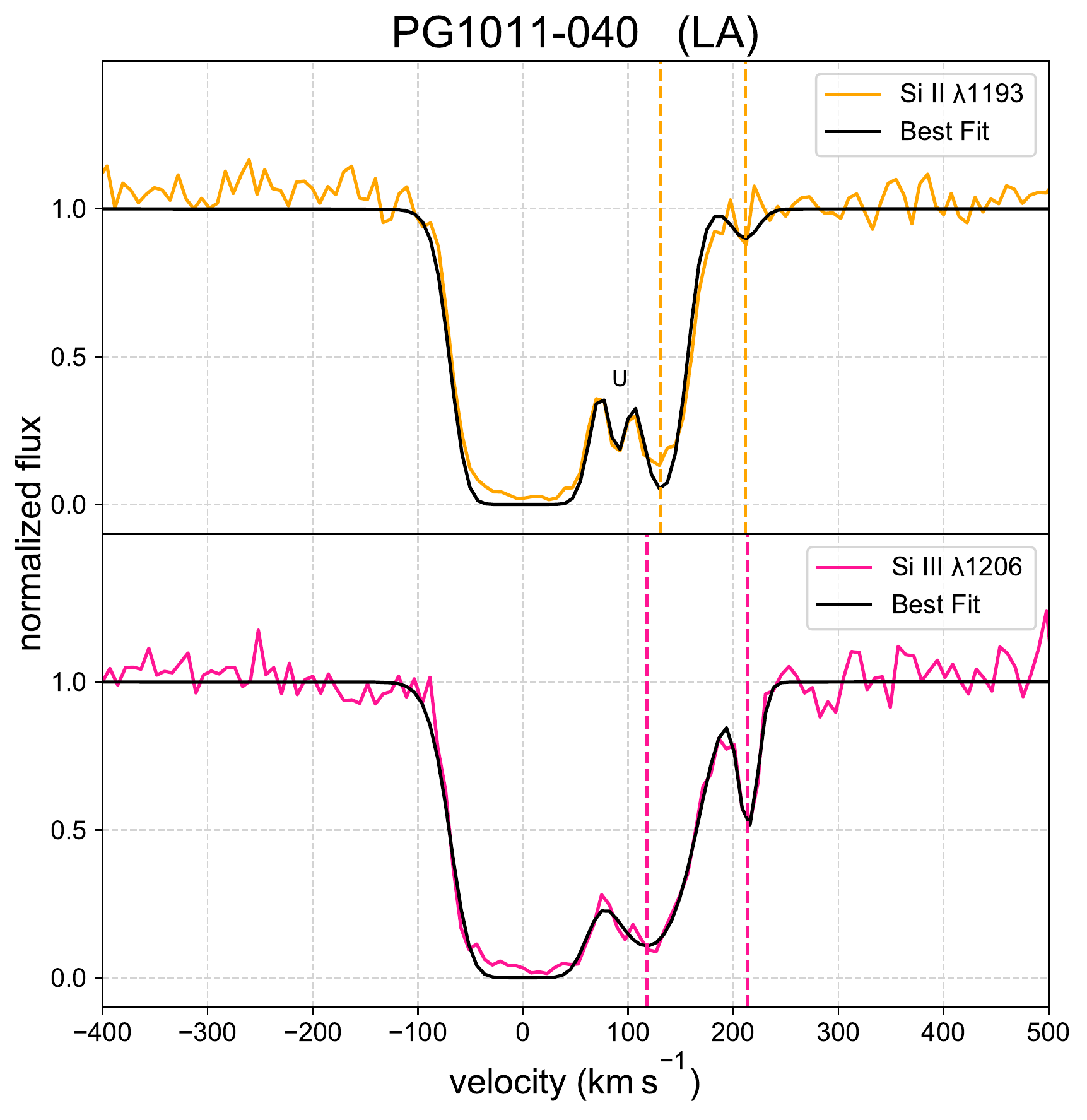}
    \includegraphics[width=0.32\textwidth]{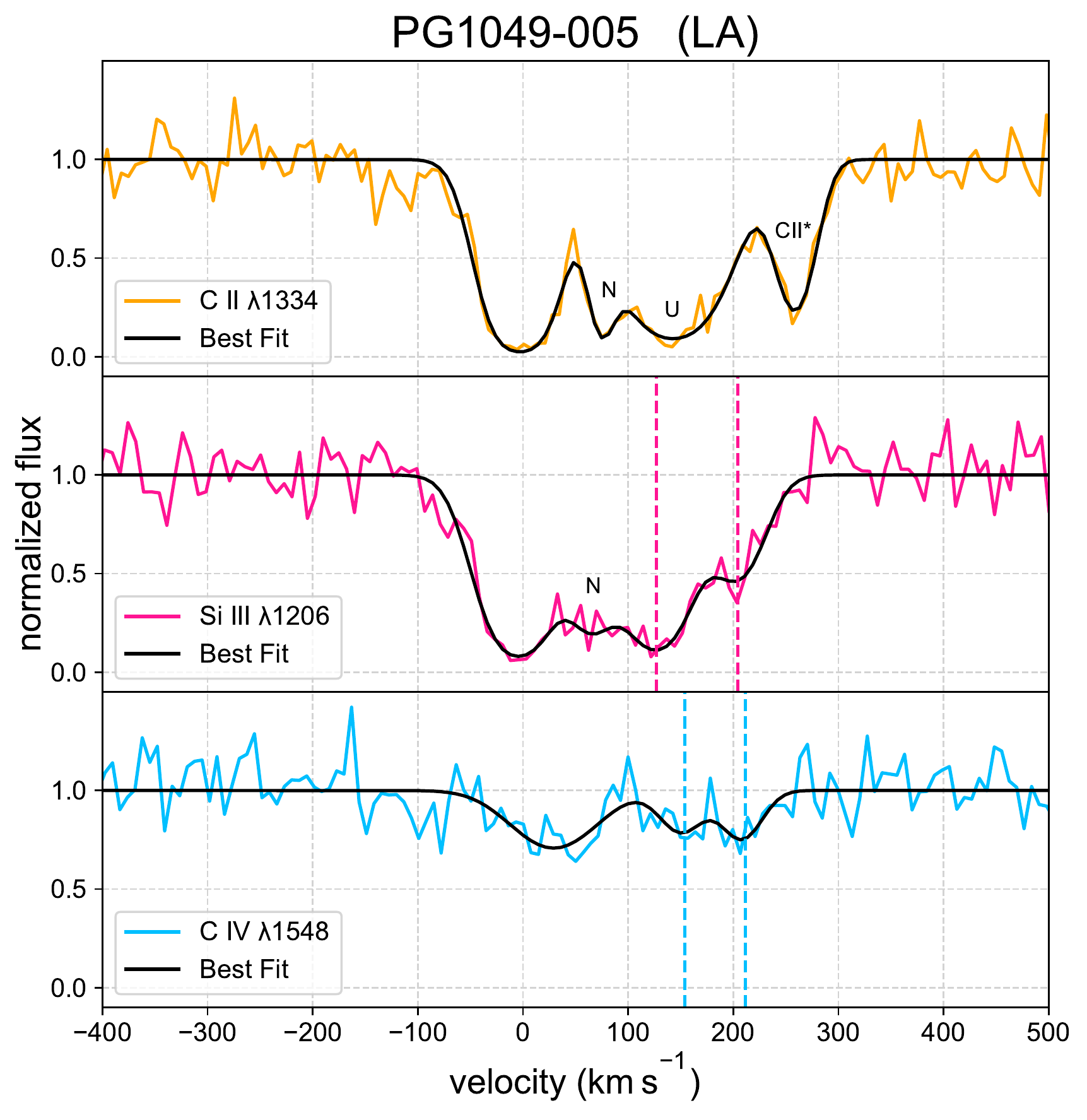}

    \includegraphics[width=0.32\textwidth]{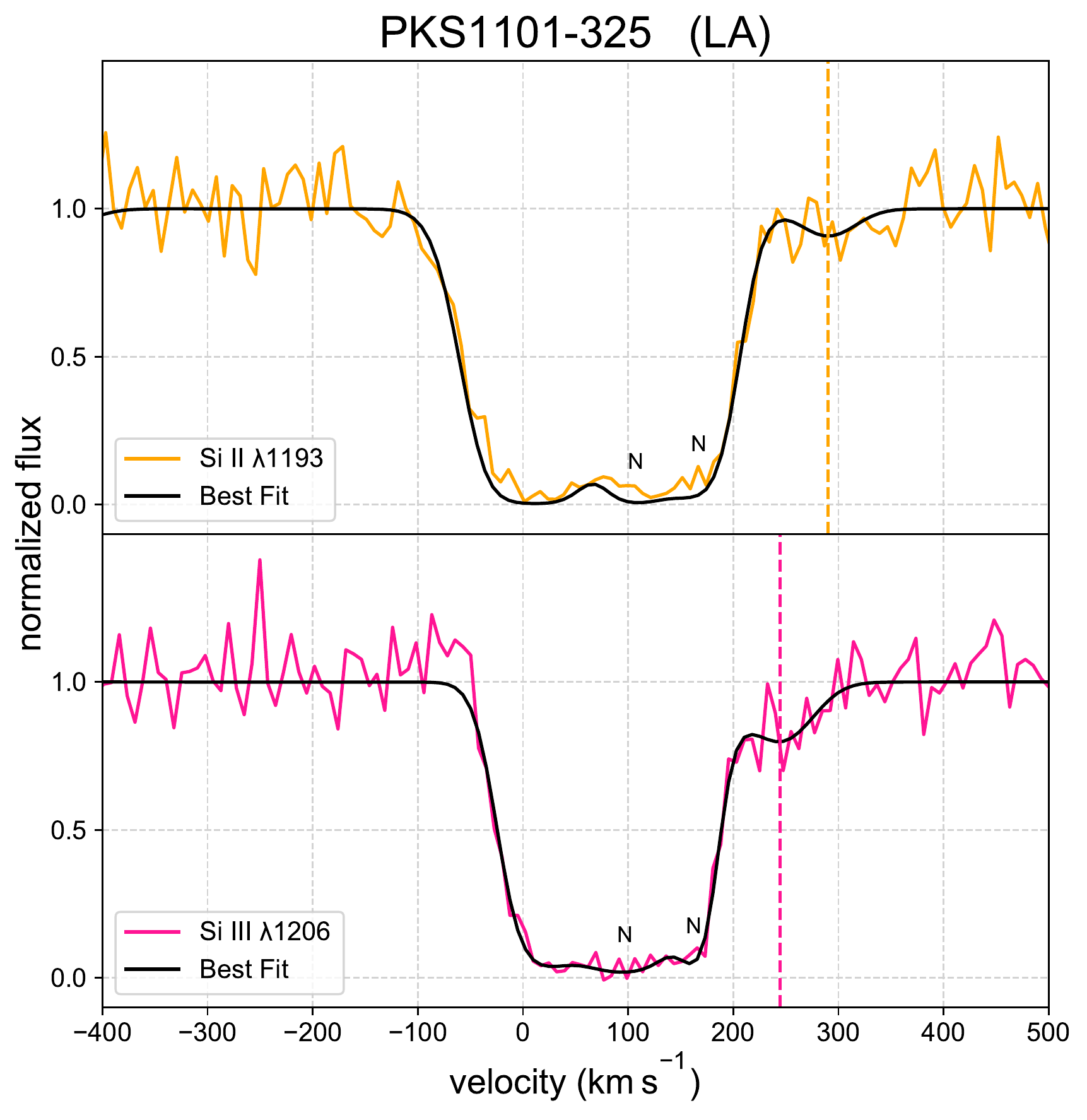}
    \includegraphics[width=0.32\textwidth]{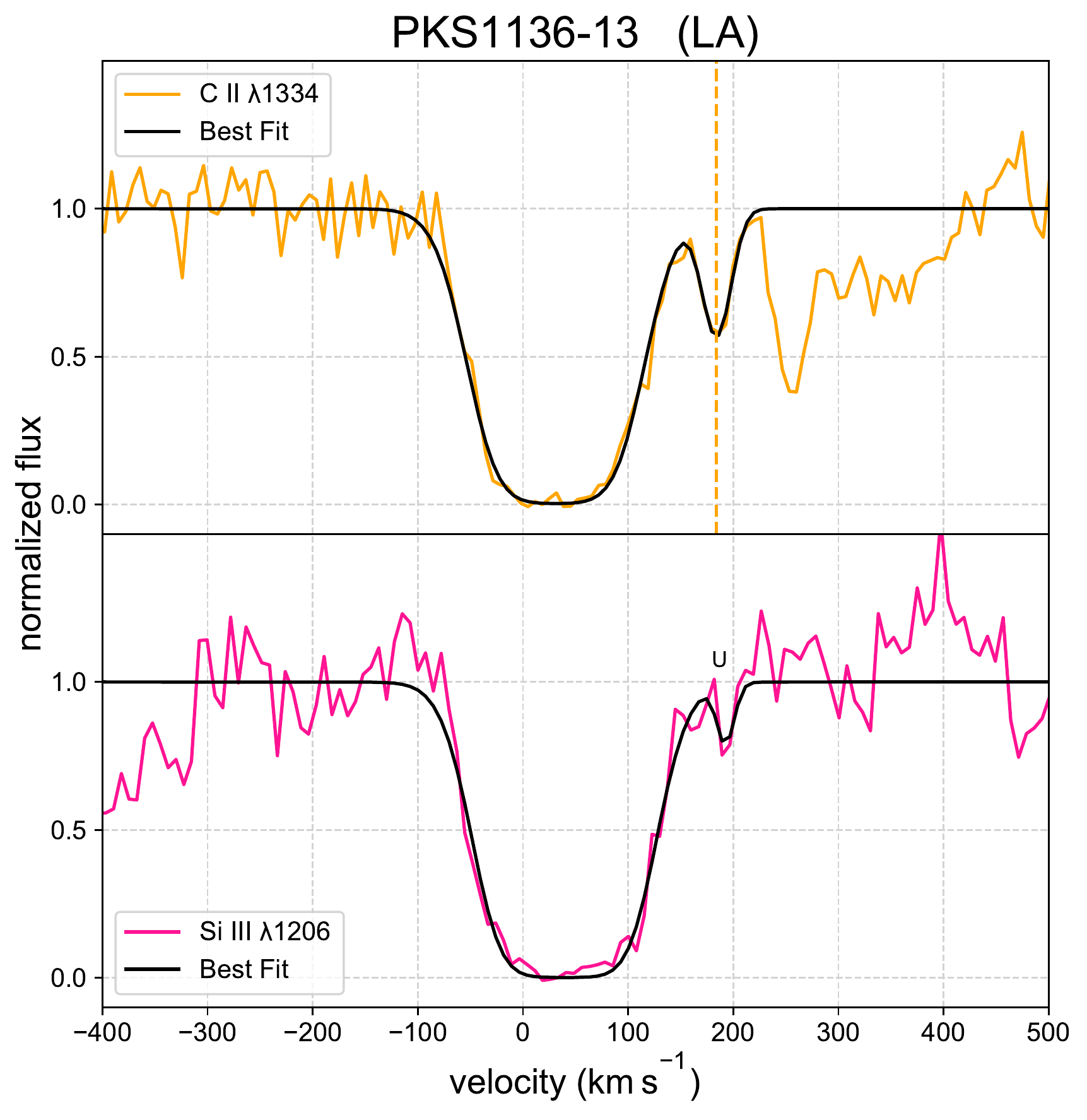}
    \includegraphics[width=0.32\textwidth]{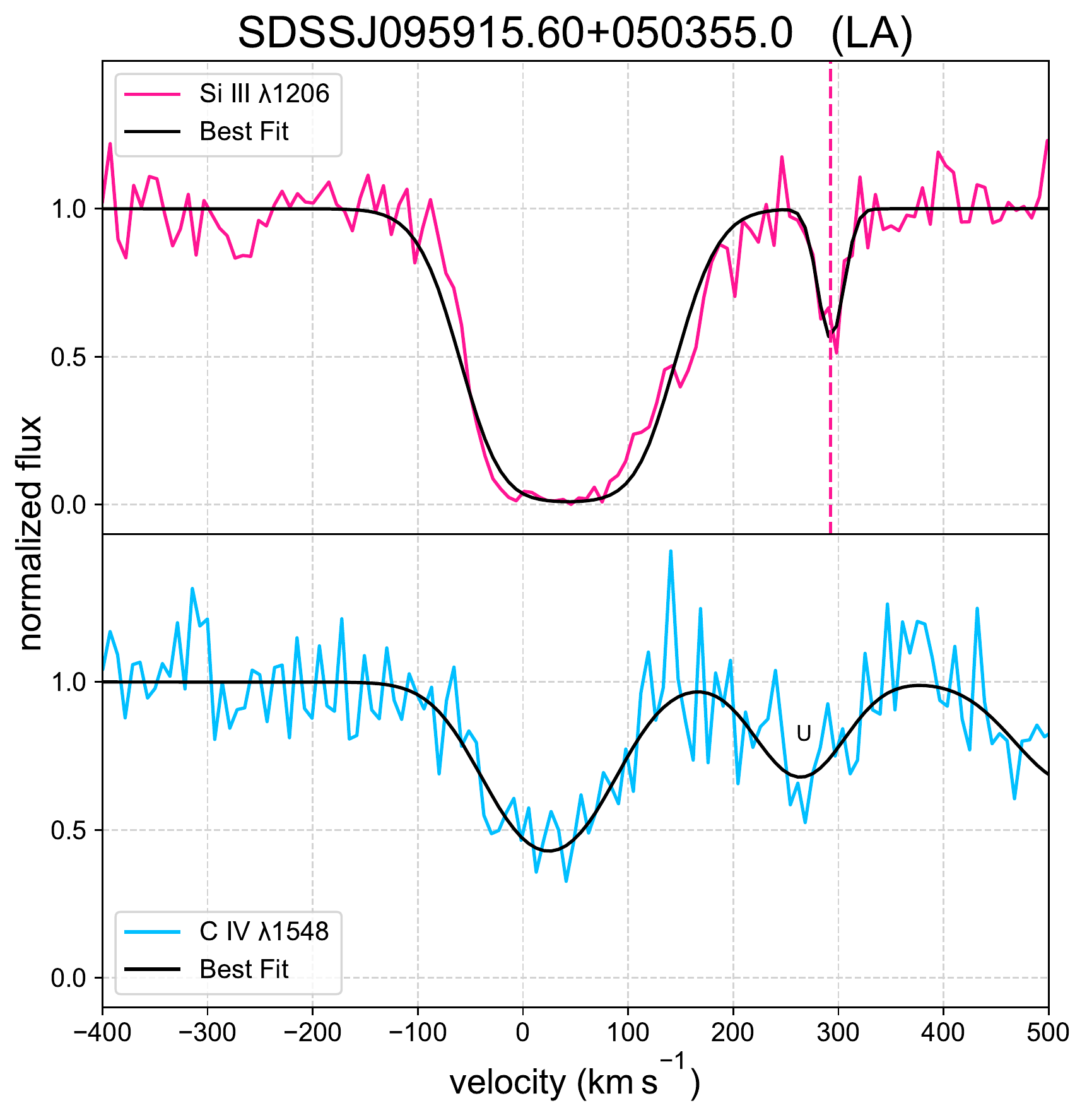}
    
    \includegraphics[width=0.32\textwidth]{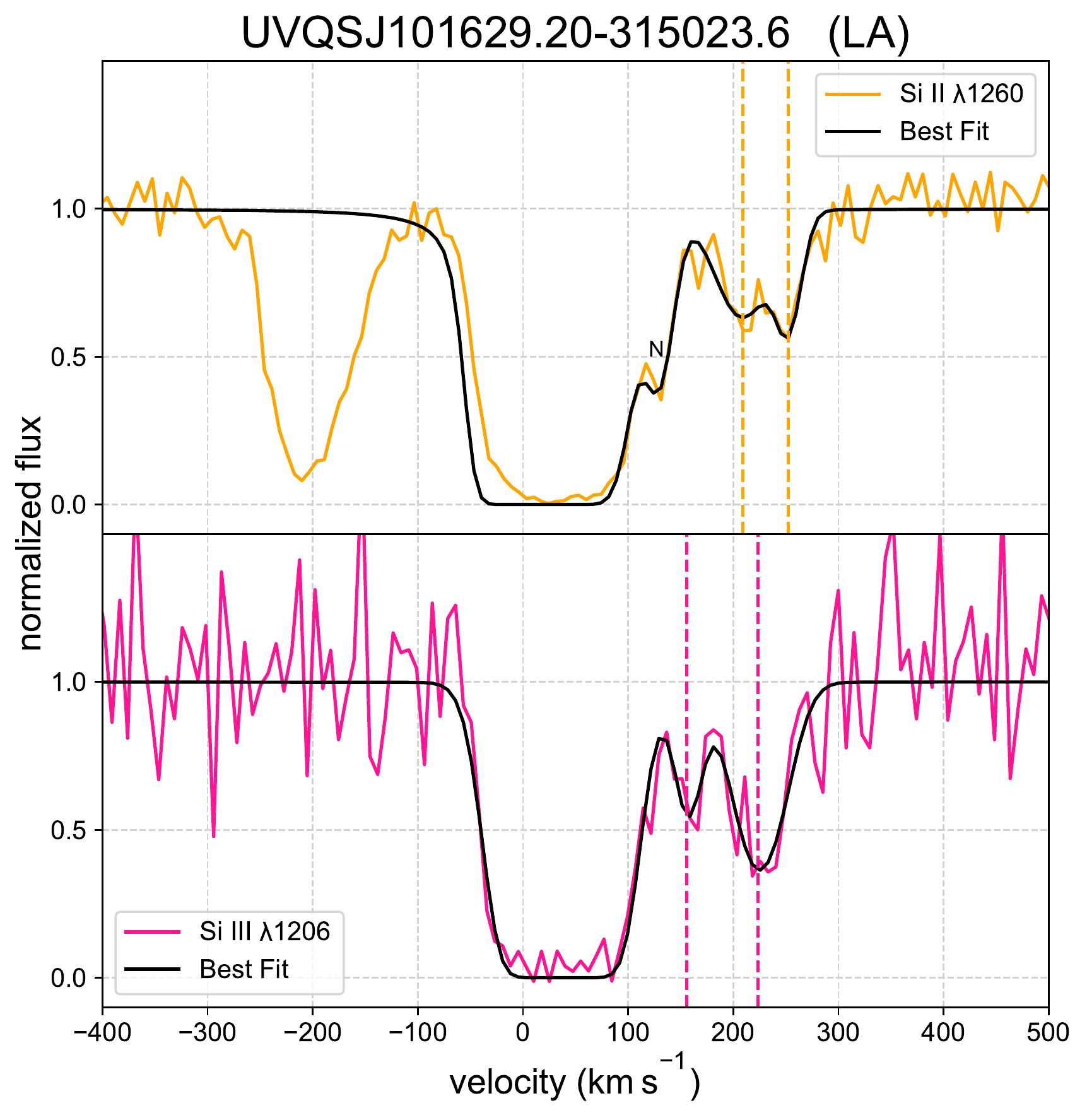}
    
    \caption{(continued). In some sightlines, only two metal lines are 
    shown, depending on which lines are covered and detected.}
    \label{fig:montage_ms3}
\end{figure*}

The COS data presented in this paper were reduced using the customized reduction and
alignment steps described in \citetalias{Fo14} and the appendix of \citet{Wa15}.
We used the Python package {\it VoigtFit} \citep{Kr18} to perform Voigt-profile 
fits of the UV metal-line absorption profiles for each sightline in the sample. 
The transitions under study are
\siw\ $\lambda$1260,1190,1193, \sit\ $\lambda$1206, 
\sif\ $\lambda$1393,$\lambda$1402, \cw\ $\lambda$1334, and 
(when COS G160M data are available) \cf\ $\lambda$1548,1550. 
These lines were chosen since they are among the strongest UV metal lines detected
in HVCs \citep{Le12, Ri17}, and they arise from only two elements (C and Si), 
which simplifies the kinematic analysis.
For each metal line, we fit the entire absorption profile, i.e. we include both 
low-velocity clouds ($|v_{\rm LSR}|\!<\!100$\,\kms; LVCs) and 
high-velocity ($|v_{\rm LSR}|\!>\!100$\,\kms; HVCs) clouds in the {\it Voigtfit} model, 
even though our analysis is focused on the (Magellanic) HVCs. 
This is because accurately modeling 
the LVCs enhances the quality of the HVC models by improving the continuum placement.
This is particularly true for overlapping clouds that are not well 
separated in velocity.

Our fitting methodology for each metal line was as follows.
First, since the COS/FUV native pixel size is 2.5\kms\ and the spectral resolution 
is $\approx$15--20\kms\ (FWHM; depending on grating)
we rebinned the data by three pixels, so the resulting spectra are Nyquist 
sampled with $\approx$2 rebinned pixels per resolution element. 
Second, we ran the {\it Voigtfit} software using the following inputs:
the rebinned data, the spectral resolution of $R=16000$ (FWHM=18.7\kms) for 
G130M observations and $R=19000$ (FWHM=15.8\kms) for G160M observations), 
the number of components to fit, the desired size of the fitting region 
(using a default of $\pm$500\kms), 
and an initial estimate for the redshift, column density, and $b$-value 
of each component. Our fitting procedure assumed the COS/FUV line spread 
function (LSF) was a Gaussian with a full-width at half maximum equal to $c/R$\footnote{As our paper was nearing completion, 
a newer version of {\it Voigtfit} became available with the ability to handle 
non-Gaussian LSFs. In \autoref{sec:lsf} we quantify the minor effect of using 
the tabulated non-Gaussian COS LSFs 
provided by the Space Telescope Science Institute instead of using a Gaussian LSF.}, 
where $c$ is the speed of light.
We took the initial parameter estimates from \citetalias{Fo14} (Tables 1 and 2) 
and modified by 
eye as needed based on our inspection of the data, 
e.g. to separate a broad component into 
two narrower components, or to refine the velocity range of Magellanic absorption.
Third, we used {\it Voigtfit} to interactively select continuum regions on 
either side of the line, to model the continuum either with a linear fit or a 
spline function, and then to normalize the spectra. Finally, we masked any 
contaminated (blended) portions of the spectrum, then ran the code to 
simultaneously fit the components and return the $\chi^2$-minimized values of 
redshift, column density, and $b$-value for every component. 

Once the line fitting was complete, we classified the absorption 
components into Galactic (low velocities), Magellanic 
(high velocities of interest to our analysis), or unrelated HVC 
(high velocities not of interest). This step made use of the known kinematic 
structure of the Stream \citep[][]{Ni08} and the velocity integration ranges of Magellanic 
absorption as listed in \citetalias{Fo14}. However, \citetalias{Fo14} did 
not sub-divide the Magellanic absorption into multiple components, as we occasionally 
did here, so the classifications of which absorbers are Magellanic have been updated 
in some cases.

We plot the \hst/COS absorption-line spectra for each sightline in the sample 
in \autoref{fig:montage_ms} together with our Voigt-profile fits.
Each figure shows a low-ion (\siw\ or \cw), intermediate-ion (\sit), and 
high-ion (\sif, \cf) absorption profile, including both the data, the 
\emph{VoigtFit} model, and dashed vertical lines identifying the Magellanic components.
In a small number of cases, only two lines are shown, depending on which
lines are covered and detected. The figures illustrate the complex component 
structure and diversity of the COS absorption profiles. 
In several LA directions, the differences between the low- and high-ion 
absorption can be seen visually as velocity centroid offsets and line-width 
differences; we explore these differences 
quantitatively in \autoref{sec:kinematics}. In many directions, 
multiple components are seen within the Magellanic velocity interval, 
revealing sub-structure that was not quantified in \citetalias{Fo14}.

\section{Kinematics} \label{sec:kinematics}

\subsection{Distribution of b-values} \label{subsec:bvalues}

\begin{figure*}[!ht]
\centering 
\subfloat{\includegraphics[width=0.43\textwidth]{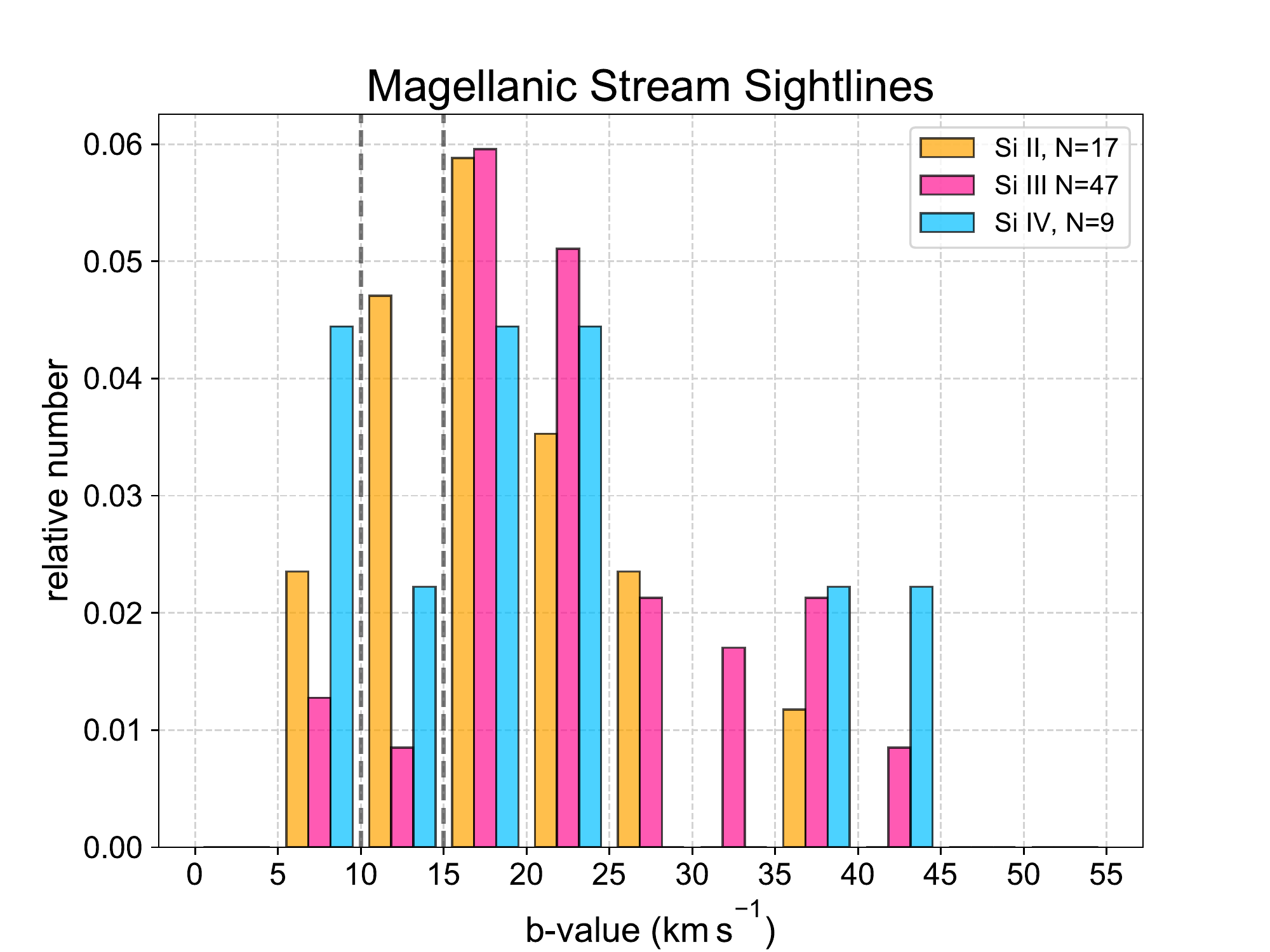}}\qquad
\subfloat{\includegraphics[width=0.43\textwidth]{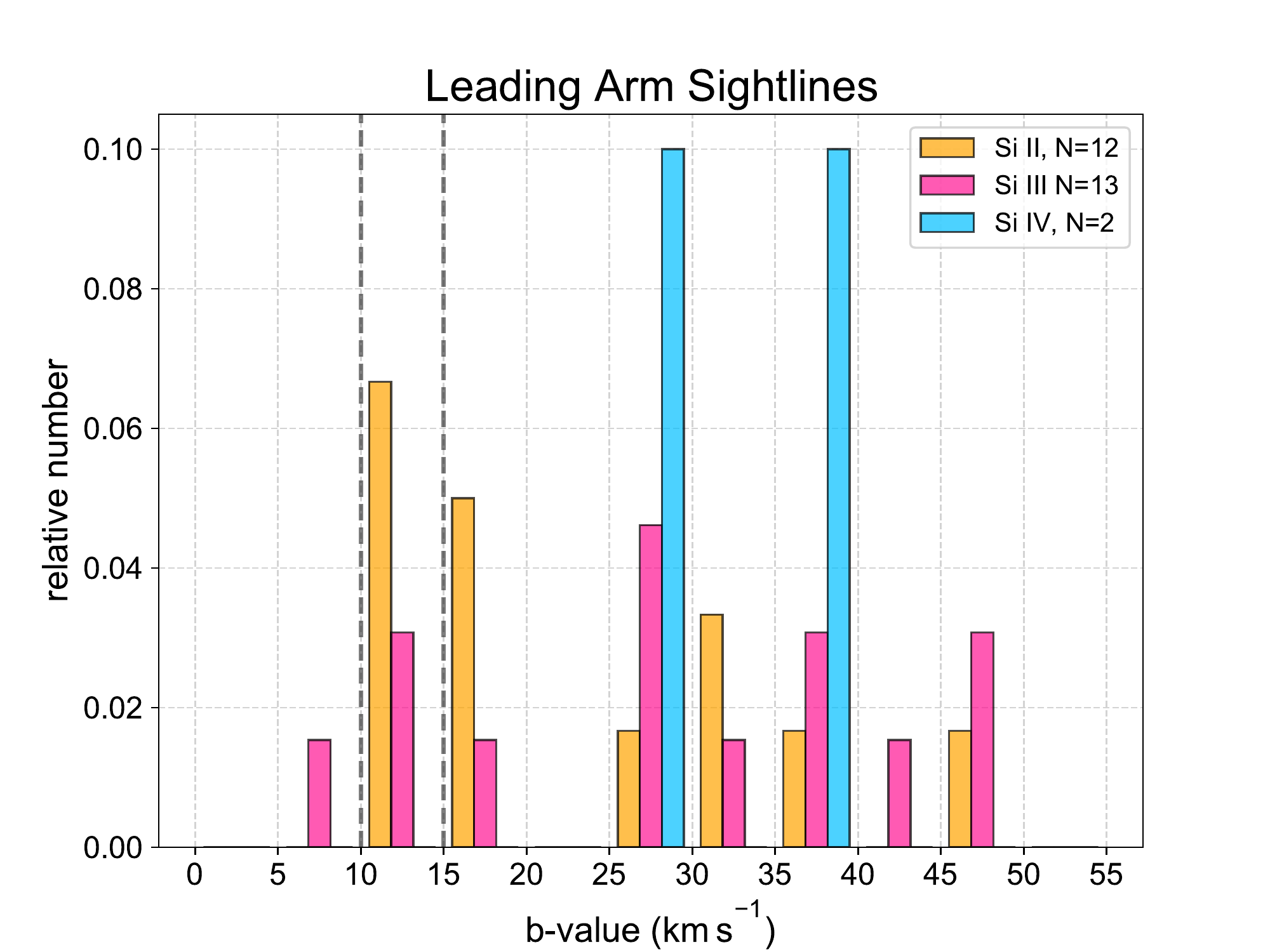}}\qquad
\subfloat{\includegraphics[width=0.43\textwidth]{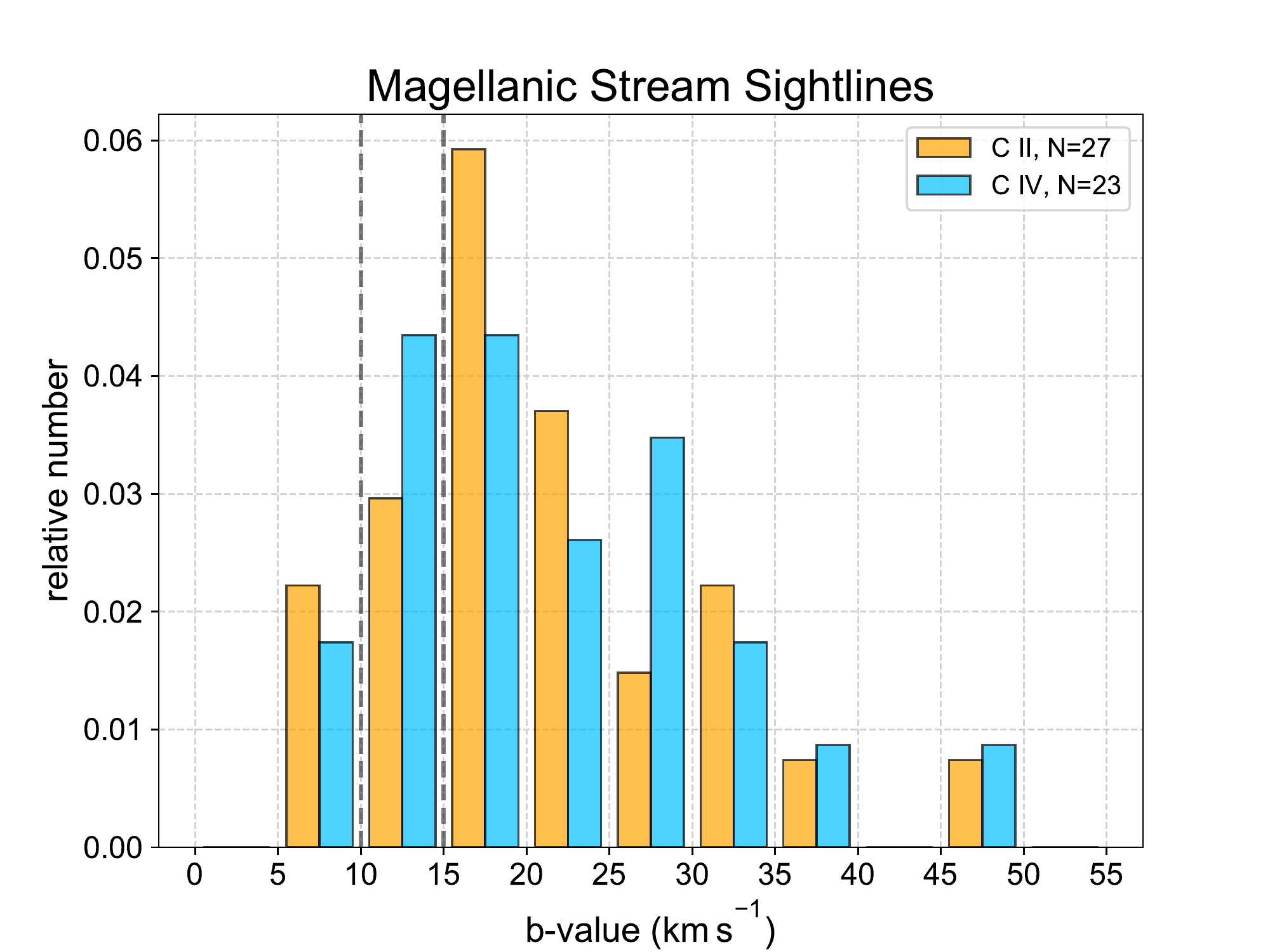}}\qquad
\subfloat{\includegraphics[width=0.43\textwidth]{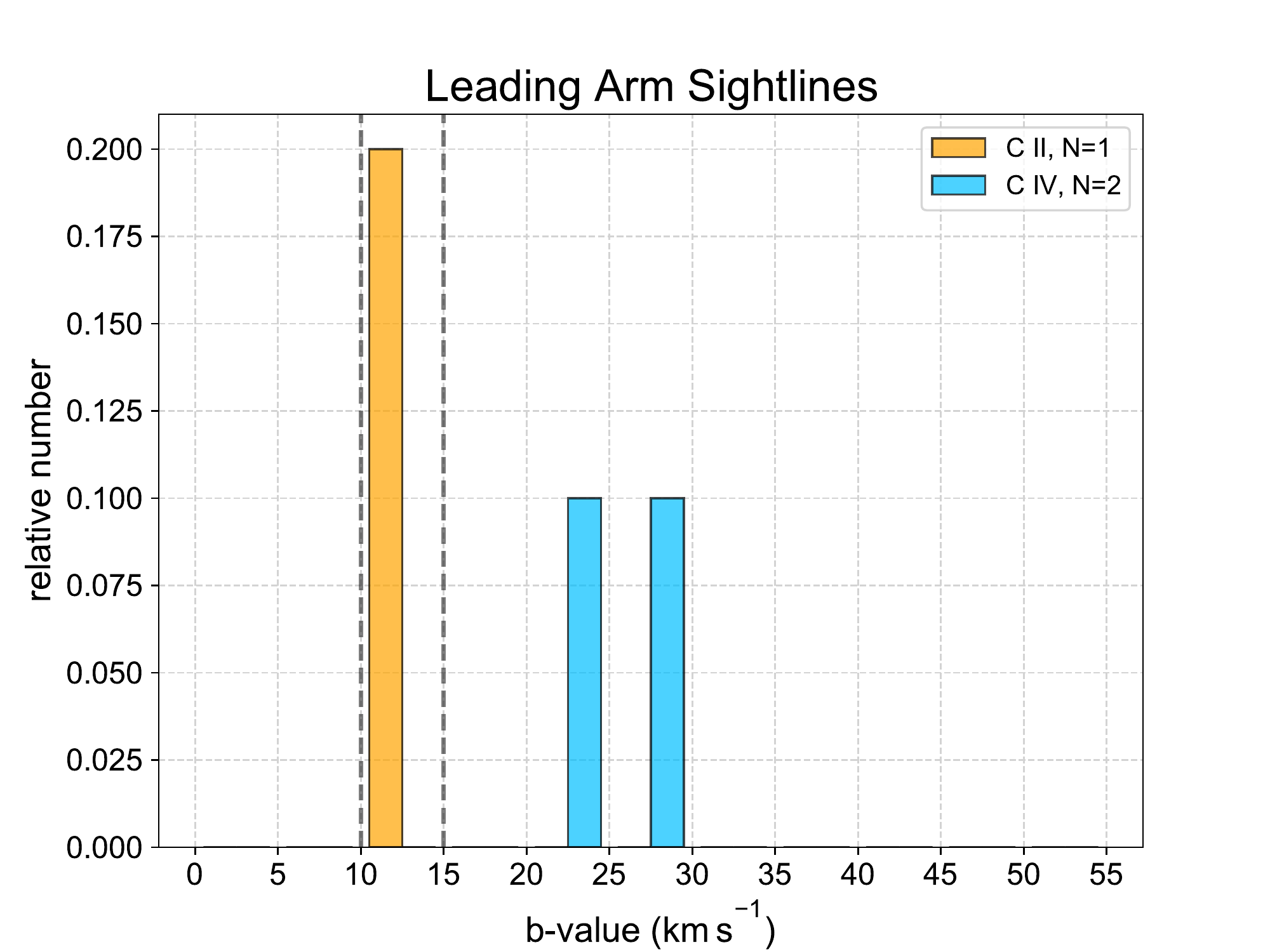}}
\caption{\textit{Top-left}: The distribution of $b$-values for \siw, \sit, and \sif\ 
components in the Stream sightlines. The histograms are normalized 
(given as relative number) and the sample size 
for each ion is indicated in the legend. 
The two dashed vertical lines shows the COS FUV G160M and G130M
instrumental resolution ($b\!\approx\!10$\kms\ and $approx$15\kms).
\textit{Top-right}: same but for \siw, \sit, and \sif\ in the LA sightlines. 
\textit{Bottom-left}: same but for the \cw\ and \cf\ components in the Stream.
\textit{Bottom-right}: same but for the \cw\ and \cf\ components in the LA.
There is no statistically 
significant difference between any of the Stream distributions; the distributions are 
indistinguishable, both for silicon and carbon. 
}
\label{fig:hist_b1}
\end{figure*}

\begin{figure*}[!ht]
\centering 
\subfloat{\includegraphics[width=0.45\textwidth]{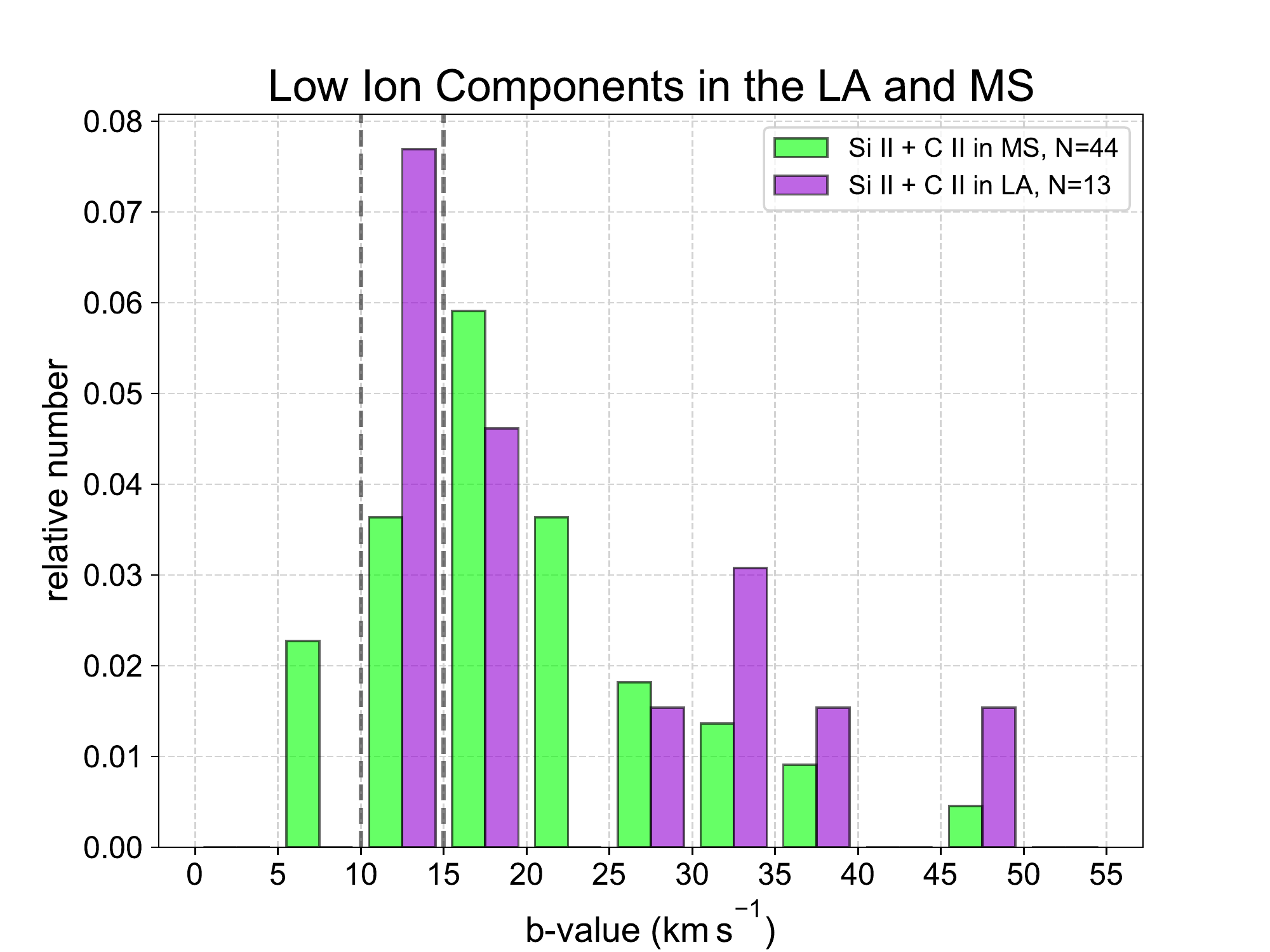}}\qquad
\subfloat{\includegraphics[width=0.45\textwidth]{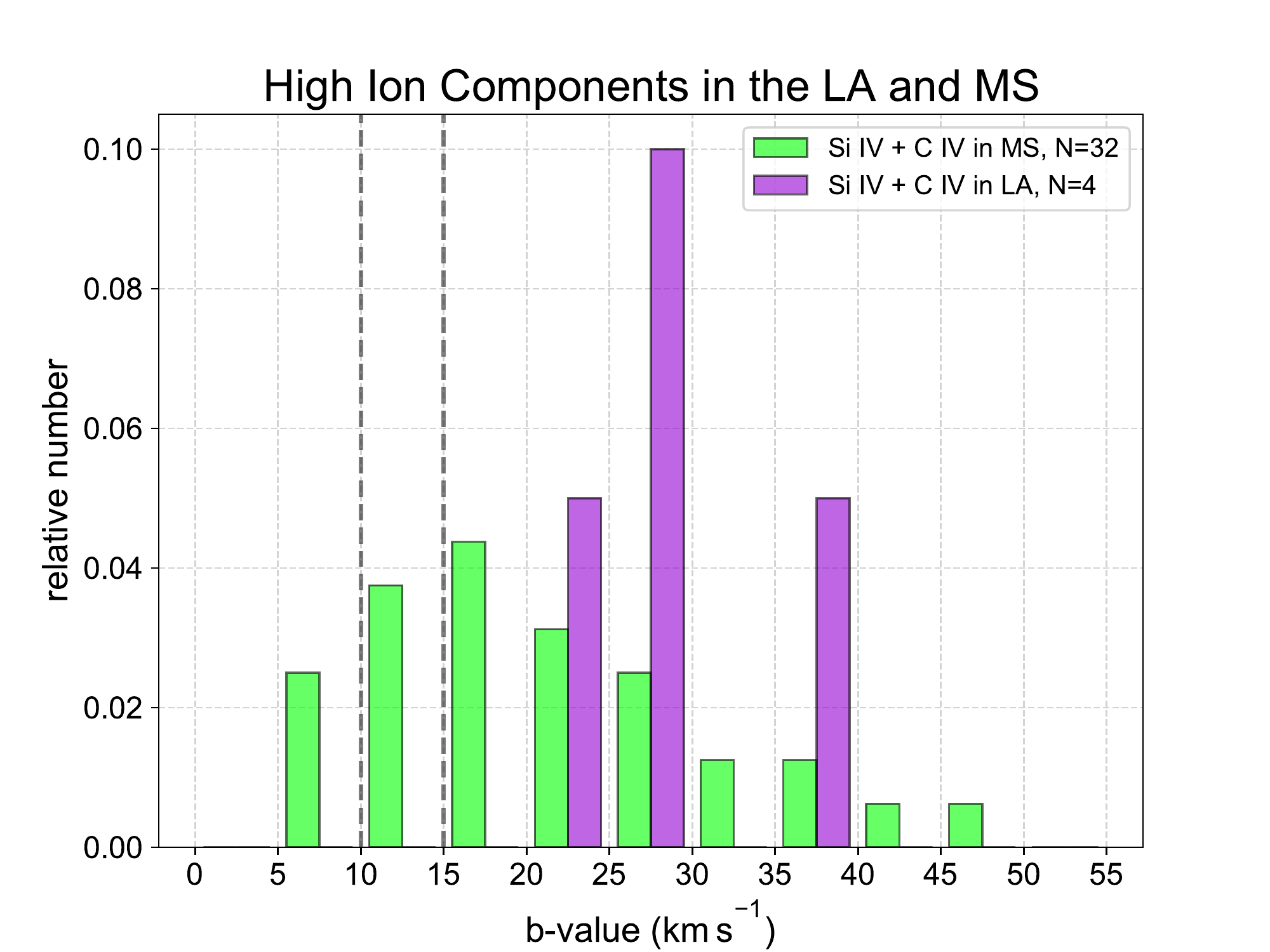}}
\caption{Same as \autoref{fig:hist_b1} but comparing all {\it low-ion} components 
between the Stream and LA in the left panel and all {\it high-ion} components
between the Stream and LA in the right panel. 
The low-ion $b$-values distribute similarly in the Stream and LA, with no evidence for 
any statistical difference. In contrast, a difference in the high-ion $b$-value 
distributions exists between the Stream and LA, with the Stream showing mostly narrow 
($b<25$\kms) high-ion components whereas no such narrow components are seen in the LA.}
\label{fig:hist_b2}
\end{figure*}

The Doppler $b$-parameter is a measure of the line width of an absorption component. 
It encodes information on both the thermal broadening, 
$b_{\rm th}=\sqrt(2kT/Am_{\rm H})$, 
and the non-thermal broadening, $b_{\rm nt}$, which add together in quadrature 
to produce 
the observed line width, $b^2=b_{\rm th}^2+b_{\rm nt}^2$. Here $A$ is the atomic number 
of the absorbing ion, $k$ is the Boltzmann constant, $m_{\rm H}$ is the mass of the 
hydrogen atom, and $T$ is the temperature. Comparing the $b$-value distributions of 
different ions allows differences in their kinematics to be quantified, 
which constrains the co-spatiality of the different species.

In \autoref{fig:hist_b1}, we present the $b$-value distributions for the low-ion (\siw),
intermediate-ion (\sit), and high-ion (\sif) components in the Stream 
(left) and LA (right). 
We focus first on all lines of silicon, because they all 
have the same atomic number, and so their thermal broadening is a function of 
temperature only (top two panels of \autoref{fig:hist_b1}). 
We then repeat this for the lines of carbon (\cw\ and \cf; lower two panels of \autoref{fig:hist_b1}).
These distributions were formed by combining all the HVC components with 
Magellanic identifications (either Stream or LA)
and then making two ``quality-control" cuts to form a reliable sample: 
(1) we only retain significantly-detected components, defined as those $b\!>\!1.5\sigma_b$; 
(2) we only retain components with line widths in the range $5<b<50$\kms, 
since a small number of components outside this range were inspected visually
and determined to be unreliable, based on saturation or low S/N.
The COS/FUV G130M instrumental resolution corresponds to $b\!\approx\!12$\kms, but we choose
to retain components down to 5\kms\ since we would otherwise be excluding narrow components from the sample,  
and although these narrow components are difficult to measure accurately, they are still real.
The sample size varies for the different ions, because the data quality (sensitivity) and 
wavelength coverage varies between systems. 


\begin{deluxetable*}{l lllll lll ccccc}[!ht]
\tabcolsep=4.0pt
\tablewidth{0pt}
\tablecaption{Comparison of $b$-value distributions} 
\tablehead{Sample & $\langle b$(\siw)$\rangle$ & $\langle b$(\sit)$\rangle$ & 
$\langle b$(\sif)$\rangle$ & $\langle b$(\cw)$\rangle$ & $\langle b$(\cf)$\rangle$ &
\multicolumn{2}{c}{\underline{\sit--\siw}} & \multicolumn{2}{c}{\underline{\sif--\siw}} & 
\multicolumn{2}{c}{\underline{\sif--\sit}} & \multicolumn{2}{c}{\underline{\cf--\cw}}\\
 & (\kms) & (\kms) & (\kms) & (\kms) & (\kms) & $D_{\rm KS}$ & $p_{\rm KS}$ & $D_{\rm KS}$ & 
 $p_{\rm KS}$ & $D_{\rm KS}$ & $p_{\rm KS}$ & $D_{\rm KS}$ & $p_{\rm KS}$ }
\startdata
MS   & 18.7$\pm$8.0 & 23.5$\pm$8.6  & 21.2$\pm$11.5 & 21.2$\pm$9.1 & 21.7$\pm$9.6 & 
 0.33 & 0.11 & 0.16 & 0.98 & 0.32 & 0.33 & 0.15 & 0.88\\ 
LA   & 23.5$\pm$11.8 & 29.2$\pm$13.3 & 32.7$\pm$3.7 & $\approx$14.5\tm{a} & $\approx$25.1\tm{a} & 
 0.28 & 0.59 & 0.67 & 0.33 & 0.38 & 0.91 & 1.00 & 0.67\\
\enddata
\tablecomments{Columns 2-6 give the mean and standard deviation of the $b$-value distributions for \siw, \sit, \sif, and \cf\ (\autoref{fig:hist_b1}). Columns 6-14 give the $D$- and $p$-values from two-sided KS tests comparing the distributions of different ion pairs.}
\tn{a}{In the LA there is only one \cw\ component and two \cf\ components, so we do not present a standard deviation.}
\label{tab:b-stats}
\end{deluxetable*}

Inspection of \autoref{fig:hist_b1} shows  
several interesting features of the $b$-value distributions in the Stream and LA. 
In the Stream, the \siw, \sit, and \sif\ all have statistically 
indistinguishable $b$-value distributions.
This is supported by two-sided Kolmogorov-Smirnov (KS) tests between each pair of ions 
[$b$(\siw) vs $b$(\sit), $b$(\siw) vs $b$(\sif), $b$(\sit) vs $b$(\sif)]
which all yield small $D$ statistics with large $p$-values, 
indicating that we cannot rule out the null hypothesis that
the \siw, \sit, and \sif\ $b$-values are all drawn from the same parent population 
(see \autoref{tab:b-stats}, which reports the results from the KS tests).
The three ions all have a peak near $b$=20\kms\ and a tail extending to 
$\approx$50\kms, with mean values of 18.7, 23.5, and 21.2\kms, respectively. 
The \cw\ and \cf\ components in the Stream have a mean $b$-value of 21.2 and 
21.7\kms, respectively (lower panels of \autoref{fig:hist_b1}), i.e. the carbon and 
silicon lines independently provide the same result that the low-ion and high-ion 
kinematics agree.

In contrast, for LA directions there is a suggestion that
the \sif\ tends to be broader than the \siw, with
a mean $b$(\sif)=32.7\kms\ compared to a mean $b$(\siw) of 23.5\kms.
However, a two-sided K-S test shows that the significance of this difference is low 
($D_{\rm KS}$=0.67, $p_{\rm KS}$=0.33), because of the small sample size.
To improve the statistics, in \autoref{fig:hist_b2} we compare the $b$-value distributions 
of {\it all low ions} (\cw\ and \siw) in the Stream and LA on the left 
and {\it all high ions} (\cf\ and \sif) in the Stream and LA on the right. 
The K-S statistics for the Stream vs LA comparison now become
$D_{\rm KS}$=0.69, $p_{\rm KS}$=0.04, showing significant evidence for a statistical 
difference between the two regions. 

\begin{figure*}[!ht]
\centering 
\subfloat{\includegraphics[width=0.45\textwidth]{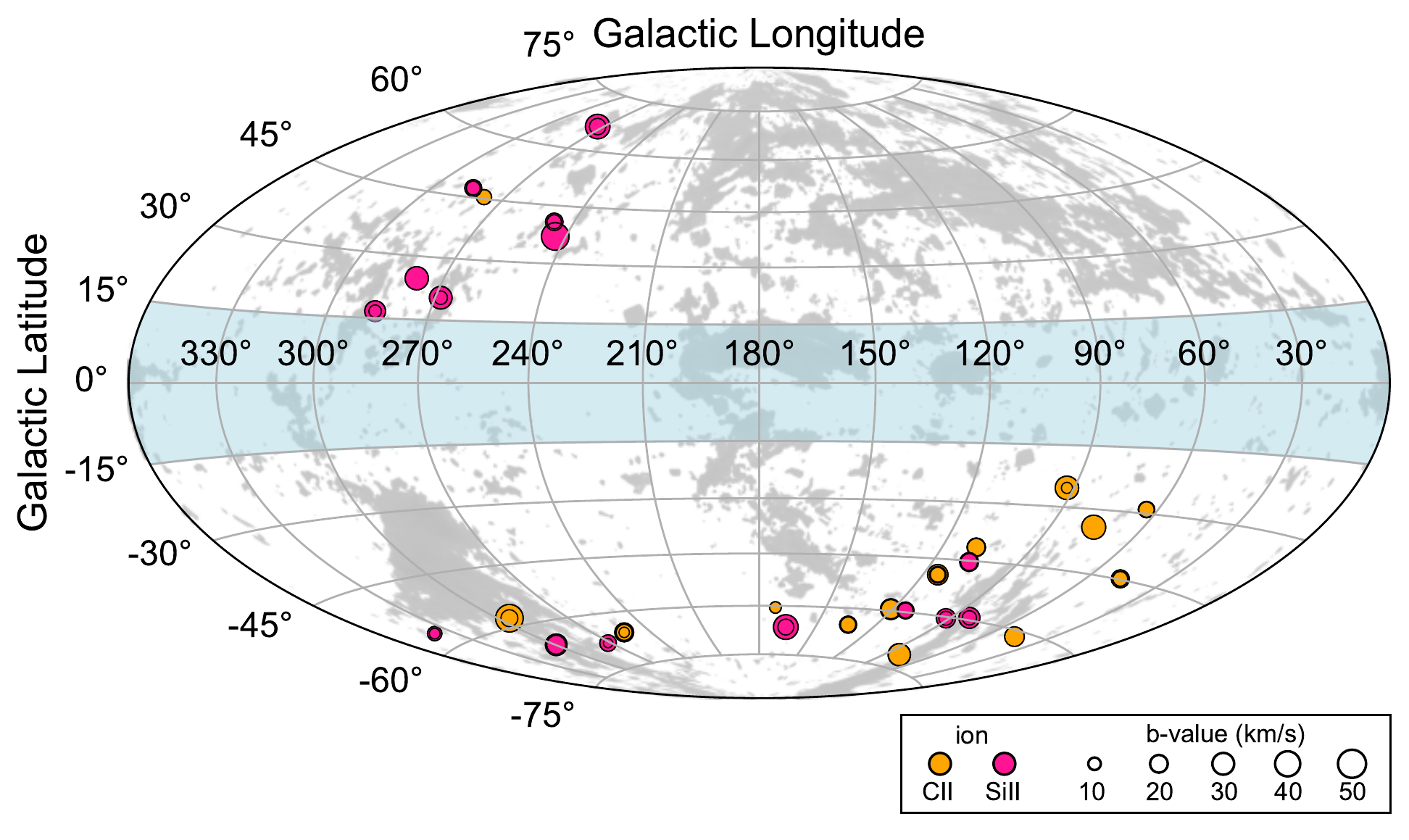}}\qquad 
\subfloat{\includegraphics[width=0.45\textwidth]{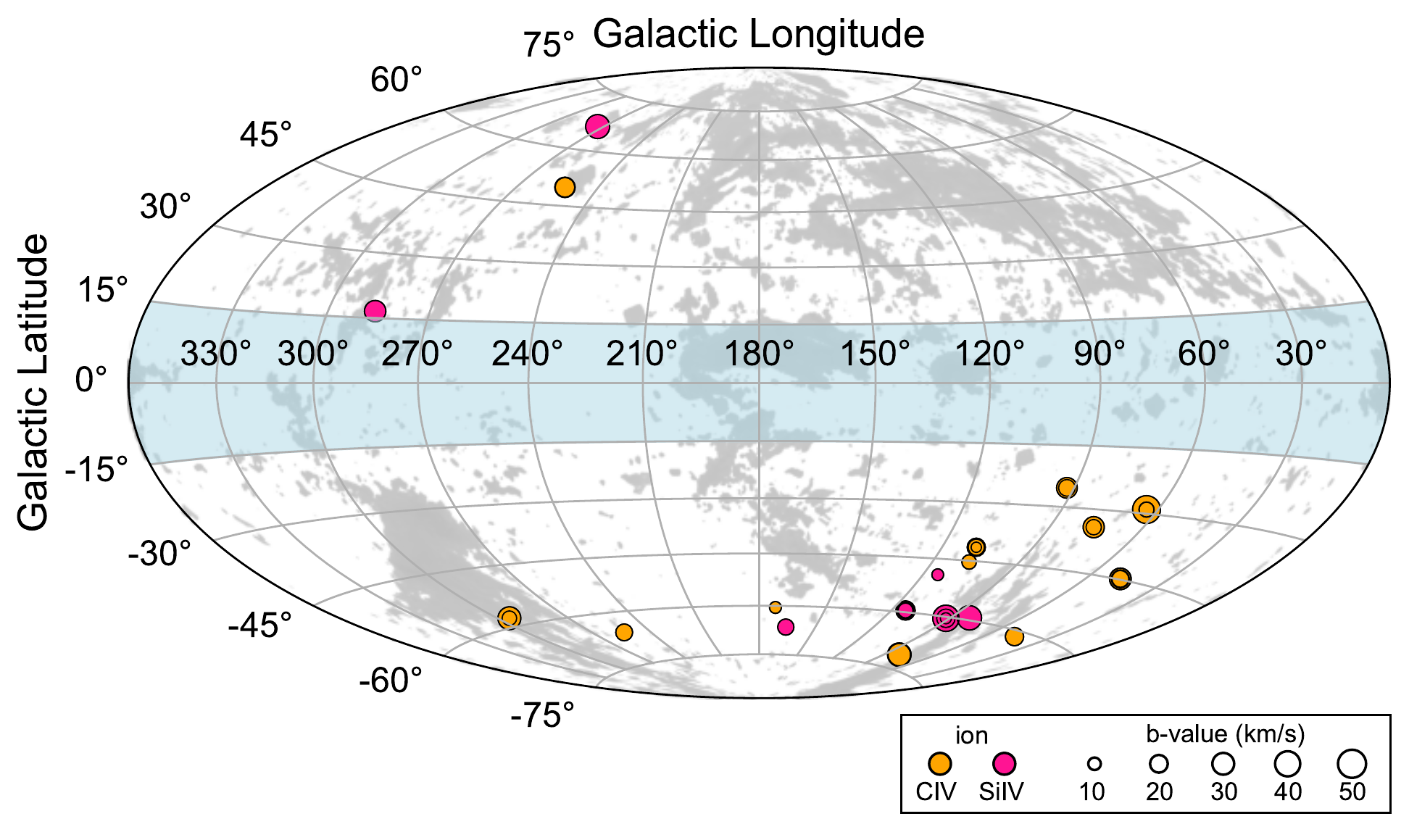}}       
\caption{All-sky maps illustrating the Stream and Leading Arm kinematics for the low ions (left panel) and high ions (right panel). Each Magellanic component is shown as a circle, with the size of the circle proportional to the line width ($b$-value). Concentric circles reflect sightlines with multiple Magellanic components. The grayscale represents the all-sky \hi\ HVC map of \citet{We18}, as in Figure 1.}
\label{fig:bmap}
\end{figure*}

We visualize the $b$-value distribution in the Stream and the LA using the maps shown 
in \autoref{fig:bmap}. In these maps the symbol sizes are proportional to the $b$-value of the 
Magellanic components, so broader components are shown as larger circles. 
While the effect is subtle, the tendency for the LA 
to show broader high-ion components than the Stream is seen in the right-hand map, because
larger circles are preferentially found in the upper-left (LA) region of the map.

While our results are statistical, in the sense that they are reported 
across the ensemble of absorbers in our sample,
a few individual spectra illustrate the aligned, narrow nature of the Stream absorbers
and the misaligned, broader nature of the LA absorbers.
Narrow high-ion components in the Stream are seen in the
spectra of \object{MRC2251--178}, \object{PG0026+129}, \object{PG2349--014}, 
and \object{MRK1044}. 
In contrast, broad high-ion components are seen in
a high fraction of LA directions, including the sightlines to
\object{ESO265-G23} and \object{PKS1101--325}, and (at lower S/N) toward
\object{IRAS\,F09539--0439},  \object{SDSS\,J095915.60+050355.0}
(see \autoref{fig:montage_ms} and \autoref{tab:bigtable}).
Furthermore,  narrow \cf\ and \sif\ components have been reported in 
earlier studies of the Stream and nearby HVCs using the high-resolution 
E140M grating on STIS in 
the sightlines to \object{HE0226--4110} \citep{Fo05},
\object{NGC\,7469}, and \object{Mrk 335} \citep[both in][]{Fo10}. 
Therefore, our result that the Stream has simple kinematics with narrow high-ion components 
has already been observed in high-resolution data, and so
is unlikely to be a COS instrumental broadening effect.

\subsection{Velocity Centroid Alignment} \label{subsec:alignment}

\begin{figure*}[!ht]
\centering 
\subfloat{\includegraphics[width=0.45\textwidth]{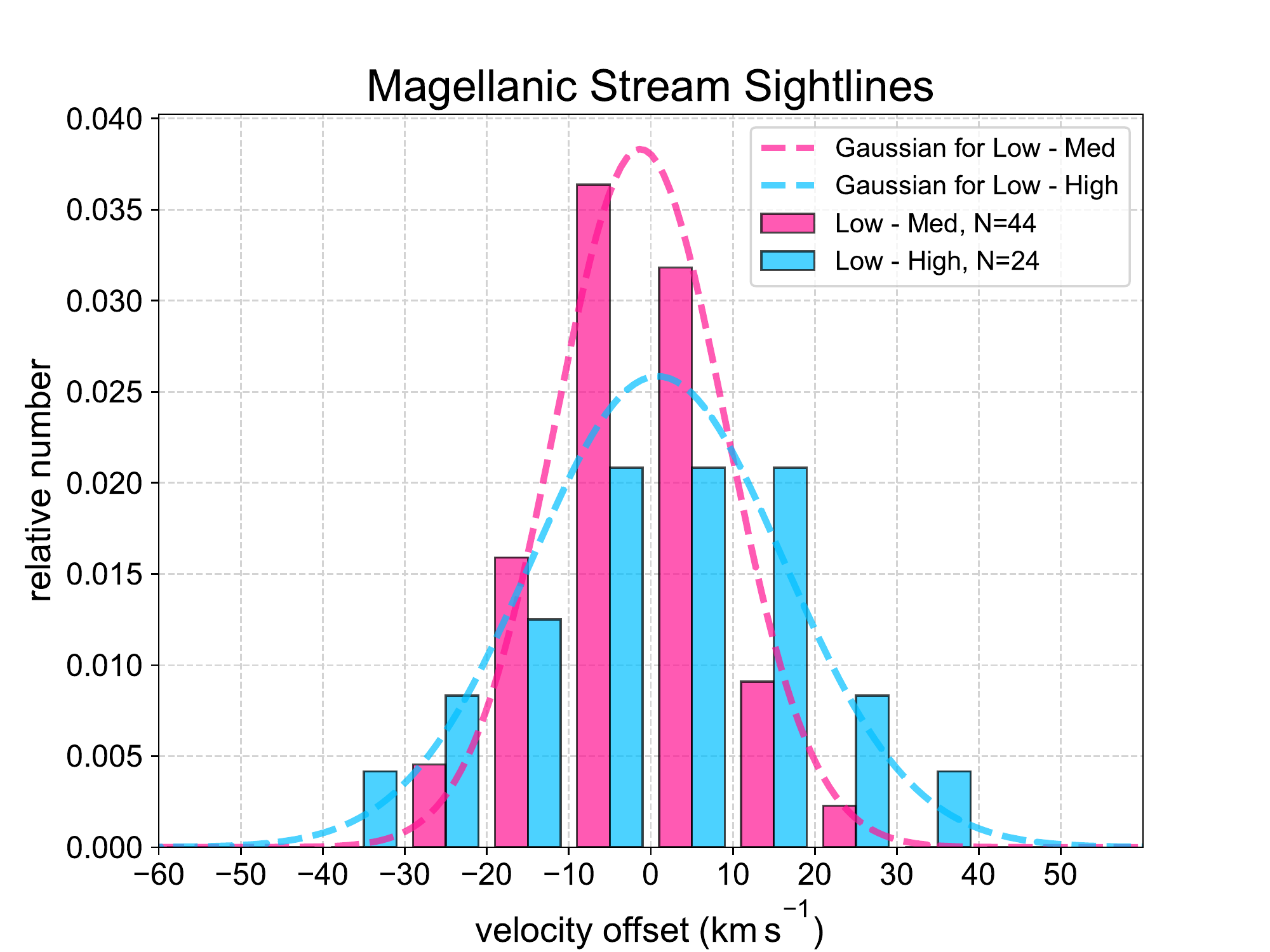}}\qquad 
\subfloat{\includegraphics[width=0.45\textwidth]{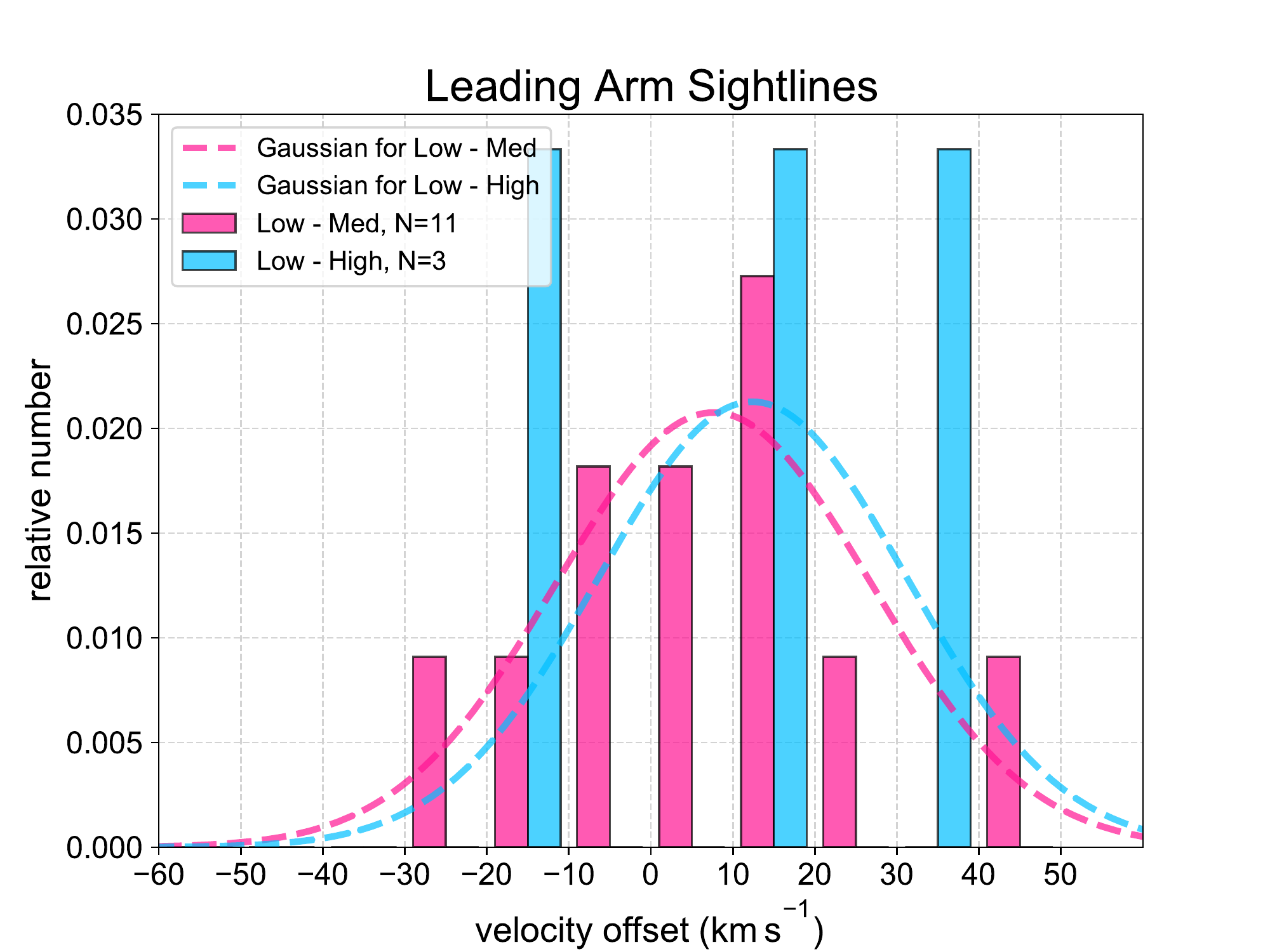}}
\caption{Distribution of velocity centroid offsets for the Stream (left panel) 
and LA (right panel). In each panel, the distributions are shown for two 
ion-pairs: high--low (either \cf--\cw\ or \sif--\siw) or medium--low 
(either \sit--\siw\ or \sit--\cw). The offsets quantify the difference 
in line centers between the two components (see \autoref{subsec:alignment}). 
Both the peak and the width of the offset distribution quantify the degree of 
alignment of the absorbers.}
\label{fig:voffset}
\end{figure*}

In addition to the $b$-values, another kinematic statistic of interest is 
the velocity centroid offset between any two ions. Single-phase gas clouds 
will show no centroid offsets between different absorbing species; 
multi-phase clouds may show significant offsets. 
A commonly invoked multi-phase scenario is an arrangement where hot, collisionally 
ionized boundary layers surround cool, photoionized cloud cores. 
Measurements of the velocity centroid offset can test this scenario
and hence diagnose the presence of collisionally ionized gas.

Here we present a new approach for conducting velocity alignment statistics: 
a nearest-neighbor analysis. In this approach, for each low-ion (\siw) 
Magellanic component we identify the high-ion (\sif) 
component that is closest in velocity. 
We then measure the velocity centroid offset $\Delta v_0=v_0$(\sif)--$v_0$(\siw) 
together with its error, which is formed by summing the 
individual errors on $v_0$(\siw) and $v_0$(\sif) errors in quadrature. 
We then make a normalized distribution of these offsets 
and measure the mean value $\langle \Delta v_0 \rangle$ and 
Gaussian width $\sigma(\Delta v_0)$ of this distribution,
excluding any outliers at $\Delta v_0>50$\kms, which are likely related to low data quality
rather than real offsets.
We then repeat this exercise using \sit\ and \siw, i.e. by analyzing the distribution of 
$v_0$(\sit)--$v_0$(\siw). For single phase clouds, one expects a narrow distribution 
centered on zero. For multi-phase clouds, one expects a broader distribution that is 
not necessarily centered on zero. This analysis is illustrated in \autoref{fig:voffset}, 
showing the distribution of velocity offsets 
for the two pairs of ions, first in the Magellanic Stream (left) and second 
in the LA (right). 
We also present a summary of the velocity alignment statistics in \autoref{tab:align}.

\begin{deluxetable*}{l ll ll}[!ht]
\tabcolsep=4.0pt
\tablewidth{0pt}
\tablecaption{Velocity Alignment Statistics: Stream vs Leading Arm}
\tablehead{Sample & \multicolumn{2}{c}{\underline{~~~~~~~Intermediate--Low~~~~~~~}} & \multicolumn{2}{c}{\underline{~~~~~~~High--Low~~~~~~~}} \\
& $\langle \Delta v_0 \rangle$ & $\sigma(\Delta v_0)$ &
$\langle \Delta v_0 \rangle$ & $\sigma(\Delta v_0)$ \\
& (\kms) & (\kms) & (\kms) & (\kms)}
\startdata
MS & $-$1.3 & 10.4 & 0.8 & 15.4\\
LA & 7.6 & 19.2 & 12.4 & 18.8\\
\enddata 
\tablecomments{This table gives the mean velocity centroid offset, $\langle \Delta v_0 \rangle$,
and its standard deviation, $\sigma(\Delta v_0$), for two pairs of ions: intermediate--low (\sit--\siw\ or \sit--\cw) and high--low (\sif--\siw\ and \cf-\cw). Small values of $\langle \Delta v_0 \rangle$ and $\sigma(\Delta v_0$) support single-phase models; larger values support multi-phase models. The absolute velocity scale uncertainty of the COS FUV channel is $\approx$7.5\kms\ \citep{Pl19} for the standard pipeline reduction.}
\label{tab:align}
\end{deluxetable*}

The velocity offset distributions shown in \autoref{fig:voffset} reveal interesting results.
In the Stream, the distribution of absolute 
velocity offsets between \sit\ and \siw\ is fairly \emph{narrow}, 
with a Gaussian width of 10.4\kms\ and a mean value of $-$1.3\kms.  
The distribution of offsets between \sif\ and \siw\ is slightly broader 
with a Gaussian width of 15.4\kms\ and a mean value of 0.8\kms.
In contrast, in the Leading Arm the corresponding \sit--\siw\ offset 
distribution has a \emph{larger width} of 19.2\kms\ and a larger mean 
value of 7.6\kms, and the \sif--\siw\ distribution also has a width of 18.8\kms\ 
(\autoref{tab:align}).
These values provide further evidence that the Leading Arm has more kinematic 
complexity than the Stream, reinforcing the
results found from the $b$-value distributions.
The Stream's offset distributions and $b$-value distributions are consistent with a single phase, 
whereas the corresponding distributions in the LA support a multi-phase structure.

Note that the velocity offset distributions are related to data quality, because
the S/N ratio in a given spectrum impacts the detectability of weak components. 
For example, a weak \sif\ component that is not detected at 3$\sigma$
significance will not be included in the \emph{VoigtFit} model, even if 
it is well-aligned with a \siw\ component; in such a case the nearest-neighbor
analysis will instead find an alternative, spurious closest \sif\ 
component to match to the \siw. This would serve to over-estimate the true velocity offset.
This effect should be minimized by out choice to exclude outliers with $\Delta v_0>50$\kms\ from the Gaussian fits 
to the offset distributions. There were only a handful of such outliers, and visual examination 
showed they were likely related to low S/N.

In summary, the velocity centroid offsets independently yield the 
same conclusion as the $b$-value distributions,
namely that the high- and low-ions show simpler kinematics (suggestive of co-spatiality) in the Stream 
but more complex kinematics (suggestive of non-cospatiality) in the Leading Arm. 
This dual finding adds to the robustness of the result.

\section{Discussion}\label{sec:discussion}

Our new results constitute the first system-wide analysis of the
UV kinematic properties of the Stream and LA. They
complement existing studies of the \hi\ 21\,cm kinematics 
\citep{Pu98, KH06, For13, For14} and high-resolution UV studies of 
a few Magellanic sightlines observed with the echelle gratings on
\hst/STIS \citep{Fo10, Ku15, Ri18}.
The distributions of $b$-values and velocity centroids of both the low and high ions, 
and their dependence on position within the Magellanic System, provide important 
information on the phase structure and origin of the ionized gas, and represent 
important observational constraints for models of the origin and evolution of 
the Magellanic System. We stress that kinematics alone can determine
whether an absorber is multi-phase, even without
ionization modeling, because complex kinematics rule out single-phase models.

\subsection{The Photoionized Magellanic Stream}

Our results provide observational evidence that the Magellanic Stream has a simple
kinematic phase structure.
Single-phase models can explain the \siw, \cw, \sit, \sif, and \cf\ absorption in the Stream 
because these ions have indistinguishable $b$-values distributions (Figure~\ref{fig:hist_b1})
and narrow velocity-centroid-offset distributions (Figure~\ref{fig:voffset}). 
In contrast, we find tentative evidence that the LA is multi-phase,
because it shows high-ion components (in \cf\ and \sif)
that are broader than the low ions (See Figure~\ref{fig:hist_b1}), 
though more high-S/N data are needed to confirm this in a larger sample.
This is suggestive of different physical conditions between the two structures;
the high-resolution STIS analysis of the LA presented by \citet{Ri18} 
supports the multi-phase picture.


The Stream's simple, single-phase kinematic structure with narrow $b$-values
indicates that it is \emph{photoionized} up to \cf\ (at least). Since \cf\ has 
an ionization potential of creation of 48\,eV (to ionize C$^{+2}$ into C$^{+3}$), 
this constrains the radiation field incident on the Stream. 
The finding that the Stream is 
photoionized up to \cf\ is an \emph{observational} result, since it
is inferred from the UV kinematic data in a model-independent manner.

The Stream's simple UV kinematics are broadly consistent with its \hi\ kinematics; 
high-resolution 21\,cm studies show components with a range of narrow velocity dispersion, 
almost all $<$20\kms\ \citep{KH06, For14}. However, the widespread presence of \ion{O}{6} 
in the Stream \citep[an even higher ionization species than \cf;][]{Se03, Fo05}
and the occasional detection of molecular gas \citep{Ri13} indicate that the Stream 
overall is multi-phase -- our finding of a single phase relates to the UV lines under study only.

The natural question to ask is what is the source of the Stream's photoionization?
Candidate sources of ionizing radiation are hot stars in the MW and Magellanic Clouds 
and the extragalactic UV background (UVB).
However, H$\alpha$ studies \citep{Ba17, BH19} have reported the inability of
hot stars or the UVB to explain the bright
observed H$\alpha$ intensity observed along the Stream, and have concluded that an additional source 
of ionization is required. This is particularly true in the region of the Stream below 
the South Galactic Pole (SGP), where several clouds with elevated \ha\ intensity are observed 
\citep{Pu03b}, although some of these clouds may be at different distance than the Stream and so may have a non-Magellanic origin.

UV studies of the Stream's ionization \citep{Se03, Fo05, Fo10, Fo13, Ku15} have also reported the inability of MW
and UVB radiation to explain the Stream's observed ionization properties.
Detailed {\it Cloudy} photoionization models that include MW and UVB radiation (but do not include a Seyfert flare)
are unable to explain the levels of high-ion absorption observed in UV studies of the Stream. 
For example, \citet{Ku15} reported that the \cf\ column densities in a Compact HVC
off the edge of the main body of the MS are under-predicted by 3\,dex by the {\it Cloudy} models. 
\citet{Fo10} reported similar findings in {\it Cloudy} ionization modeling of two Stream directions.
These models account for the non-uniform distribution of escaping radiation from the MW, 
since they include the enhanced escape fraction of starlight along the Galaxy's minor axis, but even then they 
cannot explain the observed \cf\ because the underlying stellar spectrum is not hard enough.
A different source of ionizing radiation is required.

\subsection{The Galactic Center Flare} \label{subsec:gcflare}

One potential source of ionizing photons is a Seyfert flare from 
the GC \citep{BH13, BH19}. In the Seyfert-flare model, the flare photoionized 
the Stream as it passed underneath the SGP where the flux of escaping ionizing radiation is high, 
but not the Leading Arm, which lies outside of the ionization cone.
The Seyfert-flare model naturally explains the 
simple, single-phase kinematic structure of the Stream
presented in this paper because it is a photoionization model.
It is also the only known source of radiation that is powerful 
enough to photoionize the \cf\ in the Stream.
We now discuss the evidence for this model.

An enhancement in the Stream's H$\alpha$ intensity in the region below the SGP 
was first noticed by 
\citet{Pu03b} and confirmed by later 
H$\alpha$ measurements \citep{BH13, Ba17}, 
which show emission at $\approx$500 milli-Rayleighs below 
the SGP compared to $\sim$50--100 milli-Rayleighs across the rest of the Stream. 
Although the interpretation of the H$\alpha$ enhancement is complicated by
the unknown distance to the clouds,
the enhancement can be understood as fluorescence induced by a recent GC flare, in which 
the Milky Way's central supermassive black hole (SMBH) 
Sgr A$^*$ underwent an outburst several Myr ago \citep{BH13, BH19}, releasing a burst of ionizing 
radiation and potentially creating the giant X-ray/$\gamma$-ray Fermi Bubbles 
at the same time. This burst would have preferentially ionized the polar regions 
of the Stream since they lie in the ionization cone directly underneath the GC.
The Stream would then recombine and produce the observed H$\alpha$ enhancement. 
In this scenario, the Magellanic Stream acts as a screen on which AGN-induced fluorescence occurs.

In contrast to the H$\alpha$ observations, the \cf/\cw\ ratios 
in the Stream do not show an unambiguous enhancement below the SGP \citep{BH19}. 
Instead, they are highest in the MS tip region, farthest from the Magellanic Clouds, 
where the gas is very fragmented. However, this enhancement may simply reflect the low \hi\
column density $N$(\hi) in that remote portion of the Stream. The 
\cf/\cw\ ratio depends not only on the shape and intensity of the 
radiation field, but also on $N$(\hi): gas with low $N$(\hi) is optically 
thin and so can show a high \cf/\cw\ ratio even in a weak radiation field. 
Therefore, although the \cf/\cw\ 
ratio provides important ionization information, it does not offer 
a clean diagnostic of the incident radiation field, and 
while the Stream's ion ratios are consistent with the Seyfert 
flare model, they do not require it. 

The \emph{kinematics} of the UV absorbers presented in this paper provide 
stronger evidence. Clouds photoionized by an ionizing flare will be single-phase 
and therefore show similar kinematics 
between low and high ions, with similar line widths and small velocity centroid offsets. 
Our finding that the Stream has simple, single-phase kinematics is
fully consistent with the GC flare model, 
because the Stream's orbit takes it below the SGP \citep{GN96, Be07, Be10}
where the flux of escaping ionizing radiation is highest. 
In contrast, the LA lies closer to the major axis of the disk, where it is shielded 
from the flare's ionization cone \citep{BH19}, potentially explaining the lack
of narrow high-ion components in our LA data.
We thus conclude that the Stream's UV kinematics 
are fully consistent with and provide circumstantial support to the GC flare model, 
but they do not require it, because other unknown sources of radiation could be present.
A follow-up study on the kinematics of HVCs in non-Magellanic directions 
(particularly in the northern Galactic hemisphere) would be an interesting
test of the Seyfert flare model.

The Seyfert flare model is consistent with several independent observed 
properties of the Stream, including the 
elevated H$\alpha$ intensities near the SGP \citep{Pu03b, Ba17},
the UV line ratios \citep{BH19}, and the UV kinematics (this paper).
An AGN event such as a Seyfert flare also natural explains 
many key properties of the Fermi Bubbles, 
including their spatial extent and energetics \citep{GM12, Gu12},
spatially uniform gamma-ray spectrum \citep{YR17},
X-ray emission properties \citep{MB16}, and
kinematic age based on entrained cool gas \citep{Fo15, Bo17}.
The simplest explanation of these results is that the Seyfert flare 
was the same event that created the Fermi Bubbles.


\subsection{The Kinematics and Ionization of the Leading Arm} \label{subsec:mag-group}

The complex kinematics of the LA, with broader $b$-values for 
\sif\ and \cf\ than for \siw\ and \cw\ (albeit with a small sample size), 
indicate the LA is multi-phase \citep{Ri18}. 
The LA also shows spatially variable chemical abundances, with oxygen abundances that vary 
from 4\% solar to 30\% solar between different cloud regions \citep{Lu98, Fo18, Ri18}.
These complex, multi-phase conditions provide useful clues to the origin(s) of the LA. 

The multi-phase nature of the LA suggests that a different ionization
mechanism is required for the high ions in the LA than in the Stream. 
Collisional processes including shocks \citep{BH07, BH13, TG15}, thermal conduction 
\citep{Gn10, Bo90}, and turbulent mixing of cool and hot gas \citep{Kw15, Ji18}
are all expected to be enhanced in the LA because of its proximity to the MW.
This proximity leads to an interaction with a much denser external medium than the Stream
encounters.
Both observations \citep{MG08, AD20} and models \citep{Be07, Pa18} indicate that the Leading Arm 
($d_{\rm LA}\approx$20\,kpc) 
is much closer to the MW than the Stream is ($d_{\rm MS}>55$\,kpc, and possibly 
$d_{\rm MS}\approx75-150$\,kpc).
We suggest that distance (and therefore density of the external medium) is the primary reason 
why the high-ions appear to be 
collisionally ionized in the LA but not in the Stream.

The origins of the LA remain unclear. In the classical 
picture, the LA is formed from tidally stripped Magellanic gas pulled in 
front of the orbit of the Clouds \citep[e.g.][]{Pu98, Be07, Pa18}.
However, recent work has raised the possibility of contributions from other sources.
Parts of the LA, with its highly fragmented \hi\ structure \citep{For13},
head-tail morphology \citep{Pu11}, spatially-variable metallicity \citep{Fo18, Ri18}, 
and stellar counterpart \citep{PW19, Ni19, Be19} may represent the 
debris field left over from the accretion and disruption of a forerunner (or forerunners) 
from the Magellanic Group \citep{Ha15, TG19} or a stellar cluster in the Galactic halo. 
Gas condensed from a Magellanic Corona may also contribute to the LA \citep{Lu20}, as may Galactic gas.
The LA's complex, multi-phase UV kinematics presented in this paper and in \citet{Ri18}
represent important constraints but by themselves do not allow us to distinguish between origin mechanisms.
A full investigation into the physical conditions of the gas in the LA
using ionization modeling (and ideally with higher S/N data) would be worthwhile to address these open issues.

\section{Conclusions} \label{sec:conclusions}

Using a sample of \nsamp\ \hst/COS extragalactic sightlines toward background AGN
(\nms\ through or near the Magellanic Stream and \nla\ through or near the Leading Arm),
we have presented the first detailed kinematic analysis of the UV metal-line 
absorption from the Magellanic System. We conducted Voigt-profile fits using 
the \emph{VoigtFit} software package to characterize the low-ion (\siw, \cw), 
intermediate-ion (\sit) and high-ion (\sif, \cf) component structure. 
We derived line centers, line widths, and column densities for each component 
then calculated the $b$-value distributions for each ion in both the Stream and the LA, 
as well as the velocity centroid offset distributions. We used two-sided K-S tests to explore 
whether any statistically significant differences exist between the kinematics of different ions,
and performed a comparative study of the Stream and LA kinematics.
Our main results are as follows.
   
\begin{enumerate}

    \item In the Stream, the $b$-values distributions for \siw, \sit, \sif, \cw,
    and \cf\ are statistically indistinguishable. 
    All five ions show a distribution with a peak near $b$=15--20\kms\ and a tail 
    extending to $b\approx50$\kms\
    (compared to an instrumental line width of only $\approx$12\kms). 
    Furthermore, the distribution of velocity centroid 
    offsets between intermediate- and low-ion 
    components in the Stream is narrow and centered near zero, 
    with a Gaussian width of only 10.4\kms. 
    Both these results indicate the Stream tends to show simple kinematics with 
    a predominantly single-phase structure for the ions under study.
    
    \item In contrast, the Leading Arm $b$-values for the low-ions 
    and high-ions distribute differently, although the sample size is small. 
    The \sif\ $b$-values (mean of 32.7\kms)
    tend to be broader than the \siw\ $b$-values (mean of 23.5\kms)
    and \sit\ $b$-values (mean of 29.2\kms).
    The distribution of velocity centroid offsets between \sit\ and \siw\
    components in the Leading Arm is broader than in the Stream, with a Gaussian 
    width of 19.2\kms.
    Both these results indicate that the Leading Arm has complex kinematics 
    with a multi-phase structure, as found in earlier work \citep{Ri18}.
    
    \item The finding that the Stream is predominantly single-phase suggests
    that it is photoionized up to \cf, the most highly ionized species in our dataset.
    In contrast, there is no evidence for photoionized \sif\ and \cf\
    in the Leading Arm, because its \sif\ and \cf\ components tend to be 
    broader and therefore collisionally ionized.
    The different ionization mechanism for the high ions 
    can be understood in terms of the LA's proximity to the MW 
    \citep[$d_{\rm LA}\approx$20\,kpc;][]{MG08, AD20}, which
    causes it to interact with a much denser external medium than the Stream does
    \citep[$d_{\rm MS}\approx$75--150\,kpc according to models;][]{Be07, Pa18, Lu20}.
    
    \item The simple, single-phase photoionized nature of the Stream 
    can be naturally explained 
    by the Seyfert flare model \citep{BH13, BH19}, 
    in which a flash of ionizing radiation from the GC photoionizes the Stream 
    as it passes under the south Galactic pole, where the escape fraction is highest.
    The Seyfert flare is the only known source of radiation that is both
    powerful enough to explain the H$\alpha$ intensity of the Stream and hard enough
    spectrally to photoionize \sif\ and \cf\ to the observed levels.
    \end{enumerate}

{\it Acknowledgements.}
Support for programs 12604 and 14687 was provided by NASA through a grant from the Space Telescope Science Institute, which is operated by the Association of Universities for Research in Astronomy, Inc., under NASA contract NAS5-26555.
We are grateful to Jens Krogager for his assistance with installing and implementing {\it VoigtFit}, to Max Gronke and Adam Ritchey for useful scientific discussions, and to Tobias Westmeier for assistance with HVC maps. We thank the referee for a useful report that improved the quality of the paper.
\\
\\
{\it Facilities:} HST (COS)\\   
\\
{\it Software:} VoigtFit \citep{Kr18}

\appendix
\section{Impact of the COS Line Spread Function (LSF)}
\label{sec:lsf}

\begin{deluxetable*}{llrrr}[!ht]
\tabcolsep=2.0pt
\tablewidth{0pt}
\tablecaption{Effect of COS Line Spread Function on {\it VoigtFit} Output Parameters }
\tablehead{Ion & $v_0$ & $\delta b$\tm{a} & $\delta$\,log\,$N$\tm{b} & Note\\
 & (km s$^{-1}$) & (km s$^{-1}$) & (dex) & }
\startdata
\cw\ 
&   4 & $-$4.7$\pm$1.1  & 0.13$\pm$0.03 & $*$\\
&  82 &	$-$5.0$\pm$15.5 & $-$0.18$\pm$0.30 & \\
& 113 &	2.4$\pm$8.4	    & 0.05$\pm$0.16 & \\
& 155 &	$-$2.5$\pm$3.9	& 0.02$\pm$0.06 & \\
& 205 &	$-$1.8$\pm$3.2	& 0.03$\pm$0.04 & \\
\sit\
&   3 & $-$6.0$\pm$1.1	& 0.17$\pm$0.03 & $*$\\
&  95 & $-$4.8$\pm$7.7	& $-$0.09$\pm$0.11 & \\
& 160 & $-$1.3$\pm$4.7	& 0.04$\pm$0.06 & \\
& 207 & $-$3.5$\pm$2.2	& 0.09$\pm$0.05 & \\
\cf\
&   3 & $-$6.0$\pm$1.1	& 0.17$\pm$0.03 & $*$\\
&  95 & $-$4.8$\pm$7.7	& $-$0.09$\pm$0.11 & \\
& 160 & $-$1.3$\pm$4.7	& 0.04$\pm$0.06 & \\
& 207 & $-$3.5$\pm$2.2	& 0.09$\pm$0.05 & \\
\enddata
\tablecomments{This table shows the difference in \emph{VoigtFit} output parameters ($\delta b$ and $\delta$ log\,$N$) between two sets of models for the absorption components toward the AGN \object{HE0226-4110}, one using the tabulated COS LSF and one a Gaussian LSF. The differences are calculated in the sense 
$\delta b=b_{\rm tabulated}-b_{\rm Gaussian}$ and
$\delta$log\,$N$=log\,$N_{\rm tabulated}$--log$N_{\rm Gaussian}$.
Components with significant ($>$2$\sigma$) differences are marked with a star in the final column. For all three ions, only the strong low-velocity (Galactic) component shows significant differences in $b$; the HVC results are not sensitive to the choice of LSF.}
\tn{a}{Difference in line width, with its error.}
\tn{b}{Difference in logarithmic column density, with its error.}
\label{tab:lsf-choice}
\end{deluxetable*}

The {\it VoigtFit} models presented in this paper were derived assuming the COS/FUV line spread function (LSF) is a Gaussian with a full-width at half maximum FWHM=$c/R$, where $R$=16,000 for G130M observations and $R$=19,000 for G160M observations. When our paper was nearing completion, a newer version of {\it Voigtfit} became available with the ability to handle non-Gaussian LSFs, allowing the use of the official tabulated COS LSFs, which are slightly non-Gaussian\footnote{The COS LSFs are available at \href{http://www.stsci.edu/hst/instrumentation/cos/performance/spectral-resolution}{http://www.stsci.edu/hst/instrumentation/cos/performance/spectral-resolution}.}. The COS LSFs have a Gaussian core but include extended wings due to the micro-roughness of the surface of the {\it HST} primary mirror, which transfers $\approx$3\% of the light from line center to the wings \citep{Kr11}. 

To quantify the effect of using the tabulated LSFs instead of the Gaussian LSFs, we ran a test case using the AGN HE0226-4110, which was observed at COS Lifetime Position 1 (LP1). This sightline was chosen because of its high S/N COS spectrum, good spectral resolution (LP1 has the highest resolution of all the COS FUV lifetime positions), and the presence of multiple high-velocity metal components of differing line strengths, allowing us to assess the impact of the LSF for both weak and strong lines. We considered three ions: \cw, \sit, and \cf, chosen to sample the low ions, intermediate ions, and high ions, respectively. By fitting two sets of {\it VoigtFit} models, one with the tabulated non-Gaussian LSF and one with the Gaussian LSF, we calculated the difference in the output fit parameters $b$ and log\,$N$. The results are summarized in \autoref{tab:lsf-choice}.
For all three ions, the fit parameters obtained with the two LSFs
agree within 2$\sigma$ for all high-velocity components (including all Magellanic components), but differ for the strong low velocity components, which traces the Galactic ISM. For these low-$v$ components, using the tabulated COS LSF instead of a Gaussian LSF leads to a narrower line width (by 5\,\kms) and a larger column density (by 0.15\,dex). Therefore for HVCs, there is no evidence for a significant difference in component parameters when using the tabulated LSF versus a Gaussian LSF, and so the kinematic analysis presented in this paper is unaffected by this choice. However, for strong low-velocity absorbers, using the COS LSF has a non-trivial impact on the results.

\section{VoigtFit Results}
\label{sec:voigt-results}

In \autoref{tab:bigtable} we present the full table of {\it Voigtfit} results for each Magellanic component in our sample (i.e. each HVC with a Stream or Leading Arm identification). We list the velocity centroid ($v_0$), line width ($b$) and column density (log\,$N$) of each component. The $b$-value distributions and velocity-centroid-offset distributions analyzed in the paper are based on these raw measurements. We also list the S/N per resolution element measured in the continuum next to each line. Components marked on \autoref{fig:montage_ms} with the letter ``B" (blends), ``N" (non-Magellanic HVCs), and ``U" (uncertain, low-significance HVCs) are are not included in the table. To be classed as significant, we only include components with $5<b<50$\kms\ and $b>1.5\sigma_b$, i.e. reliably measured values.

\input{final-fit-table.tex}

\end{document}

%% file: final-fit-table.tex
\startlongtable
\begin{deluxetable*}{lllrrrr}
\tablewidth{0pt}
\tablecaption{Component Parameters for Magellanic HVCs}
\tablehead{Sightline & Sample & Ion & $S/N$ & $v_0$ & $b$ & log\,$N$(ion)\\ 
                     &        &     &  (per resel)  & (km s$^{-1}$) & (km s$^{-1}$) & ($N$ in cm$^{-2}$)}
\startdata
FAIRALL9 & MS & CII & $55$ & $93.9\pm3.3$ & $  18.4\pm   6.7$ & $13.733\pm 0.142$\\ 
  &   & CII & $55$ & $175.8\pm0.9$ & $  48.2\pm   1.7$ & $14.952\pm 0.011$\\ 
  &   & SiIII & $40$ & $105.0\pm1.2$ & $   8.2\pm   3.2$ & $12.698\pm 0.071$\\ 
  &   & SiIII & $40$ & $165.9\pm1.2$ & $  43.3\pm   1.4$ & $13.580\pm 0.012$\\ 
  &   & CIV & $36$ & $126.9\pm1.8$ & $  33.0\pm   4.1$ & $13.624\pm 0.038$\\ 
  &   & CIV & $36$ & $180.9\pm2.7$ & $  13.4\pm   4.5$ & $12.953\pm 0.119$\\ 
HE0153-4520 & MS & SiII & $24$ & $91.2\pm2.4$ & $   5.5\pm   8.1$ & $12.808\pm 0.103$\\ 
  &   & SiII & $24$ & $136.2\pm5.4$ & $  17.6\pm   9.9$ & $12.702\pm 0.148$\\ 
  &   & SiIII & $27$ & $117.6\pm3.3$ & $  22.8\pm   3.7$ & $12.923\pm 0.071$\\ 
  &   & SiIII & $27$ & $194.7\pm0.9$ & $  19.4\pm   1.3$ & $12.880\pm 0.020$\\ 
HE0226-4110 & MS & CII & $47$ & $84.0\pm7.5$ & $  16.4\pm  10.7$ & $13.221\pm 0.204$\\ 
  &   & CII & $47$ & $113.1\pm2.7$ & $   7.0\pm   6.8$ & $13.345\pm 0.125$\\ 
  &   & CII & $47$ & $154.8\pm1.5$ & $  18.8\pm   3.0$ & $13.881\pm 0.045$\\ 
  &   & CII & $47$ & $204.6\pm1.5$ & $  23.0\pm   2.4$ & $13.960\pm 0.033$\\ 
  &   & SiIII & $42$ & $96.3\pm3.3$ & $  22.3\pm   5.7$ & $12.284\pm 0.080$\\ 
  &   & SiIII & $42$ & $159.3\pm2.1$ & $  24.0\pm   3.8$ & $12.882\pm 0.048$\\ 
  &   & SiIII & $42$ & $206.4\pm1.5$ & $  19.6\pm   1.6$ & $12.972\pm 0.036$\\ 
  &   & CIV & $24$ & $166.2\pm6.9$ & $  18.1\pm  11.6$ & $12.991\pm 0.184$\\ 
IO-AND & MS & SiII & $32$ & $-373.8\pm2.4$ & $  16.2\pm   3.9$ & $12.787\pm 0.064$\\ 
  &   & SiII & $32$ & $-250.2\pm3.6$ & $  14.4\pm   6.3$ & $12.550\pm 0.102$\\ 
  &   & SiII & $32$ & $-179.4\pm0.9$ & $  18.2\pm   1.6$ & $13.270\pm 0.026$\\ 
  &   & SiIII & $38$ & $-375.0\pm1.5$ & $  32.0\pm   2.0$ & $13.017\pm 0.024$\\ 
  &   & SiIII & $38$ & $-243.0\pm1.5$ & $  28.1\pm   2.1$ & $13.118\pm 0.024$\\ 
  &   & SiIII & $38$ & $-182.4\pm0.9$ & $  16.4\pm   1.6$ & $13.083\pm 0.029$\\ 
  &   & SiIV & $59$ & $-169.8\pm0.6$ & $  13.9\pm   1.3$ & $12.878\pm 0.022$\\ 
  &   & SiIV & $59$ & $-229.5\pm0.6$ & $  17.5\pm   0.0$ & $13.119\pm 0.021$\\ 
  &   & SiIV & $59$ & $-385.2\pm3.6$ & $  23.5\pm   5.6$ & $12.367\pm 0.078$\\ 
LBQS0107-0235 & MS & CII & $19$ & $-193.8\pm3.6$ & $  14.8\pm   6.7$ & $13.361\pm 0.107$\\ 
  &   & CII & $19$ & $-262.8\pm8.1$ & $  18.6\pm  13.4$ & $13.110\pm 0.200$\\ 
  &   & SiIII & $17$ & $-187.5\pm7.2$ & $  27.8\pm  11.8$ & $12.368\pm 0.138$\\ 
  &   & SiIII & $17$ & $-248.1\pm2.4$ & $   6.0\pm   7.1$ & $12.550\pm 0.301$\\ 
MRC2251-178 & MS & CII & $46$ & $-269.1\pm2.7$ & $  24.9\pm   4.3$ & $13.363\pm 0.059$\\ 
  &   & SiIII & $36$ & $-262.8\pm0.6$ & $  19.7\pm   1.1$ & $12.991\pm 0.018$\\ 
  &   & CIV & $38$ & $-269.7\pm0.6$ & $  21.4\pm   1.1$ & $13.799\pm 0.018$\\ 
MRK1044 & MS & CII & $26$ & $-211.2\pm3.9$ & $   8.6\pm   9.0$ & $13.008\pm 0.139$\\ 
  &   & SiIII & $28$ & $-209.4\pm2.7$ & $  28.1\pm   4.2$ & $12.581\pm 0.050$\\ 
  &   & CIV & $27$ & $-190.2\pm3.0$ & $   9.0\pm   6.5$ & $12.724\pm 0.125$\\ 
MRK1513 & MS & CII & $39$ & $-212.4\pm3.0$ & $  15.0\pm   5.4$ & $13.142\pm 0.086$\\ 
  &   & CII & $39$ & $-275.4\pm2.1$ & $  17.2\pm   3.8$ & $13.351\pm 0.058$\\ 
  &   & SiIII & $32$ & $-210.3\pm1.2$ & $  13.9\pm   2.3$ & $12.549\pm 0.038$\\ 
  &   & SiIII & $32$ & $-277.2\pm0.9$ & $  18.0\pm   1.6$ & $12.799\pm 0.026$\\ 
  &   & CIV & $25$ & $-207.0\pm5.1$ & $  14.0\pm   8.8$ & $12.721\pm 0.171$\\ 
  &   & CIV & $25$ & $-281.4\pm1.2$ & $  14.8\pm   2.7$ & $13.628\pm 0.078$\\ 
  &   & CIV & $25$ & $-310.5\pm3.6$ & $  49.2\pm   2.9$ & $13.981\pm 0.042$\\ 
MRK304 & MS & CII & $37$ & $-345.0\pm2.7$ & $  36.5\pm   4.2$ & $13.675\pm 0.042$\\ 
  &   & SiIII & $26$ & $-362.7\pm3.9$ & $  16.6\pm   3.5$ & $12.908\pm 0.129$\\ 
  &   & SiIII & $26$ & $-326.4\pm4.8$ & $  22.4\pm   4.3$ & $13.031\pm 0.099$\\ 
  &   & CIV & $20$ & $-309.3\pm5.1$ & $  28.3\pm   7.2$ & $13.766\pm 0.089$\\ 
  &   & CIV & $20$ & $-355.8\pm6.3$ & $  14.9\pm   9.5$ & $13.258\pm 0.251$\\ 
MRK335 & MS & CII & $38$ & $-333.9\pm2.4$ & $  21.9\pm   4.0$ & $13.423\pm 0.055$\\ 
  &   & CII & $38$ & $-411.9\pm2.7$ & $  19.7\pm   4.6$ & $13.308\pm 0.068$\\ 
  &   & SiIII & $36$ & $-250.5\pm2.4$ & $  15.5\pm   4.5$ & $12.280\pm 0.066$\\ 
  &   & SiIII & $36$ & $-297.0\pm1.8$ & $  10.0\pm   4.2$ & $12.407\pm 0.071$\\ 
  &   & SiIII & $36$ & $-339.9\pm1.8$ & $  22.4\pm   3.3$ & $12.758\pm 0.041$\\ 
  &   & SiIII & $36$ & $-412.2\pm2.1$ & $  22.0\pm   3.3$ & $12.519\pm 0.044$\\ 
  &   & CIV & $32$ & $-219.0\pm1.8$ & $   6.8\pm   4.3$ & $12.905\pm 0.077$\\ 
  &   & CIV & $32$ & $-258.0\pm2.1$ & $  17.2\pm   4.6$ & $13.211\pm 0.069$\\ 
  &   & CIV & $32$ & $-306.3\pm3.3$ & $  16.9\pm   5.8$ & $13.261\pm 0.129$\\ 
  &   & CIV & $32$ & $-345.9\pm9.6$ & $  21.0\pm  11.5$ & $12.961\pm 0.232$\\ 
PG0003+158 & MS & SiII & $26$ & $-318.6\pm1.5$ & $  18.3\pm   2.8$ & $13.142\pm 0.041$\\ 
  &   & SiII & $26$ & $-380.1\pm6.3$ & $  22.4\pm  10.9$ & $12.591\pm 0.144$\\ 
  &   & SiIII & $23$ & $-244.2\pm5.7$ & $  43.2\pm   8.9$ & $12.635\pm 0.070$\\ 
  &   & SiIII & $23$ & $-326.4\pm1.5$ & $  23.7\pm   2.5$ & $13.055\pm 0.037$\\ 
  &   & SiIII & $23$ & $-394.8\pm3.3$ & $  35.4\pm   4.7$ & $12.902\pm 0.045$\\ 
  &   & CIV & $24$ & $-230.7\pm1.2$ & $  13.2\pm   2.2$ & $13.218\pm 0.042$\\ 
PG0026+129 & MS & CII & $22$ & $-188.7\pm9.0$ & $  26.4\pm  15.3$ & $13.224\pm 0.179$\\ 
  &   & CII & $22$ & $-249.0\pm3.9$ & $  18.2\pm   8.2$ & $13.510\pm 0.117$\\ 
  &   & CII & $22$ & $-294.0\pm3.9$ & $  14.1\pm   7.9$ & $13.400\pm 0.136$\\ 
  &   & SiIII & $22$ & $-183.6\pm4.2$ & $  17.6\pm   7.5$ & $12.333\pm 0.107$\\ 
  &   & SiIII & $22$ & $-235.8\pm2.7$ & $  16.2\pm   5.8$ & $12.591\pm 0.080$\\ 
  &   & SiIII & $22$ & $-282.6\pm2.1$ & $  16.4\pm   3.7$ & $12.713\pm 0.056$\\ 
  &   & SiIV & $42$ & $-290.7\pm2.1$ & $   8.8\pm   4.8$ & $12.337\pm 0.078$\\ 
PG0044+030 & MS & CII & $11$ & $-214.2\pm6.0$ & $  27.1\pm   9.6$ & $13.755\pm 0.114$\\ 
  &   & CII & $11$ & $-292.8\pm3.9$ & $  23.4\pm   5.8$ & $13.930\pm 0.081$\\ 
  &   & SiIII & $14$ & $-203.7\pm3.0$ & $  14.6\pm   5.0$ & $12.678\pm 0.088$\\ 
  &   & SiIII & $14$ & $-287.4\pm3.0$ & $  37.4\pm   5.0$ & $13.202\pm 0.045$\\ 
PG2349-014 & MS & SiII & $24$ & $-291.6\pm1.8$ & $  23.6\pm   2.7$ & $13.421\pm 0.034$\\ 
  &   & SiII & $24$ & $-340.5\pm3.6$ & $  12.5\pm   6.1$ & $12.771\pm 0.114$\\ 
  &   & SiIII & $33$ & $-313.5\pm3.9$ & $  36.0\pm   3.5$ & $13.362\pm 0.089$\\ 
  &   & SiIV & $26$ & $-255.6\pm23.1$ & $  43.6\pm  19.7$ & $13.037\pm 0.278$\\ 
  &   & SiIV & $26$ & $-299.4\pm2.4$ & $  21.5\pm   4.5$ & $13.197\pm 0.182$\\ 
  &   & SiIV & $26$ & $-351.6\pm1.2$ & $   7.8\pm   3.2$ & $12.750\pm 0.050$\\ 
PHL1811 & MS & CII & $37$ & $-166.2\pm1.5$ & $  13.9\pm   2.1$ & $13.757\pm 0.038$\\ 
  &   & CII & $37$ & $-205.5\pm0.9$ & $  14.8\pm   1.6$ & $14.099\pm 0.022$\\ 
  &   & CII & $37$ & $-258.3\pm2.1$ & $  20.6\pm   3.7$ & $13.489\pm 0.050$\\ 
  &   & SiIII & $29$ & $-165.9\pm2.1$ & $  16.2\pm   2.0$ & $12.974\pm 0.061$\\ 
  &   & SiIII & $29$ & $-208.8\pm1.5$ & $  22.7\pm   3.6$ & $13.244\pm 0.050$\\ 
  &   & SiIII & $29$ & $-260.4\pm2.4$ & $  23.7\pm   2.6$ & $12.976\pm 0.051$\\ 
  &   & SiIII & $29$ & $-350.1\pm1.2$ & $  18.2\pm   2.1$ & $12.511\pm 0.033$\\ 
  &   & CIV & $41$ & $-162.3\pm0.6$ & $  17.9\pm   1.0$ & $13.941\pm 0.019$\\ 
  &   & CIV & $41$ & $-225.6\pm4.8$ & $  29.4\pm   7.2$ & $13.601\pm 0.110$\\ 
  &   & CIV & $41$ & $-282.3\pm9.0$ & $  28.6\pm  12.9$ & $13.299\pm 0.215$\\ 
  &   & CIV & $41$ & $-348.6\pm0.9$ & $  20.8\pm   1.5$ & $13.730\pm 0.027$\\ 
PHL2525 & MS & CII & $22$ & $-145.2\pm4.8$ & $  31.7\pm   8.2$ & $14.079\pm 0.103$\\ 
  &   & CII & $22$ & $-207.9\pm4.5$ & $  30.3\pm   4.5$ & $14.145\pm 0.075$\\ 
  &   & SiIII & $24$ & $-202.2\pm4.5$ & $  35.7\pm   3.9$ & $13.474\pm 0.060$\\ 
  &   & SiIII & $24$ & $-147.3\pm3.6$ & $  23.1\pm   4.0$ & $13.274\pm 0.094$\\ 
  &   & CIV & $20$ & $-219.0\pm3.9$ & $  31.4\pm   5.0$ & $13.738\pm 0.067$\\ 
  &   & CIV & $20$ & $-146.4\pm3.9$ & $  35.1\pm   6.0$ & $13.800\pm 0.063$\\ 
RBS144 & MS & SiII & $23$ & $105.0\pm0.9$ & $  22.8\pm   1.1$ & $13.862\pm 0.020$\\ 
  &   & SiII & $23$ & $178.8\pm4.5$ & $  29.6\pm   7.5$ & $12.945\pm 0.081$\\ 
  &   & SiIII & $28$ & $98.4\pm2.7$ & $  31.2\pm   3.0$ & $13.400\pm 0.048$\\ 
  &   & SiIII & $28$ & $179.4\pm1.2$ & $  18.9\pm   1.8$ & $12.899\pm 0.028$\\ 
RBS1897 & MS & SiII & $38$ & $86.4\pm1.8$ & $   7.4\pm   4.7$ & $11.941\pm 0.063$\\ 
  &   & SiII & $38$ & $131.4\pm1.8$ & $  13.2\pm   3.6$ & $12.068\pm 0.056$\\ 
  &   & SiIII & $63$ & $84.9\pm4.5$ & $  20.8\pm   7.2$ & $12.441\pm 0.359$\\ 
  &   & SiIII & $63$ & $124.8\pm19.2$ & $  38.7\pm  14.9$ & $12.579\pm 0.282$\\ 
\footnotesize{SDSSJ015530.02-085704.0} & MS & SiII & $13$ & $-131.7\pm4.5$ & $  38.7\pm   7.5$ & $12.969\pm 0.064$\\ 
  &   & SiII & $13$ & $-225.0\pm1.5$ & $  16.2\pm   2.5$ & $13.073\pm 0.054$\\ 
  &   & SiIII & $18$ & $-130.2\pm3.9$ & $  30.6\pm   5.4$ & $13.072\pm 0.070$\\ 
  &   & SiIII & $18$ & $-222.0\pm5.1$ & $  26.5\pm   8.0$ & $12.539\pm 0.098$\\ 
  &   & SiIV & $16$ & $-120.6\pm6.9$ & $  16.7\pm  12.2$ & $12.611\pm 0.194$\\ 
\footnotesize{SDSSJ234500.43-005936.0} & MS & SiII & $14$ & $-129.0\pm5.4$ & $  13.1\pm  10.3$ & $12.259\pm 0.166$\\ 
  &   & SiII & $14$ & $-282.6\pm2.1$ & $  27.7\pm   2.9$ & $13.399\pm 0.042$\\ 
  &   & SiIII & $9$ & $-130.8\pm3.0$ & $  22.2\pm   4.2$ & $12.976\pm 0.068$\\ 
  &   & SiIII & $9$ & $-266.1\pm3.9$ & $  34.3\pm   5.3$ & $13.499\pm 0.059$\\ 
  &   & SiIV & $10$ & $-270.0\pm4.2$ & $  37.5\pm   6.2$ & $13.373\pm 0.061$\\ 
UGC12163 & MS & CII & $15$ & $-349.8\pm2.4$ & $   7.9\pm   5.7$ & $13.489\pm 0.107$\\ 
  &   & CII & $15$ & $-429.3\pm3.0$ & $  34.8\pm   4.5$ & $14.033\pm 0.047$\\ 
  &   & SiIII & $10$ & $-348.3\pm5.4$ & $  18.0\pm   9.2$ & $12.401\pm 0.137$\\ 
  &   & SiIII & $10$ & $-425.4\pm1.8$ & $  25.7\pm   2.6$ & $13.292\pm 0.044$\\ 
  &   & CIV & $17$ & $-426.9\pm4.5$ & $  27.1\pm   6.7$ & $13.496\pm 0.087$\\ 
  &   & CIV & $17$ & $-351.0\pm8.7$ & $  17.6\pm  14.3$ & $12.933\pm 0.238$\\
ESO265-G23 & LA & SiII & $12$ & $191.4\pm2.4$ & $  10.1\pm   4.0$ & $13.351\pm 0.091$\\ 
  &   & SiII & $12$ & $237.3\pm0.0$ & $  27.4\pm  10.8$ & $13.043\pm 0.124$\\ 
  &   & SiIII & $11$ & $218.7\pm3.9$ & $  46.5\pm   5.6$ & $13.329\pm 0.043$\\ 
  &   & SiIV & $14$ & $202.5\pm7.5$ & $  29.0\pm  11.5$ & $12.820\pm 0.133$\\ 
H1101-232 & LA & SiII & $22$ & $89.4\pm1.5$ & $  17.2\pm   2.4$ & $13.832\pm 0.055$\\ 
  &   & SiII & $22$ & $143.1\pm3.6$ & $  38.1\pm   3.6$ & $13.904\pm 0.043$\\ 
  &   & SiIII & $20$ & $87.6\pm6.3$ & $  27.0\pm  12.3$ & $13.355\pm 0.331$\\ 
  &   & SiIII & $20$ & $136.8\pm18.0$ & $  49.9\pm  10.9$ & $13.563\pm 0.198$\\ 
  &   & SiIV & $21$ & $75.9\pm4.2$ & $  36.4\pm   7.2$ & $13.205\pm 0.064$\\ 
HE1159-1338 & LA & SiII & $12$ & $162.3\pm3.3$ & $  18.3\pm   5.8$ & $13.106\pm 0.080$\\ 
  &   & SiII & $12$ & $208.5\pm2.7$ & $  12.2\pm   4.9$ & $13.087\pm 0.078$\\ 
  &   & SiIII & $11$ & $147.3\pm3.6$ & $  29.1\pm   6.2$ & $13.125\pm 0.063$\\ 
  &   & SiIII & $11$ & $210.0\pm4.2$ & $  17.0\pm   6.6$ & $12.710\pm 0.107$\\ 
IRAS\_F09539-0439 & LA & SiII & $24$ & $158.4\pm2.1$ & $  48.2\pm   2.5$ & $14.221\pm 0.023$\\ 
PG1011-040 & LA & SiII & $35$ & $131.4\pm0.6$ & $  19.5\pm   0.7$ & $13.830\pm 0.023$\\ 
  &   & SiII & $35$ & $211.5\pm2.4$ & $  12.8\pm   4.7$ & $12.233\pm 0.078$\\ 
  &   & SiIII & $31$ & $118.2\pm1.8$ & $  42.8\pm   2.2$ & $13.508\pm 0.019$\\ 
  &   & SiIII & $31$ & $214.2\pm1.2$ & $   5.8\pm   3.4$ & $12.572\pm 0.185$\\ 
PG1049-005 & LA & SiIII & $15$ & $126.9\pm6.3$ & $  36.9\pm   9.2$ & $13.442\pm 0.095$\\ 
  &   & SiIII & $15$ & $204.6\pm5.7$ & $  30.2\pm  10.9$ & $12.888\pm 0.081$\\ 
  &   & CIV & $13$ & $154.2\pm12.6$ & $  26.0\pm  19.4$ & $13.176\pm 0.260$\\ 
  &   & CIV & $13$ & $211.5\pm10.5$ & $  24.1\pm  14.6$ & $13.224\pm 0.222$\\ 
PKS1101-325 & LA & SiII & $20$ & $290.4\pm14.7$ & $  34.1\pm  22.8$ & $12.517\pm 0.237$\\ 
  &   & SiIII & $17$ & $244.5\pm0.0$ & $  38.7\pm  19.9$ & $12.457\pm 0.176$\\ 
PKS1136-13 & LA & CII & $22$ & $184.2\pm1.8$ & $  14.5\pm   3.3$ & $13.614\pm 0.054$\\ 
\footnotesize{SDSSJ095915.60+050355.0} & LA & SiIII & $18$ & $292.8\pm2.4$ & $  12.1\pm   4.3$ & $12.513\pm 0.074$\\ 
\footnotesize{UVQSJ101629.20-315023.6} & LA & SiII & $24$ & $209.1\pm5.7$ & $  32.5\pm   0.0$ & $12.757\pm 0.098$\\ 
  &   & SiII & $24$ & $252.6\pm3.0$ & $  11.6\pm   5.6$ & $12.472\pm 0.119$\\ 
  &   & SiIII & $8$ & $155.7\pm5.7$ & $  13.7\pm  10.4$ & $12.552\pm 0.166$\\ 
  &   & SiIII & $8$ & $223.5\pm4.8$ & $  29.5\pm   7.6$ & $13.020\pm 0.087$\\ 
\enddata
\label{tab:bigtable}
\end{deluxetable*}